\documentclass[leqno,draft,11pt,a4paper]{article}
\usepackage{amssymb,amsmath,theorem}
\usepackage[includeheadfoot,margin=1in]{geometry}

\allowdisplaybreaks[2]

\newcommand\M{\mathit}
\newcommand\delim[4]{\ifx X#3X\left#1#4\right#2\else\csname#3l\endcsname#1#4\csname#3r\endcsname#2\fi}

\newcommand\eq{\leftrightarrow}
\newcommand\model{\vDash}
\newcommand\lang{\mathcal L}

\newcommand\fii{\varphi}
\newcommand\ep{\varepsilon}

\newcommand\p[1]{\langle#1\rangle}
\newcommand\abs[1]{\lvert#1\rvert}
\newcommand\Abs[2][]{\delim||{#1}{#2}}
\newcommand\dlh[1]{\lVert#1\rVert}
\newcommand\Dlh[2][]{\delim\lVert\rVert{#1}{#2}}
\newcommand\bez{\smallsetminus}
\newcommand\sset{\subseteq}
\newcommand\Sset{\supseteq}
\newcommand\res{\mathbin\restriction}

\newcommand\fl[1]{\lfloor#1\rfloor}
\newcommand\Fl[2][]{\delim\lfloor\rfloor{#1}{#2}}
\newcommand\cl[1]{\lceil#1\rceil}
\newcommand\CL[2][]{\delim\lceil\rceil{#1}{#2}}
\newcommand\fcl[1]{\lfloor#1\rceil}
\newcommand\FCL[2][]{\delim\lfloor\rceil{#1}{#2}}
\newcommand\rsuv{\M{RSUV}}
\newcommand\fafac[2]{{#1}^{\underline{#2}}_{\mathstrut}}

\newcommand\down{{\downarrow}}

\DeclareMathOperator\dom{dom}
\DeclareMathOperator\im{im}
\DeclareMathOperator\bit{bit}
\DeclareMathOperator\card{card}
\DeclareMathOperator\E{E}
\DeclareMathOperator\sgn{sgn}
\DeclareMathOperator\dist{dist}
\let\Re\relax
\let\Im\relax
\DeclareMathOperator\Re{Re}
\DeclareMathOperator\Im{Im}
\DeclareMathOperator\sech{sech}
\DeclareMathOperator\csch{csch}
\DeclareMathOperator\arccot{arccot}
\DeclareMathOperator\arcsec{arcsec}
\DeclareMathOperator\arccsc{arccsc}
\DeclareMathOperator\arsinh{arsinh}
\DeclareMathOperator\arcosh{arcosh}
\DeclareMathOperator\artanh{artanh}
\DeclareMathOperator\arcoth{arcoth}
\DeclareMathOperator\arsech{arsech}
\DeclareMathOperator\arcsch{arcsch}
\DeclareMathOperator\cod{cod}

\newcommand\cxt[1]{\mathrm{#1}}
\newcommand\tc{\cxt{TC}^0}
\newcommand\LT{\cxt{DLOGTIME}}

\newcommand\thry[1]{\mathsf{#1}}
\newcommand\sig{\Sigma^b_}
\newcommand\Sig{\Sigma^B_}
\newcommand\io{\thry{IOpen}}
\newcommand\dicr{\thry{\Delta^b_1\text{-}CR}}
\newcommand\vtc{\thry{VTC^0}}
\newcommand\comp{\thry{COMP}}

\newcommand\N{\mathbb N}
\newcommand\Q{\mathbb Q}
\newcommand\Z{\mathbb Z}
\newcommand\RR{\mathbb R}
\newcommand\strc[1]{\mathbf{#1}}
\newcommand\MN{\strc N}
\newcommand\ML{\strc L}
\newcommand\MZ{\strc Z}
\newcommand\MQ{\strc Q}
\newcommand\MR{\strc R}
\newcommand\MC{\strc C}
\newcommand\MZL{\MZ_\ML}
\newcommand\MQL{\MQ_\ML}
\newcommand\MRL{\MR_\ML}
\newcommand\MCL{\MC_\ML}
\newcommand\DML{\down\ML}

\newcommand\sM{\mathfrak M}

\newcommand\ob[1]{\overline{#1}}
\newcommand\txto{${}\to{}$}

\newcommand\bme{\hskip.75em\relax}
\newcommand\noproof{\leavevmode\unskip\bme\vadjust{}\nobreak\hfill$\qed$\par}
\newcommand\qed{\Box}
\newenvironment{Pf}[1][]
  {\par\noindent\textit{Proof\optpar{#1}:}\bme\ignorespaces}
  {\noproof\pagebreak[2]\vskip\medskipamount\ignorespacesafterend}
\newcommand\optpar[1]{\ifx\relax#1\relax\else\ #1\fi}
\newcommand\qedhere{\relax\ifmmode\eqno\qed\expandafter\aftergroup
                   \else\noproof\fi\noqed}
\newcommand\noqed{\let\noproof\relax}

\theoremstyle{plain}
\newtheorem{Thm}{Theorem}[section]
\newtheorem{Prop}[Thm]{Proposition}
\newtheorem{Cor}[Thm]{Corollary}
\newtheorem{Lem}[Thm]{Lemma}
\newtheorem{Que}[Thm]{Question}
\theorembodyfont\upshape
\newtheorem{Def}[Thm]{Definition}
\newtheorem{Rem}[Thm]{Remark}

\author{Emil Je\v r\'abek\\[\medskipamount]
Institute of Mathematics, Czech Academy of Sciences\\
\small \v Zitn\'a 25,
115\:67 Praha 1,
Czech Republic,
email: \texttt{jerabek@math.cas.cz}}

\title{Elementary analytic functions in $\vtc$}

\begin{document}
\maketitle

\begin{abstract}
It is known that rational approximations of elementary analytic functions ($\exp$, $\log$, trigonometric, and
hyperbolic functions, and their inverse functions) are computable in the weak complexity class $\tc$. We show how to
formalize the construction and basic properties of these functions in the corresponding theory of bounded arithmetic,
$\vtc$.


\end{abstract}

\section{Introduction}

The complexity class%
\footnote{Originally defined by Hajnal et al.~\cite{tc0} in a non-uniform setting, but in this paper we always mean the
$\LT$-uniform version of the class, which gives a robust notion of ``fully uniform'' $\tc$ with several equivalent
definitions across various computational models (cf.~\cite{founif}).}
$\tc$ is a weak subclass of polynomial time and logarithmic space; we can think of~$\tc$, conflated with the
corresponding function class, as the complexity class of elementary integer arithmetic operations: $+$, $-$, $\cdot$,
$\fl{x/y}$, and~$<$ are $\tc$-computable, with $\cdot$ and~$\fl{x/y}$ being $\tc$-complete (under $\cxt{AC}^0$ Turing
reductions). Iterated addition $\sum_{i<n}x_i$ and multiplication $\prod_{i<n}x_i$ are also $\tc$-complete. (The
$\tc$-computability of $\prod_{i<n}x_i$ and $\fl{x/y}$ was a difficult problem, finally settled by Hesse, Allender, and
Barrington~\cite{hab}.) Apart from integers, $\tc$ can compute the corresponding operations in various related
structures: $\Q$, $\Q(i)$, and other number fields, or polynomial rings. Using iterated sums and products, $\tc$ can
compute approximations of analytic functions given by power series with $\tc$-computable coefficients
\cite{reif,reif-tate,mac-the:ser,hab}, such as the \emph{elementary analytic functions} \cite{abr-steg,nist:fun}: $\exp$, $\log$, trigonometric,
hyperbolic, inverse trigonometric, and inverse hyperbolic functions.

One of the basic themes in proof complexity is that for many complexity classes~$C$, we can associate to~$C$ a theory of
bounded arithmetic~$T$ whose reasoning power is captured by~$C$: the axiom schemata of~$T$ that provide the bulk of its
deductive capabilities (induction, comprehension, minimization, \dots) are postulated for formulas that express
predicates computable in~$C$, while the provably total computable functions of~$T$ (of suitable syntactic shape) are
exactly the $C$-functions. We may consider $T$ to be a formalization of \emph{feasible reasoning} of complexity~$C$:
what properties of concepts from~$C$ are derivable if we restrict our deductions to only use $C$-predicates and
$C$-computable objects, shunning any higher-level reasoning?

In this paper, we are interested in feasible reasoning of complexity~$\tc$. The basic theory of bounded arithmetic
corresponding to~$\tc$ is the Zambella-style two-sorted theory $\vtc$ introduced by Nguyen and Cook~\cite{ngu-cook},
or equivalently (up to the $\rsuv$ isomorphism), the Buss-style one-sorted theory%
\footnote{Earlier, Johannsen and Pollett~\cite{joh-pol:c02} defined a theory~$\thry{C^0_2}$ that might be more
convenient to work with; $\thry{C^0_2}$ is a $\forall\exists\sig1$-conservative extension of $\dicr$.
Johannsen~\cite{joh:c02div} introduced an extension $\thry{C^0_2[div]}$ of~$\thry{C^0_2}$, which is however essentially identical to~$\thry{C^0_2}$ by results of~\cite{ej:vtcimul}.}
$\dicr$ of Johannsen and Pollett~\cite{joh-pol:d1cr}. It turns out that $\vtc$ is quite powerful when it
comes to proving properties of $\tc$-computable arithmetic operations, even though this can be rather challenging
to prove. Notably, as shown in Je\v r\'abek~\cite{ej:vtcimul} by formalizing a variant of the
Hesse--Allender--Barrington algorithm, $\vtc$ proves the existence of iterated products $\prod_{i<n}x_i$ satisfying the
defining recurrence
\[\prod_{i<0}x_i=1,\qquad\prod_{i<n+1}x_i=x_n\prod_{i<n}x_i,\]
and of integer division satisfying $y\fl{x/y}\le x<y\bigl(\fl{x/y}+1\bigr)$. Earlier, Je\v r\'abek~\cite{ej:vtc0iopen}
formalized in $\vtc$ (augmented with an iterated multiplication axiom, redundant by~\cite{ej:vtcimul}) approximation of
complex roots of constant-degree univariate polynomials; as a consequence, $\vtc$ includes $\io$ (quantifier-free
induction in the language of ordered rings) on the binary number sort, and even the $\rsuv$ translation of $\sig0$
induction and minimization in Buss's language. 

We continue the investigation of the power of $\vtc$, shifting the focus from integer (or rational) operations to real
and complex analytic functions. As we already mentioned, a vast number of such functions can be approximated by
$\tc$~functions, and it's not clear to what extent we can develop a general theory of such functions in~$\vtc$; in this
paper, we start with the most notorious examples---the elementary analytic functions: $\exp$, trigonometric
functions ($\sin$, $\cos$, $\tan$, \dots), hyperbolic functions ($\sinh$, $\cosh$, $\tanh$, \dots), and their inverse
functions ($\log$, $\arcsin$, $\arsinh$, \dots). We mostly concentrate on complex $\exp$ and $\log$, as the other
functions can be defined in terms of these.

It would be extremely laborious to work directly in the language of~$\vtc$ all the time, expressing everything in terms
of rational approximations. We follow a different approach---we present the constructions and arguments
model-theoretically, considering an extension of a given model $\sM\model\vtc$ to a larger structure where we can
define the elementary analytic functions properly as bona fide functions: the model itself gives the ordered ring of
``integers'' $\MZ^\sM$, its fraction field is the ordered field of ``rationals'' $\MQ^\sM$, and the \emph{completion}
of $\MQ^\sM$ (in the sense of ordered, topological, or valued field theory) gives the ``reals'' $\MR^\sM$ and ``complex numbers''
$\MC^\sM=\MR^\sM(i)$. (The completion $\MR^\sM$ was already used as a technical tool in~\cite{ej:vtc0iopen}, but here
we make it the central structure of interest, along with $\MC^\sM$.) We still need to consider rational approximations
so that we have a way of translating our results back into the language of~$\vtc$, and in particular, so that we can
refer to the newly constructed functions in more sophisticated arguments that employ induction or related axiom schemata
of~$\vtc$, which only hold for properties expressible by $\tc$~formulas in~$\sM$.

The very fact that rational or Gaussian rational (i.e., $\Q(i)$) approximations of $\exp$ and $\log$ on suitable
domains are $\tc$-computable ensures that they are representable as provably total computable functions in~$\vtc$. But
by itself, this only means that for each standard rational input, $\vtc$ proves that the function has the right
value, which is a very low bar to clear: e.g., it does not even imply that the approximations converge to a unique
real or complex value. What we really need is that the functions can be represented in such a way that $\vtc$ proves
the most fundamental properties they have in the real world.

What these properties are is a judgement call. We consider the most salient properties of $\exp$ to be the identity
$\exp(z+w)=\exp z\exp w$, and the shape of its domain, codomain, and preimages: real $\exp$ is an increasing
bijection from $\MRL^\sM$ (the logarithmically bounded reals) onto $\MR_{>0}^\sM$; complex $\exp$ maps
$\MRL^\sM+i\MR^\sM$ to $\MC_{\ne0}^\sM$, and there is a constant $\pi$ such that $\exp$ is $2\pi i$-periodic and
maps $\MRL^\sM+i(-\pi,\pi]$ bijectively onto $\MC_{\ne0}^\sM$. (Actually, we also define $\exp z$ when $\Re z$ is
negative, but not logarithmically bounded, putting $\exp z=0$.) Of these, the most difficult to prove will be the
surjectivity of $\exp$, including the existence of~$\pi$; we will need to construct $\log$ to prove this. The main
properties of $\log$ are that it is a bijection from $\MC_{\ne0}^\sM$ onto $\MRL^\sM+i(-\pi,\pi]$, and a right inverse
of $\exp$, i.e., $\exp\log z=z$ (which implies the surjectivity of $\exp$, as mentioned).

Our construction of $\exp$ is fairly straightforward, using the common power series (though we will need the existence
of~$\pi$ and the $2\pi i$-periodicity of $\exp$ to extend its domain from $\MRL^\sM+i\MRL^\sM$ to $\MRL^\sM+i\MR^\sM$).
The proof of $\exp(z+w)=\exp z\exp w$ is not very difficult either. The construction of $\log$ is much more
complicated, as a power series only defines it on a neighbourhood of~$1$; we will need to extend it in several stages
to eventually define it on all of $\MC_{\ne0}^\sM$. It will also take us a lot of work to prove the key right-inverse
property $\exp\log z=z$: the basic strategy of our argument is to show $\log zw=\log z+\log w$ under suitable
restrictions on $z$ and~$w$, which ensures that $\log\exp z$ obeys Cauchy's functional equation $\log\exp(z+w)=\log\exp
z+\log\exp w$ (again, under certain conditions on $z,w$);
coupled with the asymptotic expansion of $\exp$ near~$0$ and $\log$ near~$1$, we will derive $\log\exp z=z$ for small
enough~$z$, and use the injectivity of $\log$ to infer $\exp\log z=z$.

After we finish with $\exp$ and $\log$, we proceed to define and show basic properties of complex powering $z^w$ (with
$\sqrt[n]z$ as a special case), iterated multiplication $\prod_{j<n}z_j$ for sequences of Gaussian
rationals~$z_j\in\MQ^\sM(i)$, and last but not least, the promised hyperbolic, trigonometric, inverse hyperbolic, and
inverse trigonometric functions.

Our principal motivation for developing the theory of $\exp$, $\log$, and other elementary analytic functions in $\vtc$
is that it is intrinsically interesting. However, we also have one specific application concerning models of arithmetic
in mind. Recall that an \emph{integer part} of an ordered field~$R$ is a discretely ordered subring~$D$ such that all
elements of~$R$ can be approximated within distance~$1$ in~$D$; Shepherdson~\cite{sheph} proved that a model of
arithmetic is an integer part of a real-closed field iff it satisfies $\io$.

An (ordered) \emph{exponential field} is an ordered field $\p{R,+,\cdot,{<}}$ endowed with an ordered group isomorphism
${\exp}\colon\p{R,+,{<}}\to\p{R_{>0},\cdot,{<}}$. Ressayre~\cite{ress:eip} introduced the notion of an
\emph{exponential integer part} (EIP) of an exponential field $\p{R,\exp}$, which is essentially an integer part of $R$
whose positive part is closed under~$\exp$ (here, we should think of $\exp$ as $2^x$ rather than the usual~$e^x$). In
view of Shepherdson's characterization, we may wonder when a model of arithmetic is an EIP of a real-closed exponential
field (RCEF), and in particular, whether this implies $\thry{EXP}$ (the totality of the usual $2^n$ function) or at
least some nontrivial consequences of $\thry{EXP}$. Note that the definition of EIP does not require $\exp$ to extend
the usual~$2^n$.

Using our results on the construction of $\exp$, we can show that the property of being an EIP of a RCEF has few (if
any) first-order consequences: every countable model of $\vtc$ is an EIP of a RCEF, and every model of $\vtc$ has an
elementary extension to an EIP of a RCEF. Notice that this is still nontrivial: while an $\sM\model\vtc$ is an integer
part of $\MR^\sM$ which is a real-closed field, the natural $\exp$ or $2^x$ function we construct is an isomorphism
$\p{\MRL^\sM,+,<}\simeq\p{\MR_{>0}^\sM,\cdot,<}$ rather than $\p{\MR^\sM,+,<}\simeq\p{\MR_{>0}^\sM,\cdot,<}$, hence
there is additional work needed. Since this is somewhat tangential to the main part of the present---already
long---article, we relegated these results to the follow-up paper~\cite{ej:vtceip}.

This paper is organized as follows. After this Introduction, Section~\ref{sec:preliminaries} includes preliminaries on
$\vtc$, its models, and approximation of real-valued functions. Section~\ref{sec:expon-logar} is the core of the paper
in which we construct $\exp$ and $\log$ and prove their fundamental properties: it starts with a summary of the main
results, followed by a construction of $\exp$ in Section~\ref{sec:exponential} and a construction of $\log$ in several
steps in Sections \ref{sec:logarithm-near-1}--\ref{sec:full-compl-logar}. In Section~\ref{sec:compl-powers-iter}, we
introduce complex powering and iterated multiplication of Gaussian rationals, and in
Section~\ref{sec:trig-hyperb-funct}, we treat trigonometric, hyperbolic, inverse trigonometric, and inverse hyperbolic
functions. Concluding remarks are presented in Section~\ref{sec:conclusion}. Appendix~\ref{sec:deta-constr-tc} gives
the formal details of proofs of the existence of $\tc$~approximations of $\exp$ and~$\log$.

\section{Preliminaries}\label{sec:preliminaries}

We work with two-sorted (second-order) theories of bounded arithmetic in the style of Zambella~\cite{zamb:notes}. Our
main reference for these theories is Cook and Nguyen~\cite{cook-ngu}, including a detailed treatment of~$\vtc$,
however we present the main definitions here in order to fix notation.

The language $\lang_2=\p{0,S,+,\cdot,\le,\in,\dlh\cdot}$ of two-sorted bounded arithmetic is a first-order language with
equality with two sorts of variables, one for natural numbers (called \emph{small} or \emph{unary} numbers), and one
for finite sets of small numbers, which can also be interpreted as \emph{large} or \emph{binary} numbers so that $X$
represents $\sum_{u\in X}2^u$. The standard convention is that variables of the first sort are written with lowercase
letters $x,y,z,\dots$, and variables of the second sort with uppercase letters $X,Y,Z,\dots$; while we adhere to this
convention in the introductory material here, we will not follow it in the rest of the paper (we will mostly work with
binary numbers of various kind, and generally write them all in lower case in accordance with common mathematical
practice). The symbols $0,S,+,\cdot,\le$ of $\lang_2$ denote the usual arithmetic operations and relation on the unary
sort; $x\in X$ is the elementhood predicate, also written as $X(x)$, and the intended meaning of the $\dlh X$ function
is the least unary number strictly greater than all elements of~$X$. This function is usually denoted as~$|X|$, however
we reserve the latter symbol for the absolute value function, which we will use much more often. We write $x<y$ as an
abbreviation for $x\le y\land x\ne y$.

Bounded quantifiers are introduced by
\begin{align*}
\exists x\le t\:\fii&\iff\exists x\:(x\le t\land\fii),\\
\exists X\le t\:\fii&\iff\exists X\:\bigl(\dlh X\le t\land\fii\bigr),
\end{align*}
where $t$ is a term of unary sort not containing $x$ or~$X$ (respectively), and similarly for universal bounded quantifiers. A
formula is $\Sig0$ if it contains no second-order quantifiers, and all its first-order quantifiers are bounded. A
formula is $\Sig i$ if it consists of $i$ alternating blocks of bounded quantifiers, the first of which is existential,
followed by a $\Sig0$ formula.

The theory~$\thry{V^0}$ in~$\lang_2$ can be axiomatized by the basic axioms
\begin{align*}
&x+0=x&&x+Sy=S(x+y)\\
&x\cdot0=0&&x\cdot Sy=x\cdot y+x\\
&Sy\le x\to y<x&&\dlh X\ne0\to\exists x\:\bigl(x\in X\land\dlh X=Sx\bigr)\\
&x\in X\to x<\dlh X&&\forall x\:(x\in X\eq x\in Y)\to X=Y
\end{align*}
and the comprehension schema
\[\tag{$\fii\text{-}\comp$} \exists X\le x\:\forall u<x\:\bigl(u\in X\eq\fii(u)\bigr)\]
for $\Sig0$ formulas~$\fii$, possibly with parameters not shown (but with no occurrence of~$X$). We denote the set~$X$
whose existence is postulated by $\fii\text{-}\comp$ as $\{u<x:\fii(u)\}$. Using $\comp$, $\thry{V^0}$ proves the (unary number)
induction and minimization schemata
\begin{align}
\tag{$\fii\text{-}\thry{IND}$} &\fii(0)\land\forall x\:\bigl(\fii(x)\to\fii(x+1)\bigr)\to\forall x\:\fii(x),\\
\tag{$\fii\text{-}\thry{MIN}$} &\fii(x)\to\exists y\:\bigl(\fii(y)\land\forall z<y\:\neg\fii(z)\bigr)
\end{align}
for $\Sig0$ formulas $\fii$.

Following~\cite{cook-ngu}, a set~$X$ codes a sequence (indexed by small numbers) of sets whose $u$th element is
$X^{[u]}=\bigl\{x:\p{u,x}\in X\bigr\}$, where $\p{x,y}=\frac12(x+y)(x+y+1)+y$. Likewise, we can code sequences of small
numbers using $X^{(u)}=\dlh{X^{[u]}}$. While we stick to the official notation in formal contexts such as when stating
axioms, elsewhere we will generally write $X=\p{X_i:i<n}$ to indicate that $X$ codes a sequence of length~$n$ whose
$i$th element is~$X_i$. The theory $\vtc$ extends $\thry{V^0}$ by the axiom
\[\forall n,X\:\exists Y\:\bigl[Y^{(0)}=0\land\forall i<n\:\bigl((i\notin X\to Y^{(i+1)}=Y^{(i)})
     \land(i\in X\to Y^{(i+1)}=Y^{(i)}+1)\bigr)\bigr],\]
asserting that for any set~$X$, there is a sequence~$Y$ supplying the counting function
$Y^{(i)}=\card(X\cap\{0,\dots,i-1\})$.

$\tc$ was originally introduced by Hajnal et al.~\cite{tc0} as a non-uniform class, but we define it as the class of
languages $L\sset\{0,1\}^*$ recognizable by a $\LT$-uniform family of polynomial-size constant-depth circuits
using~$\neg$ and unbounded fan-in $\land$, $\lor$, and Majority gates; equivalently, it consists of languages
computable by $O(\log n)$-time threshold Turing machines with $O(1)$ thresholds, or by constant-time TRAM with
polynomially many processors \cite{par-sch}. In terms of descriptive complexity, a language is in~$\tc$ iff the
corresponding class of finite structures is definable in $\thry{FOM}$, first-order logic with majority
quantifiers~\cite{founif}.

In connection with bounded arithmetic, it is convenient to consider not just the complexity of languages, but of
predicates $P(X_1,\dots,X_n,x_1,\dots,x_m)$ with several inputs, where $X_i\in\{0,1\}^*$ as usual, and $x_i\in\N$ are
written in unary. It is straightforward to generalize $\tc$ and similar classes to this context, see \cite[\S
IV.3]{cook-ngu} for details. Likewise, we consider computability of functions. A function
$F\colon(\{0,1\}^*)^n\times\N^m\to\{0,1\}^*$ is a $\tc$ function if
$\dlh{F(X_1,\dots,X_n,x_1,\dots,x_m)}\le p\bigl(\dlh{X_1},\dots,x_1,\dots\bigr)$ for some polynomial $p$, and the
bit-graph $\bigl\{\p{\vec X,\vec x,i}:\bit\bigl(F(\vec X,\vec x),i\bigr)=1\bigr\}$ is a $\tc$ predicate; a unary number
function $f\colon(\{0,1\}^*)^n\times\N^m\to\N$ is a $\tc$ function if
$f(\vec X,\vec x)\le p\bigl(\dlh{X_1},\dots,x_1,\dots\bigr)$, and the graph
$\bigl\{\p{\vec X,\vec x,y}:f(\vec x,\vec X)=y\bigr\}$ is $\tc$. The class of $\tc$ functions is denoted $\cxt{FTC}^0$.
We note that by results of~\cite{hab}, class~$\mathcal K$ of Constable~\cite{const} consists exactly of $\tc$ functions
$(\{0,1\}^*)^n\to\{0,1\}^*$ where the inputs and output are interpreted as natural numbers written in binary.

All $\tc$ functions have provably total $\Sig1$ definitions in $\vtc$. More precisely, as shown in \cite[\S
IX.3]{cook-ngu}, $\vtc$ has a conservative extension $\ob\vtc$ by $\Sig1$-definable functions such that every $\tc$
function is represented by a function symbol in $\ob\vtc$, and $\ob\vtc$ proves comprehension, induction, and
minimization for $\Sig0$ formulas in the expanded language of $\ob\vtc$; we will call such formulas \emph{$\tc$
formulas}, and identify $\ob\vtc$ with $\vtc$, using $\tc$ functions freely when working in $\vtc$ or in its models.

$\vtc$ can define (as $\tc$ functions) $+$, $-$, $\cdot$, and $<$  on binary
natural numbers, and prove that they form a non-negative part of a discretely ordered ring; in fact, they satisfy $\io$
(induction for open formulas in the language of ordered rings, which entails integer division) by the results of
\cite{ej:vtc0iopen,ej:vtcimul}.

If $\sM\model\vtc$, we denote by $\p{\MN^\sM,0,1,{+},{\cdot},{<}}$ the second sort of~$\sM$ interpreted as a set of
binary natural numbers along with its arithmetic structure. We extend it with negative numbers to form the
discretely ordered ring $\p{\MZ^\sM,0,1,{+},{\cdot},{<}}$ (the \emph{integers of~$\sM$}). Let
$\p{\MQ^\sM,0,1,{+},{\cdot},{<}}$ (the \emph{rationals of~$\sM$}) be the fraction field of $\MZ^\sM$, let
$\p{\MR^\sM,0,1,{+},{\cdot},{<}}$ (the \emph{reals of~$\sM$}) be the \emph{completion} (see below for more details) of
$\MQ^\sM$, which is a real-closed field by~\cite{ej:vtc0iopen,ej:vtcimul}, and let $\p{\MC^\sM,0,1,{+},{\cdot}}$ (the
\emph{complex numbers of~$\sM$}) be its algebraic closure, i.e., $\MC^\sM=\MR^\sM(i)$ where $i^2=-1$. We also consider
the field $\MQ^\sM(i)$ of \emph{Gaussian rationals of~$\sM$}. The structures $\MZ^\sM$, $\MQ^\sM$, and $\MQ^\sM(i)$ are
interpretable in~$\sM$, by formulas independent of~$\sM$ (albeit with non-absolute equality in the cases of $\MQ^\sM$ and
$\MQ^\sM(i)$, as we do not know how to reduce fractions to lowest terms in~$\vtc$), but in general, $\MR^\sM$ and
$\MC^\sM$ are not (e.g., it is easy to show that $\MR^\sM$ is always uncountable; moreover, if $\sM$ has countable
cofinality, then $\abs{\MR^\sM}=\abs\sM^\omega$).

The completion of an ordered field $\p{F,+,\cdot,<}$ can be described in several equivalent ways. One way using only
the basic structure of ordered fields is as follows (cf.~\cite{scott-cof}). A \emph{cut} in $F$ is a pair $\p{A,B}$ of
sets such that $F=A\cup B$, $\inf\{b-a:b\in B,a\in A\}=0$, and $A$ has no largest element; $F$ is \emph{complete} if
$\min B$ exists for every cut $\p{A,B}$. The \emph{completion} of $F$ is a complete ordered field
$\p{\hat F,+,\cdot,<}$ such that $F$ is a dense subfield of $\hat F$ (i.e., every non-degenerate interval of $\hat F$
intersects $F$). The completion of $F$ is unique up to $F$-isomorphism; it can be explicitly constructed by endowing
the set of all cuts of~$F$ with suitable structure.

We will most often use a topological description of~$\hat F$ (see~\cite{wer:topf}). The interval topology makes $F$ a
topological field, and therefore a uniform space\footnote{We require all uniform spaces and topological groups to be
Hausdorff.} with a fundamental system of entourages $\mathcal U=\{U_\ep:\ep\in F_{>0}\}$, where
$U_\ep=\{\p{x,y}\in F^2:\abs{x-y}\le\ep\}$. $F$ is \emph{complete} as a uniform space if every Cauchy net in $F$
converges. Here, a \emph{net} is an indexed set $A=\{a_i:i\in I\}\sset F$ where $\p{I,\le}$ is a directed poset; $A$ is
a \emph{Cauchy net} if for every $U\in\mathcal U$, there exists $i_0\in I$ such that $\p{a_i,a_j}\in U$ for all
$i,j\ge i_0$, and $A$ \emph{converges} to $a\in F$, written $a=\lim_{i\in I}a_i$, if for every $U\in\mathcal U$, there
exists $i_0\in I$ such that $\p{a_i,a}\in U$ for all $i\ge i_0$. (In our applications, $I$ will usually be a totally
ordered set such as $\p{\ML^\sM,\le}$.) The \emph{completion} of $F$ is a complete uniform space $\hat F$ such that $F$
is a (topologically) dense subspace of~$\hat F$; it is again unique up to $F$-isomorphism. The key property of $\hat F$
is that every uniformly continuous function from $F$ to a complete uniform space~$S$ extends uniquely to a uniformly
continuous function $\hat F\to S$. The ring operations on $F$ extend to continuous operations on~$\hat F$ that make it
a topological ring. For ordered fields~$F$, the completion $\hat F$ is in fact an ordered field, and coincides with the
order-theoretic completion of~$F$ as above.

Apart from the ordered fields $\MQ^\sM$ and~$\MR^\sM$, we will also consider $\MQ^\sM(i)$ and $\MC^\sM$ as topological
fields; in particular, $\MC^\sM$ is the completion of $\MQ^\sM(i)$. Consequently, a Cauchy net in $\MQ^\sM(i)$ has a
unique limit in~$\MC^\sM$, and any uniformly continuous function $D\to\MC^\sM$, $D\sset\MQ^\sM(i)$, has a unique
uniformly continuous extension to a function $\ob D\to\MC^\sM$; we will commonly use these facts.

If an ordered field $F$ is archimedean (which for our $\MQ^\sM$ happens only when $\sM$ is the standard model), it
embeds in~$\RR$, and its completion is just~$\RR$. Otherwise, $F$ is a \emph{valued field} with valuation ring
$\{x\in F:\exists n\in\N\,\abs x\le n\}$, and $\hat F$ can be described as the valued field completion of~$F$;
see~\cite{engl-pres} and \cite[\S6]{ej:vtc0iopen} for details.

The unary number sort of~$\sM$ embeds (via a $\tc$ function) into $\MN^\sM$ as an initial segment of \emph{logarithmic
numbers}, which we denote $\ML^\sM$. We define the \emph{logarithmically bounded} integers, reals, etc., by
\[\MCL^\sM=\{z\in\MC^\sM:\exists n\in\ML^\sM\:\abs z\le n\},\]
$\MRL^\sM=\MR^\sM\cap\MCL^\sM$, $\MQL^\sM=\MQ^\sM\cap\MCL^\sM$, and $\MZL^\sM=\MZ^\sM\cap\MCL^\sM$. Here, the complex
absolute value $\abs z=\sqrt{x^2+y^2}=\sqrt{z\ob z}$ for $z=x+iy\in\MC^\sM$ is a well-defined element of~$\MR^\sM$
as the latter is real-closed. However, notice that the definition of $\MCL^\sM$ would not change if we used $\abs x+\abs
y$, $\max\bigl\{\abs x,\abs y\bigr\}$, or $x^2+y^2$ in place of $\abs z$.

We also write
$\MR^\sM_{>0}=\{x\in\MR^\sM:x>0\}$, $\MC^\sM_{\ne0}=\{z\in\MC^\sM:z\ne0\}$, etc.
We define the open and closed disks $D_r^\sM(z_0)=\{z\in\MC^\sM:\abs{z-z_0}<r\}$, $\ob
D_r^\sM(z_0)=\{z\in\MC^\sM:\abs{z-z_0}\le r\}$ for $z_0\in\MC^\sM$, $r\in\MR^\sM_{>0}$.

We will usually work with a fixed model $\sM\model\vtc$, in which case we will omit the $\sM$ superscripts to simplify
the notation; we do this for the rest of this section as well.

When manipulated (as inputs or outputs) by $\tc$ functions, elements of $\MZ$, $\MQ$, and $\MQ(i)$ are represented in
binary in the expected way (i.e., rationals are represented by fractions of binary integers, and Gaussian rationals by
their real and imaginary parts), while elements of $\ML$ or $\MZL$ are represented as unary integers. However, we do not
introduce a standard representation for elements of $\MQL$ or $\MQL(i)$; they will be treated as elements of $\MQ$ or
$\MQ(i)$, and if needed, the function will explicitly take another input enforcing the logarithmic restriction (such as
an element of $\ML$ bounding the absolute value).

Some of our results will state the existence of $\tc$ functions with certain properties. Even though we otherwise work
relative to a fixed model $\sM\model\vtc$, such statements are always meant to be \emph{uniform}: i.e., interpreted as
``there is a specific function symbol of $\ob\vtc$ such that for every model $\sM\model\vtc$, etc.''

If $n$ is a unary natural number, $2^n$ is represented as a binary number by the set $\{n\}$. Thus, we can define a
$\tc$ function $2^n\colon\ML\to\MN$ satisfying $2^1=2$ and $2^{n+m}=2^n2^m$. Much more generally, given a sequence
$\p{x_j:j<n}$ coded in~$\sM$, where $x_j\in\MZ$ and $n\in\ML$, there is a $\tc$ definition of $\sum_{j<n}x_j$ and (due
to \cite{ej:vtcimul}) $\prod_{j<n}x_j$ satisfying
\begin{align*}
\sum_{j<0}x_j&=0,&
\sum_{j<n+1}x_j&=x_n+\sum_{j<n}x_j,\\
\prod_{j<0}x_j&=1,&
\prod_{j<n+1}x_j&=x_n\cdot\prod_{j<n}x_j.
\end{align*}
We can extend these operations to coded sequences of rational fractions by
\begin{align*}
\sum_{j<n}\frac{p_j}{q_j}&=\frac{\sum_{j<n}p_j\prod_{l\ne j}q_l}{\prod_{j<n}q_j},\\
\prod_{j<n}\frac{p_j}{q_j}&=\frac{\prod_{j<n}p_j}{\prod_{j<n}q_j}.
\end{align*}
We can further extend $\sum$ to coded sequences of Gaussian rationals with
\[\sum_{j<n}(x_j+iy_j)=\sum_{j<n}x_j+i\sum_{j<n}y_j.\]
Defining $\prod$ for such sequences is problematic, as the obvious formula requires a sum of $2^n$ terms; we will see
later how to do it using $\exp$ and $\log$, but at this point, we can at least define powering using
\[(x+iy)^n=\sum_{m\le n/2}\binom n{2m}(-1)^mx^{n-2m}y^{2m}+i\sum_{m<n/2}\binom n{2m+1}(-1)^mx^{n-2m-1}y^{2m+1}.\]
We can extend $z^n$ to a function $\MC\times\ML\to\MC$ as follows: for a fixed $n\in\ML$, $z^n$ is uniformly continuous
on $\ob D_r(0)\cap\MQ(i)$ for each $r\in\MR_{>0}$, as
\[\abs{z-w}\le\delta\implies\abs{z^n-w^n}=\Abs[Big]{(z-w)\sum_{j<n}z^jw^{n-1-j}}\le(n-1)r^{n-1}\delta\]
using Lemma~\ref{lem:abs-bd} below. Thus, it has a unique continuous extension to each $\ob D_r(0)$, and therefore a
unique continuous extension to~$\MC$, which we still denote $z^n$. For $z\ne0$, we also define $z^{-n}=1/z^n$. Powering
satisfies the basic identities $z^0=1$, $z^1=z$, $z^{n+m}=z^nz^m$, $z^{nm}=(z^n)^m$, and $(zw)^n=z^nw^n$: for
$z,w\in\MQ(i)$, this either holds immediately, or can be proved by induction on~$m$; then we use the density of $\MQ(i)$
in~$\MC$, observing that both sides of each identity are continuous in~$z$. (For $(zw)^n=z^nw^n$, we do it in two
steps: first as a function of $z\in\MC$ with fixed $w\in\MQ(i)$, then as a function of $w\in\MC$ for fixed $z\in\MC$.)
It is also easy to check that $x^n$ is increasing on~$\MR_{>0}$ for $n>0$, and decreasing for $n<0$.

For any $z\in\MC$, the absolute value $\abs z$ is a well-defined element of~$\MR$, but if $z\in\MQ(i)$, we do not
necessarily have $\abs z\in\MQ$. This is a hindrance to its use in arguments by induction, bounded sums and products,
etc. For this reason, we consider the predicate $\abs z\le r$ for $z=x+iy\in\MQ(i)$ and $r\in\MQ$, which can be
equivalently defined without reference to~$\MR$ by
\[\abs z\le r\iff r\ge0\land r^2\ge x^2+y^2(=z\ob z).\]
The following lemma summarizes its basic properties (some of which hold even for real arguments, as indicated).
\begin{Lem}\label{lem:abs-bd}\
\begin{enumerate}
\item\label{item:1}
Let $z,w\in\MC$ and $r,s\in\MR$. If $\abs z\le r$ and $\abs w\le s$, then $\abs{z+w}\le r+s$ and $\abs{zw}\le rs$.
\item\label{item:2}
Let $\p{z_j:j<n}$ and $\p{r_j:j<n}$ be sequences of elements of $\MQ(i)$ and $\MQ$ (respectively) coded in~$\sM$. If
$\abs{z_j}\le r_j$ for each $j<n$, then $\Abs{\sum_{j<n}z_j}\le\sum_{j<n}r_j$.
\item\label{item:3}
Let $z\in\MC$ and $r\in\MR$. If $\abs z\le r$, then $\abs{z^n}\le r^n$ for each $n\in\ML$.
\end{enumerate}
\end{Lem}
\begin{Pf}
\ref{item:1}: $\abs{zw}\le rs$ follows immediately from $(zw)\ob{zw}=(z\ob z)(w\ob w)$. Write $z=x+iy$ and $w=u+iv$.
Since
\[(xu+yv)^2\le(xu+yv)^2+(xv-yu)^2=(x^2+y^2)(u^2+v^2)\le r^2s^2,\]
we have $xu+yv\le rs$, thus
\[(x+u)^2+(y+v)^2=x^2+y^2+u^2+v^2+2(xu+yv)\le r^2+s^2+2rs=(r+s)^2,\]
which means $\abs{z+w}\le r+s$.

\ref{item:2} follows from~\ref{item:1} by induction on~$n$.

\ref{item:3}: If $z\in\MQ(i)$ and $r\in\MQ$, then $\abs z\le r$ implies $\abs{z^n}\le r^n$ by induction on~$n$
using~\ref{item:1}. For general $z$ and~$r$, we then use the density of $\MQ(i)$ in~$\MC$ and the continuity of $z^n$
and~$r^n$.
\end{Pf}

It will be most convenient for us to present constructions and arguments in a model-theoretic way, working directly
with functions $f\colon\MC\to\MC$ and the like. However, we need to keep in mind that $\MR$ and $\MC$ are not definable
in~$\sM$, and most of their elements cannot be represented as objects of $\sM$. Since we are ultimately interested in
what is provable in the theory $\vtc$, we need a way of restating properties of $\MC$-valued functions as first-order
properties of~$\sM$. Moreover, we want these properties to be definable by low-complexity ($\tc$) formulas so that they
can be used in induction arguments, comprehension instances, etc. We will accomplish this by means of
\emph{approximation} by $\tc$~functions. We formalize this concept as follows.

Consider $f\colon D\to\MC$, where $D\sset\MC$ is such that $D\cap\MQ(i)$ is dense in $D$. An \emph{additive
$\tc$ approximation} of $f$ is a $\tc$ function $f_+\colon\MQ(i)\times\ML\to\MQ(i)$ such that
\[\abs{f_+(z,n)-f(z)}\le2^{-n}\]
for all $z\in D\cap\MQ(i)$ and $n\in\ML$. A \emph{multiplicative $\tc$ approximation} of $f$ is
$f_\times\colon\MQ(i)\times\ML\to\MQ(i)$ such that
\[\abs{f_\times(z,n)-f(z)}\le2^{-n}\abs{f(z)}\]
for all $z\in D\cap\MQ(i)$ and $n\in\ML$; i.e., if $f(z)=0$, then $f_\times(z,n)=0$, and if $f(z)\ne0$, then
\[\Abs{\frac{f_\times(z,n)}{f(z)}-1}\le2^{-n}.\]
Multiplicative approximation is stronger than additive approximation in the following sense.
\begin{Lem}\label{lem:add-mult-apx}
For any function $f\colon D\to\MC$, $D\sset\MC$, the following are equivalent.
\begin{enumerate}
\item\label{item:48}
$f$ has a multiplicative $\tc$ approximation $f_\times$.
\item\label{item:49}
$f$ has an additive $\tc$ approximation $f_+$, and there exists a $\tc$ function $h\colon\MQ(i)\to\ML$ (with unary
output) such that
\[f(z)\ne0\implies\abs{f(z)}\ge2^{-h(z)}\]
for all $z\in D\cap\MQ(i)$.
\end{enumerate}
\end{Lem}
\begin{Pf}

\ref{item:48}\txto\ref{item:49}: Observe that $\abs{f(z)}\le2\abs{f_\times(z,1)}$. This allows us to define a
$\tc$~function $t\colon\MQ(i)\to\ML$ such that $\abs{f(z)}\le2^{t(z)}$ for all $z\in D\cap\MQ(i)$: given $z$, we
compute $\abs{f_\times(z,1)}^2\in\MQ_{\ge0}$, and using integer division and the length function, we compute
$t'=\Dlh{\cl{\abs{f_\times(z,1)}^2}}\in\ML$ so that $2^{t'}>\abs{f_\times(z,1)}^2$; then $t(z)=1+\cl{t'/2}$ works. Thus,
$f_+(z,n)=f_\times\bigl(z,n+t(z)\bigr)$ is an additive approximation of~$f$.

If $f(z)\ne0$, then $f_\times(z,1)\ne0$, and a similar argument as above gives us a $\tc$~function $h'$ such that
$2^{h'(z)}\ge\abs{f_\times(z,1)}^{-2}$; then $h(z)=1+\cl{h'/2}$ satisfies $\abs{f(z)}\ge2^{-h(z)}$, using the fact that
$\abs{f_\times(z,1)}\le\frac32\abs{f(z)}$.

\ref{item:49}\txto\ref{item:48}: First, given $z\in D\cap\MQ(i)$, we can decide in $\tc$ whether $f(z)=0$, as
\begin{align*}
f(z)=0&\implies\Abs{f_+\bigl(z,h(z)+2\bigr)}\le\tfrac142^{-h(z)},\\
f(z)\ne0&\implies\Abs{f_+\bigl(z,h(z)+2\bigr)}\ge\abs{f(z)}-\tfrac142^{-h(z)}\ge\tfrac342^{-h(z)}.
\end{align*}
Thus,
\[f_\times(z,n)=\begin{cases}0&\text{if $f(z)=0$,}\\
f_+\bigl(z,h(z)+n\bigr)&\text{otherwise}
\end{cases}\]
gives a multiplicative $\tc$ approximation of~$f$.
\end{Pf}

In practice, $\tc$ approximation functions will often need additional inputs. For example, to compute a $\tc$
approximation of $\exp z$, it is not enough to have $z$ (in binary) as input, as the output may be exponentially large;
we will also require a bound $r\in\ML$ (in unary) such that $\abs{z}\le r$, or at least $\Re z\le r$. We will employ
the following terminology.

For a function $f\colon D\to\MC$ as above, and a property $P(z,r)$, we say that a $\tc$ function $f_+(z,r,n)$ is
an \emph{additive approximation of $f(z)$ parametrized by~$r$ such that $P(z,r)$} if
\[P(z,r)\implies\abs{f_+(z,r,n)-f(z)}\le2^{-n}\]
for all $z\in D\cap\MQ(i)$ and $r,n\in\ML$; analogously for multiplicative approximation. We also require that for
every $z\in D\cap\MQ(i)$, there exists $r\in\ML$ such that $P(z,r)$. 

The following facts are useful for basic manipulation of multiplicative approximations.
\begin{Lem}\label{lem:mult-apx}
Let $z,w\in\MC$ and $\ep,\delta\in\MR_{\ge0}$.
\begin{enumerate}
\item\label{item:32} If $\abs{z-1}\le\ep$ and $\abs{w-1}\le\delta$, then $\abs{zw-1}\le\ep+\delta+\ep\delta$.
\item\label{item:33} If $\abs{z^2-1}\le2\ep-\ep^2$ and $\Re z\ge0$, then $\abs{z-1}\le\ep$.
\item\label{item:34} If $\abs{z-1}\le\ep/(1+\ep)$, then $\abs{z^{-1}-1}\le\ep$.
\item\label{item:35} Let $z=x+iy$ and $w=u+iv$ with $x,y,u,v\in\MR$. Assuming $x,y\ne0$,
\[\Abs{\frac ux-1}\le\ep\land\Abs{\frac v{\smash y}-1}\le\ep\implies\Abs{\frac wz-1}\le\ep.\]
\end{enumerate}
\end{Lem}
\begin{Pf}

\ref{item:32}: $\abs{zw-1}\le\abs{zw-w}+\abs{w-1}\le\abs w\ep+\delta\le(1+\delta)\ep+\delta$ using Lemma~\ref{lem:abs-bd}.

\ref{item:33}:
Put $r=\abs{1-z}$. We have $\abs{1+z}\ge2-r$ by Lemma~\ref{lem:abs-bd}, hence
\[2\ep-\ep^2\ge\Abs{1-z^2}=\Abs{1-z}\Abs{1+z}\ge r(2-r).\]
Thus, $(1-\ep)^2\le(1-r)^2$, i.e., $r\le\min\{\ep,2-\ep\}$ or $r\ge\max\{\ep,2-\ep\}$. Since $\Re z\ge0$, we have
$\abs{1+z}\ge r$; thus, if $r\ge2-\ep$ and $r>\ep$, we have $\Abs{1-z^2}=r\abs{1+z}>\ep\abs{1+z}\ge\ep(2-\ep)$, a
contradiction. Hence the only possibility is $r\le\ep$.

\ref{item:34}: We have $\abs z\ge1-\ep/(1+\ep)=1/(1+\ep)$, hence $\abs{z^{-1}-1}=\abs{z-1}/\abs z\le\ep$.

\ref{item:35}: $\abs{w-z}^2=\abs{u-x}^2+\abs{v-y}^2\le\ep^2\abs x^2+\ep^2\abs y^2=\ep^2\abs z^2$.
\end{Pf}

\section{Exponential and logarithm}\label{sec:expon-logar}

In this section, which is the main part of the paper, we will construct functions $\exp$ and $\log$ on suitable subsets
of $\MC^\sM$, and verify their basic properties. Since especially the construction of $\log$ will be somewhat
complicated, proceeding in several stages, we will need many technical lemmas along the way, which will be mixed with
bits and pieces of the intended end results. To help the reader not lose track of what is going on, we start by
collecting the most useful results and stating them in one place upfront.

Let us fix a model $\sM\model\vtc$ for the duration of this section. Put
\[\MR_{\DML}=\{x\in\MR:\exists n\in\ML\:x\le n\}=\MRL\cup\MR_{<0}.\]
We are going to define functions
\[\exp\colon\MR_{\DML}+i\MR\to\MC\]
(in Lemma~\ref{lem:exp-ir}, following up on
Definition~\ref{def:exp-ser} and Lemmas \ref{lem:exp-cauch} and~\ref{lem:exp-CL}) and
\[\log\colon\MC_{\ne0}\to\MCL\]
(in Definition~\ref{def:log-c}, following up on Definition~\ref{def:log-ser}, Lemmas \ref{lem:log-cauch}, \ref{lem:log-disc},
and~\ref{lem:log-r}, and Definitions \ref{def:sqrt} and~\ref{def:log-s},
and renamed to $\log$ in view of Lemma~\ref{lem:log-all}), as well as a constant $\pi\in\MR_{>0}$ and the argument
function $\arg\colon\MC_{\ne0}\to\MRL$ (in Definition~\ref{def:arg}).
\begin{Thm}\label{thm:main}
The functions $\exp$ and $\log$ have the following properties.
\begin{enumerate}
\item\label{item:50}
For all $z,w\in\dom(\exp)$, $\exp(z+w)=\exp z\exp w$, and $\exp\ob z=\ob{\exp z}$.
\item\label{item:51}
${\exp}\res\MRL+i\MR$ is a surjective group homomorphism $\p{\MRL+i\MR,+,0,-}\to\p{\MC_{\ne0},\cdot,1,{}^{-1}}$ with
kernel $2\pi i\MZ$.
\item\label{item:52}
$\exp z=0$ iff $\Re z\in\MR_{\DML}\bez\MRL$.
\item\label{item:53}
$\log$ maps $\MC_{\ne0}$ onto $\MRL+i(-\pi,\pi]$, and it is a right inverse of $\exp$, i.e., $\exp\log z=z$
for all $z\in\MC_{\ne0}$. Also, $\log\exp z=z$ for all $z\in\MRL+i(-\pi,\pi]$.
\item\label{item:54}
${\exp}\res\MRL$ is an ordered group isomorphism
$\p{\MRL,+,0,-,{<}}\to\p{\MR_{>0},\cdot,1,{}^{-1},{<}}$ whose inverse is ${\log}\res\MR_{>0}$.
\item\label{item:55}
$\exp$ is continuous, and it is uniformly continuous on $(-\infty,r]+i\MR$ for each $r\in\ML$;
$\log$ is continuous on $\MC\bez\MR_{\le0}$ and on $\{z\ne0:\Im z\ge0\}$, and for each $\ep\in\MR_{>0}$, it is
uniformly continuous on $\bigl\{z:\abs z\ge\ep\land(\Re z\ge0\lor\Im z\ge0\lor\Im z\le-\ep)\bigr\}$.
\item\label{item:56}
$\abs{\exp z}=\exp\Re z$ for all $z\in\dom(\exp)$, and $\log z=\log\abs z+i\arg z$ for all
$z\in\MC_{\ne0}$.
\item\label{item:57}
$\arg$ maps the quadrant $\{z\ne0:\Re z,\Im z\ge0\}$ to $\bigl[0,\frac\pi2\bigr]$,
$\{z\ne0:\Re z\le0,\Im z\ge0\}$ to $\bigl[\frac\pi2,\pi\bigr]$,
$\{z\ne0:\Re z\ge0,\Im z\le0\}$ to $\bigl[-\frac\pi2,0\bigr]$,
and $\{z:\Re z\le0,\Im z<0\}$ to $\bigl(-\pi,-\frac\pi2\bigr]$. In each quadrant, it increases or decreases in tandem
with $\Re\sgn z$ and $\Im\sgn z$ as determined in Lemma~\ref{lem:arg-mon}, where $\sgn z=z/\abs z$.
\item\label{item:58}
If $\abs z\le\frac32$, then $\Abs[big]{\exp z-(1+z)}\le\abs z^2$. If $\abs z\le\frac12$, then
$\Abs[big]{\log(1+z)-z}\le\abs z^2$.
\item\label{item:60}
If $z\in\MCL$ and $n\in\MZL$, then $\exp nz=(\exp z)^n$.
\item\label{item:61}
If $z\in\MC$ and $n\in\ML_{>0}$ is such that $n\ge\max\bigl\{2\abs z,\abs z^2\bigr\}$, then
\[\Abs{\frac{\left(1+\frac zn\right)^n}{\exp z}-1}\le\frac{2\abs z^2}n.\]
\item\label{item:62}
For all $x\in\MR_{\DML}$, $\exp x\ge1+x$. Consequently, ${\exp}\res\MR_{\DML}$ is convex: for all $x,y\in\MR_{\DML}$
and $t\in[0,1]$,
\begin{align*}
(y-x)\exp x\le\exp y&-\exp x\le(y-x)\exp y,\\
\exp\bigl((1-t)x+ty\bigr)&\le(1-t)\exp x+t\exp y.
\end{align*}
\item\label{item:63}
$\exp z$ has $\tc$ additive approximation $E_+(z,r,n)$ for $z\in\MQ_{\DML}+i\MQ$, parametrized by $r\in\ML$ such
that $\Re z\le r$, and $\tc$ multiplicative approximation $E_\times(z,r,n)$ for $z\in\MQL+i\MQ$, parametrized by
$r\in\ML$ such that $\abs{\Re z}\le r$. For $z\in\MQ(i)\bez\{0\}$, $\log z$ has $\tc$ additive approximation
$L_+(z,n)$ and $\tc$ multiplicative approximation $L_\times(z,n)$.
\end{enumerate}
\end{Thm}
While Theorem~\ref{thm:main} is for the most part a summary of various lemmas that appear separately throughout the course
of Section~\ref{sec:expon-logar}, we will formally prove it at the end of the section.

We omit listing some useful properties that can be inferred from the above: in particular, \ref{item:51}
and~\ref{item:53} imply Lemma~\ref{lem:log-homo-all}; we mention though that they also imply a variant
of Lemma \ref{lem:log-c-homo} \ref{item:21} and~\ref{item:22} with a perhaps clearer geometric meaning:
\begin{Cor}\label{cor:log-homo-arg}
If $z,w\in\MC_{\ne0}$ satisfy $\arg z+\arg w\in(-\pi,\pi]$, then $\log zw=\log z+\log w$.
\noproof\end{Cor}
Further facts of interest that will be proved in this section are the
bounds on $\pi$ in Proposition~\ref{prop:pi-apx}, bounds on $e=\exp1$ in Lemma~\ref{cor:e-apx}, and properties of the complex
square root function in \S\ref{sec:complex-square-root}.

We should comment on the decision to put $\exp z=0$ for $\Re z\in\MR_{\DML}\bez\MRL$, rather than leaving it undefined.
This violates the basic property that $\exp z\ne0$ for all $z$. However, it retains other fundamental properties of
$\exp$, in particular $\exp(z+w)=\exp z\exp w$, and the monotonicity of $\exp$ on the reals. Conceptually, it seems
to be the right thing to do, as $\abs{\exp z}$ drops down to~$0$ as $\Re z\to-\infty$ for $\Re z\in\MRL$. (By the same
reasoning, we could also define $\exp z=\infty$ when $\Re z>\MRL$, but we prefer to keep functions finite.)

Perhaps the best technical reason for this definition stems from point \ref{item:63}. Recall that our overarching goal
is to explore what is provable in the theory $\vtc$, which cannot directly talk about $\MC$-valued functions such as
$\exp$ or $\log$; from this viewpoint, the $\tc$ approximations of these functions are more fundamental than the
functions themselves, which are just figments of our imagination. Now, additive approximation $E_+(z,r,n)$ of $\exp z$
is most naturally presented with a parameter $r\in\ML$ such that $\Re z\le r$ as indicated, as this is exactly what is
needed to keep the approximation efficiently computable. However, there is no way for a $\tc$ function to distinguish
inputs $z$ with $\Re z\in\MQ_{\DML}\bez\MQL$ from those where $\Re z\in\MQL$ is merely very small; the approximating
function is bound to output (approximately) $0$ for both. Thus, the exponential function determined by
$\lim_{n\to\infty}E_+(z,r,n)$ will be defined even for $\Re z\in\MQ_{\DML}\bez\MQL$, assigning such $z$ the value~$0$. We
chose to make our official exponential function agree with this.

In any case, a skeptical reader is free to restrict the $\exp$ function to $\MRL+i\MR$.

\subsection{Exponential}\label{sec:exponential}

Our first task is relatively straightforward: define a function $\exp\colon\MCL\to\MC_{\ne0}$ using the power series
\[\sum_n\frac{z^n}{n!},\]
and (among other basic properties) prove the homomorphism property
\[\exp(z+w)=\exp z\exp w\]
by means of the standard argument exploiting the binomial theorem.

We start by defining partial sums of this power series, which we will then use to define $\exp$ on $\MQL(i)$.
\begin{Def}\label{def:exp-ser}
We define a function $e\colon\MQ(i)\times\ML\to\MQ(i)$ by
\[e(z,n)=\sum_{j<n}\frac{z^j}{j!}.\]
\end{Def}
\begin{Lem}\label{lem:fact-lb}
$n!\ge2\bigl(\frac14(n+1)\bigr)^n$ for all $n\in\ML$, $n\ge1$.
\end{Lem}
\begin{Pf}
By induction on~$n$. The statement holds for $n=1$. If $n\ge2$, the induction hypothesis for $m=\fl{n/2}$
gives
\[n!\ge m!\,(m+1)^{n-m}\ge2\frac{(m+1)^n}{4^m}=2\frac{(2m+2)^n}{4^m2^n}\ge2\frac{(n+1)^n}{4^n}.\]
\end{Pf}
\begin{Lem}\label{lem:exp-cauch}
If $z\in\MQL(i)$, then $\{e(z,n):n\in\ML\}$ is a Cauchy net. Thus, we can define a function $\exp_{\MQL(i)}\colon\MQL(i)\to\MC$
by
\[\exp_{\MQL(i)}z=\lim_{\substack{n\in\ML\\n\to\infty}}e(z,n).\]
\end{Lem}
\begin{Pf}
Assume $\abs z\le r\in\ML$. If $2r\le n\le m\in\ML$, we have
\[\Abs[big]{e(z,m)-e(z,n)}=\Abs[bigg]{\sum_{j=n}^{m-1}\frac{z^j}{j!}}
\le\sum_{j=n}^{m-1}\frac{r^j}{j!}
\le\frac{r^n}{n!}\sum_{j<m-n}\left(\frac rn\right)^j
\le\frac{r^n}{n!}\sum_{j<m-n}2^{-j}
\le2\frac{r^n}{n!},\]
using Lemma~\ref{lem:abs-bd} and $(n+j)!\ge n!\,n^j$. Thus, if $n,m\ge\max\{8r,t\}$, then
\[\Abs[big]{e(z,m)-e(z,n)}\le\left(\frac{4r}n\right)^n\le2^{-n}\le2^{-t}\]
by Lemma~\ref{lem:fact-lb}.
\end{Pf}

We are going to extend the domain of $\exp_{\MQL(i)}$ first to $\MCL$, and later to $\MR_{\DML}+i\MR$. Formally, we
distinguish these functions by subscripts, but since they give the same values whenever they are defined, we may write
just $\exp$ unless the distinction becomes important.
\begin{Lem}\label{lem:exp-bdabs}
Let $z\in\MQ(i)$ and $r\in\MQL$. If $\abs z\le r$, then $\abs{\exp z}\le\exp r$.
\end{Lem}
\begin{Pf}
For any $n\in\ML$, we have
\[\abs{e(z,n)}=\Abs[bigg]{\sum_{j<n}\frac{z^j}{j!}}\le\sum_{j<n}\frac{r^j}{j!}\le\exp r\]
by Lemma~\ref{lem:abs-bd}. The result follows by taking the limit $n\to\infty$.
\end{Pf}
\begin{Lem}\label{lem:exp-small}
If $z\in\ob D_{3/2}(0)\cap\MQ(i)$, then
\[\Abs[big]{\exp z-(1+z)}\le\abs z^2.\]
\end{Lem}
\begin{Pf}
For any $n\in\ML$, $n\ge2$,
\[\Abs{\frac{e(z,n)-(1+z)}{z^2}}=\Abs[bigg]{\sum_{j=2}^{n-1}\frac{z^{j-2}}{j!}}
\le\sum_{j=2}^{n-1}\frac{3^{j-2}}{2^{j-2}j!}
\le\sum_{j=2}^{n-1}2^{1-j}\le1\]
using Lemma~\ref{lem:abs-bd} and $j!\ge2\cdot3^{j-2}$ for $j\ge2$, thus
\[\Abs[big]{e(z,n)-(1+z)}\le\abs z^2.\]
Taking the limit, a similar inequality holds for $\exp z$.
\end{Pf}
\begin{Lem}\label{lem:exp-homo}
For any $z,w\in\MQL(i)$,
\[\exp(z+w)=\exp z\exp w.\]
\end{Lem}
\begin{Pf}
For any $n\in\ML$, we have
\[e(z,n)e(w,n)=\sum_{j,k<n}\frac{z^jw^k}{j!\,k!},\]
while
\[e(z+w,2n)=\sum_{l<2n}\frac{(z+w)^l}{l!}
=\sum_{l<2n}\sum_{j+k=l}\binom lj\frac{z^jw^k}{l!}
=\sum_{j+k<2n}\frac{z^jw^k}{j!\,k!},\]
hence
\[e(z+w,2n)-e(z,n)e(w,n)
=\sum_{j=n}^{2n-1}\frac{z^j}{j!}\sum_{k<2n-j}\frac{w^k}{k!}
+\sum_{k=n}^{2n-1}\frac{w^k}{k!}\sum_{j<2n-k}\frac{z^j}{j!}.\]
Fix $r\in\ML$ such that $\abs z,\abs w\le r$. Then Lemma~\ref{lem:abs-bd} gives
\[\Abs[big]{e(z+w,2n)-e(z,n)e(w,n)}
\le2\sum_{j=n}^{2n-1}\frac{r^j}{j!}\sum_{k<2n-j}\frac{r^k}{k!}
\le2\exp(r)\sum_{j=n}^{2n-1}\frac{r^j}{j!}.\]
Thus, for all $n\ge8r$, we have
\[\Abs[big]{e(z+w,2n)-e(z,n)e(w,n)}\le2^{1-n}\exp r\]
by the proof of Lemma~\ref{lem:exp-cauch}. The result follows by taking the limit $n\to\infty$.
\end{Pf}
\begin{Lem}\label{lem:exp-CL}
The restrictions ${\exp_{\MQL(i)}}\res\ob D_r(0)\cap\MQ(i)$ are uniformly continuous for all $r\in\ML$. Thus,
$\exp_{\MQL(i)}$ has a unique extension to a continuous function $\exp_{\MCL}\colon\MCL\to\MC$.
\end{Lem}
\begin{Pf}
Let $\delta\in\MQ$, $0<\delta\le1$. Then any $z,w\in\ob D_r(0)\cap\MQ(i)$ such that $\abs{w-z}\le\delta$ satisfy
\[\abs{\exp w-\exp z}=\Abs{\bigl(\exp(w-z)-1\bigr)\exp z}\le(\delta+\delta^2)\exp r\le2\delta\exp r\]
by Lemmas \ref{lem:exp-bdabs}, \ref{lem:exp-small}, and~\ref{lem:exp-homo}, thus ${\exp_{\MQL(i)}}\res\ob D_r(0)\cap\MQ(i)$ is indeed
uniformly continuous. It follows that it has a unique continuous extension $\exp_r\colon\ob D_r(0)\to\MC$. Uniqueness
ensures that $\exp_r={\exp_s}\res\ob D_r(0)$ whenever $r\le s$, hence $\exp_{\MCL}=\bigcup_r\exp_r$ is a well-defined
function $\MCL\to\MC$, and it is continuous as its restrictions to all $\ob D_r(0)$ are continuous. Conversely, any
continuous extension of $\exp_{\MQL(i)}$ to $\MCL$ must coincide with $\exp_r$ on $\ob D_r(0)$ for each~$r$, hence it
equals $\exp_{\MCL}$.
\end{Pf}

The main take-away from Section~\ref{sec:exponential} is the next summary lemma.
\begin{Lem}\label{thm:exp-homo}\
\begin{enumerate}
\item\label{item:4}
The function $\exp_{\MCL}$ is a group homomorphism $\p{\MCL,+,0,-}\to\p{\MC_{\ne0},\cdot,1,{}^{-1}}$ commuting with~$\ob
z$.
\item\label{item:5}
The restriction $\exp_{\MRL}={\exp_{\MCL}}\res\MRL$ is an embedding of ordered groups
$\p{\MRL,+,0,-,{<}}\to\p{\MR_{>0},\cdot,1,{}^{-1},{<}}$.
\item\label{item:42}
For all $z\in\MCL$, $\abs{\exp_{\MCL} z}=\exp_{\MRL}\Re z$.
\item\label{item:6}
If $z\in\ob D_{3/2}(0)$, then $\Abs[big]{\exp z-(1+z)}\le\abs z^2$.
\end{enumerate}
\end{Lem}
\begin{Pf}

\ref{item:4}: For any $w\in\MQL(i)$, the set
\[H_w=\bigl\{z\in\MCL:\exp(z+w)=\exp z\exp w\bigr\}\]
is closed due to the continuity of $\exp$, and includes $\MQL(i)$ by Lemma~\ref{lem:exp-homo}. Since $\MQL(i)$ is dense
in~$\MCL$, we see that $H_w=\MCL$, i.e.,
\begin{equation}\label{eq:1}
\exp(z+w)=\exp z\exp w
\end{equation}
for all $z\in\MCL$ and $w\in\MQL(i)$. Using the same density argument once more, $\MQL(i)\sset H_w$ for each $w\in\MCL$
by~\eqref{eq:1} (with arguments swapped), hence $H_w=\MCL$, i.e., \eqref{eq:1} holds for all $z,w\in\MCL$. This shows
that $\exp$ is a group homomorphism as indicated in~\ref{item:4}, provided that the codomain is right,
i.e., $\exp z\ne0$ for all $z\in\MCL$. This, too, follows from~\eqref{eq:1}, as $\exp z\exp(-z)=\exp0=1$.

We have $e(\ob z,n)=\ob{e(z,n)}$ for all $z\in\MQL(i)$ and $n\in\ML$, hence $\exp\ob z=\ob{\exp z}$ by taking
limits. Using density of $\MQL(i)\sset\MCL$ again, the same holds for all $z\in\MCL$.

\ref{item:5}: Let $x\in\MRL$. Since $\ob{\exp x}=\exp x$ by~\ref{item:4}, we have $\exp x\in\MR$. If $x\in\MQL(i)$,
$x>0$, then $e(x,n)$ is non-decreasing in~$n$, hence $\exp x\ge e(x,2)=1+x$. By density, $\exp x\ge1+x$ for all
$x\in\MRL$, $x>0$, hence $\exp x>1$. Since $\exp x\exp(-x)=1$ by~\ref{item:4}, this ensures $\exp x>0$ for all
$x\in\MRL$, and it implies that $\exp$ is strictly increasing on~$\MRL$: if $x<y$, we have
\[\exp(y)=\exp(y-x)\exp x>\exp x\]
as $\exp(y-x)>1$ and $\exp x>0$. In particular, $\exp_{\MRL}$ is injective, and it is an ordered group homomorphism.

\ref{item:42}: $\abs{\exp z}^2=\exp z\mathop{\ob{\exp z}}=\exp(z+\ob z)=\exp(2\Re z)=(\exp\Re z)^2$
using~\ref{item:4}.

\ref{item:6}: By Lemma~\ref{lem:exp-small} and the density of $\ob D_{3/2}(0)\cap\MQ(i)$ in $\ob D_{3/2}(0)$.
\end{Pf}

The main remaining problem now is to prove that $\exp_{\MCL}$ is \emph{surjective} (onto $\MC_{\ne0}$); consequently,
$\exp_{\MRL}$ is an ordered group isomorphism, and there is a constant $\pi$ such that $\exp_{\MCL}$ is $2\pi i$-periodic,
which will enable its extension to $\MRL+i\MR$. Let us first mention a few failed approaches so that we understand that
the problem is nontrivial.

Considering the real case for simplicity, the most obvious idea how to find for a given $x\in\MR_{>0}$ a preimage
$y\in\MRL$ such that $\exp y=x$ is to show, for any integer $n>0$, that there is an integer $m$ such that
$\exp(m/n)\le x\le\exp\bigl((m+1)/n\bigr)$, using the monotonicity of $\exp$. On closer inspection, this argument
amounts to induction on $m$ for the formula $\exp(m/n)\le x$; thus, to make it work in $\vtc$, we actually need to use
rational approximations of $\exp$ rather than the function itself, and even so, it only works for $n\in\ML$, which
implies $m\in\MZL$. (Note that binary number induction for $\tc$~predicates implies $\thry{VPV}$ over~$\vtc$. Here,
$\thry{VPV}$ is a theory of bounded arithmetic corresponsing to polynomial-time functions, and as such it is geenrally
assumed to be stronger than~$\vtc$; cf.~\cite{cook-ngu}.) Thus, we can only determine logarithmically many most
significant bits of $y$, which is insufficient to construct it as an element of $\MR$.

We could use binary search to determine $y$ with precision $2^{-n}$ rather than $n^{-1}$, but this is an inherently
sequential algorithm taking us outside $\tc$; likewise for more sophisticated iterative methods such as Newton
iteration, which parallelize better, but still need a non-constant number of sequential iterations. In
\cite{ej:vtc0iopen}, we formalized a form of the Lagrange inversion theorem, which can in principle be used to invert
any function $f$ given by power series, such as $\exp$; however, the core argument in \cite[Thm.~5.1]{ej:vtc0iopen} (or
even the definition of the inverse series) only works when $f$ is a constant-degree polynomial, as it relies on bounded
sums with $n^{O(d)}$ terms, where $d=\deg f$.

We will solve the problem by constructing in an ad hoc way a function $\log\colon\MC_{\ne0}\to\MCL$, and proving its
various properties, eventually showing $\exp\log z=z$. This will take us the next few subsections.

But before we leave, let us present some bounds on $\exp_{\MRL}$ that express its convexity.
\begin{Lem}\label{prop:exp-conv}
\ \begin{enumerate}
\item\label{item:7}
For all $x\in\MRL$, $\exp x\ge1+x$.
\item\label{item:8}
For all $x,y\in\MRL$, $(y-x)\exp x\le\exp y-\exp x\le(y-x)\exp y$.
\item\label{item:9}
For all $x,y\in\MRL$ and $t\in[0,1]$, $\exp\bigl((1-t)x+ty\bigr)\le(1-t)\exp x+t\exp y$.
\end{enumerate}
\end{Lem}
\begin{Pf}

\ref{item:7}: By density, it suffices to prove the result for $x\in\MQL$. We have shown $\exp x\ge1+x$ for $x\ge0$ in
the proof of Lemma \ref{thm:exp-homo}~\ref{item:5}. Moreover, if $0\le x<1$, then
\[e(x,n)=\sum_{j<n}\frac{x^j}{j!}\le\sum_{j<n}x^j\le\frac1{1-x},\]
hence $\exp x\le(1-x)^{-1}$, and $\exp(-x)\ge1-x$. Thus, $\exp x\ge1+x$ also holds when $-1<x\le0$; if $x\ge-1$, then
$\exp x\ge0\ge1+x$.

\ref{item:8}: We have $\exp y-\exp x=\bigl(\exp(y-x)-1\bigr)\exp x\ge(y-x)\exp x$ by~\ref{item:7}; the other
inequality follows by swapping $x$ and~$y$.

\ref{item:9}: Put $w=(1-t)x+ty$. We have $\exp w-\exp x\le(w-x)\exp w=t(y-x)\exp w$ by~\ref{item:8}, hence
$\exp x\ge\bigl(1-t(y-x)\bigr)\exp w$. Likewise, $\exp y-\exp w\ge(y-w)\exp w$ implies
$\exp y\ge\bigl(1+(1-t)(y-x)\bigr)\exp w$. Thus,
$$(1-t)\exp x+t\exp y\ge\bigl[(1-t)\bigl(1-t(y-x)\bigr)+t\bigl(1+(1-t)(y-x)\bigr)\bigr]\exp w=\exp w.\qedhere$$
\end{Pf}

\subsection{Logarithm near $1$}\label{sec:logarithm-near-1}

We intend to construct a logarithm function which is a right inverse of $\exp$, implying that $\exp$ is surjective.
Defining $\log$ will be more complicated than $\exp$, largely due to the fact that $\exp$ is entire, whereas $\log$ has
a branching singularity at the origin. Thus, a power series will only give us $\log$ in a circular neighbourhood
of~$1$: this will be the topic of the present subsection. We will then extend it to $\MC_{\ne0}$ (with a branch cut
along the negative real axis) in several stages:
\begin{itemize}
\item Using the function $2^n\colon\ML\to\MN$, we extend $\log$ to $\MR_{>0}$ (Section~\ref{sec:real-logarithm}).
\item Combining $\MR_{>0}$ with the neighbourhood of~$1$, we extend $\log$ to a sector $\{x+iy:\abs y<cx\}$ for a
suitable~$c$ (Section~\ref{sec:logarithm-sector}).
\item Using $\sqrt z$ (treated in Section~\ref{sec:complex-square-root}), we can increase the angle of the sector. We
iterate this a few times to cover $\MC_{\ne0}$ (Section~\ref{sec:full-compl-logar}).
\end{itemize}
We will rely on restricted forms of the identity $\log zw=\log z+\log w$ (which does not quite hold, due to the
branch cut) to make sure that the successive extensions fit together well, and to eventually derive $\exp\log z=z$.

We start with the power series for $\log$, or rather, for the function $-\log(1-z)$.
\begin{Def}\label{def:log-ser}
We define a function $\lambda\colon\MQ(i)\times\ML\to\MQ(i)$ by
\[\lambda(z,n)=\sum_{j=1}^n\frac{z^j}j.\]
We write $x<^*y$ if $x\le y-h^{-1}$ for some $h\in\ML_{>0}$, and we put $D_r^*(z_0)=\{z\in\MC:\abs{z-z_0}<^*r\}$,
$(a,b)^*=\{x\in\MR:a<^*x<^*b\}$, $[a,b)^*=\{x\in\MR:a\le x<^*b\}$, etc.
\end{Def}
\begin{Lem}\label{lem:exp-est}
If $h\in\ML_{>0}$, then $(1-h^{-1})^h\le\frac12$.
\end{Lem}
\begin{Pf}
$h^h=\sum_{j\le h}\binom hj(h-1)^j\ge(h-1)^h+h(h-1)^{h-1}\ge2(h-1)^h$.
\end{Pf}
\begin{Lem}\label{lem:log-cauch}
If $z\in D^*_1(0)\cap\MQ(i)$, then $\{\lambda(z,n):n\in\ML\}$ is a Cauchy net. Thus, we
can define a function $\Lambda\colon D^*_1(0)\cap\MQ(i)\to\MC$ by
\[\Lambda(z)=\lim_{\substack{n\in\ML\\n\to\infty}}\lambda(z,n).\]
\end{Lem}
\begin{Pf}
Assume $\abs z\le1-h^{-1}$, where $h\in\ML$, and let $n\le m\in\ML$. Then
\begin{align*}
\Abs[big]{\lambda(z,m)-\lambda(z,n)}=\Abs[bigg]{\sum_{j=n+1}^m\frac{z^j}j}
&\le\sum_{j=n+1}^m\frac{(1-h^{-1})^j}j
\le\frac{(1-h^{-1})^{n+1}}{n+1}\sum_{j<m-n}(1-h^{-1})^j\\
&\le\frac h{n+1}(1-h^{-1})^{n+1}\le2^{-t}
\end{align*}
if $n+1\ge ht$, using Lemmas \ref{lem:abs-bd} and~\ref{lem:exp-est}.
\end{Pf}
\begin{Lem}\label{lem:log-small}
If $z\in\ob D_{1/2}(0)\cap\MQ(i)$, then
\[\abs{\Lambda(z)-z}\le\abs z^2.\]
\end{Lem}
\begin{Pf}
For any $n\in\ML$, $n\ge2$,
\[\Abs{\frac{\lambda(z,n)-z}{z^2}}=\Abs[bigg]{\sum_{j=2}^{n-1}\frac{z^{j-2}}j}
\le\sum_{j=2}^{n-1}\frac1{2^{j-2}j}
\le\sum_{j=2}^{n-1}2^{1-j}\le1\]
using Lemma~\ref{lem:abs-bd}, which gives the result by taking the limit $n\to\infty$.
\end{Pf}

We are now heading to prove the identity $\Lambda(z)+\Lambda(w)=\Lambda(z+w-zw)$, which will yield
$\log(z)+\log(w)=\log(zw)$; this is the most technical part of the construction of $\log$. We will need the next lemma
as an ingredient in the proof; it effectively means that $\nabla^nf=0$ for any polynomial $f$ of degree $<n$ (expressed
as a linear combination of the falling factorials $\fafac xh$, $h<n$, rather than the usual monomials $x^h$), where
$(\nabla f)(x)=f(x)-f(x-1)$ is the backwards difference operator.
\begin{Lem}\label{lem:difop-nul}
For all $h<n\in\ML$ and $x\in\MQ$,
\[\sum_{k\le n}\binom nk(-1)^k\fafac{(x-k)}h=0,\]
where $\fafac xh=\prod_{j<h}(x-j)$.
\end{Lem}
\begin{Pf}
Fix $x\in\MQ$ and $m\in\ML_{>0}$; we will prove
\begin{equation}\label{eq:2}
\sum_{k\le m+h}\binom{m+h}k(-1)^k\fafac{(x+h-k)}h=0
\end{equation}
by induction on $h\in\ML$. For $h=0$, we have
\[\sum_{k\le m}\binom mk(-1)^k\fafac{(x-k)}0=\sum_{k\le m}\binom mk(-1)^k=(1-1)^m=0.\]
Assuming \eqref{eq:2} holds for~$h$, and writing $m'=m+h$, $x'=x+h$, we obtain
\begin{align*}
\sum_{k\le m'+1}&\binom{m'+1}k(-1)^k\fafac{(x'+1-k)}{h+1}\\
&=\sum_{k\le m'}\binom{m'}k(-1)^k\fafac{(x'+1-k)}{h+1}+\sum_{k\le m'}\binom{m'}k(-1)^{k+1}\fafac{(x'-k)}{h+1}\\
&=\sum_{k\le m'}\binom{m'}k(-1)^k(x'+1-k)\fafac{(x'-k)}h-\sum_{k\le m'}\binom{m'}k(-1)^k\fafac{(x'-k)}h(x'-k-h)\\
&=(h+1)\sum_{k\le m'}\binom{m'}k(-1)^k\fafac{(x'-k)}h=0
\end{align*}
using $\binom{m'+1}k=\binom{m'}k+\binom{m'}{k-1}$ and the induction hypothesis.
\end{Pf}
\begin{Lem}\label{lem:log-ser-hom}
Let $r,s\in\MQ_{>0}$ be such that $(1+r)(1+s)<^*2$. Then
\[\Lambda(z)+\Lambda(w)=\Lambda(z+w-zw)\]
for all $z\in\ob D_r(0)\cap\MQ(i)$ and $w\in\ob D_s(0)\cap\MQ(i)$.
\end{Lem}
\begin{Pf}
Since $\abs{z+w-zw}\le r+s+rs<^*1$ by Lemma~\ref{lem:abs-bd}, $\Lambda(z+w-zw)$ is defined. For any $n\in\ML_{>0}$, we
have
\begin{align*}
\lambda(z,n)+\lambda(w,n)-{}&\lambda(z+w-zw,n)\\
&=\sum_{j=1}^n\frac{z^j}j+\sum_{k=1}^n\frac{w^k}k
-\sum_{\substack{j,k,l\\0<j+k+l\le n}}\binom{j+k+l}{j,k,l}\frac{(-1)^lz^{j+l}w^{k+l}}{j+k+l}\\
&=-\sum_{\substack{j,k,l\\0<j+l,k+l\\j+k+l\le n}}\binom{j+k+l}{j,k,l}\frac{(-1)^lz^{j+l}w^{k+l}}{j+k+l}.
\end{align*}
We claim that
\[\sum_{\substack{j,k,l\\0<j+l,k+l\le n}}\binom{j+k+l}{j,k,l}\frac{(-1)^lz^{j+l}w^{k+l}}{j+k+l}
=\sum_{a,b=1}^nz^aw^b\sum_{l\le a,b}\binom{a+b-l}{a-l,b-l,l}\frac{(-1)^l}{a+b-l}=0:\]
indeed, if w.l.o.g.\ $a\le b$, Lemma~\ref{lem:difop-nul} gives
\begin{align*}
\sum_{l\le a,b}\binom{a+b-l}{a-l,b-l,l}\frac{(-1)^l}{a+b-l}
&=\sum_{l\le a}\frac{(a+b-1-l)!}{(a-l)!\,(b-l)!\,l!}(-1)^l\\
&=\frac1{a!}\sum_{l\le a}\binom al(-1)^l\fafac{(a+b-1-l)}{a-1}=0.
\end{align*}
Thus,
\[\lambda(z,n)+\lambda(w,n)-\lambda(z+w-zw,n)
=\sum_{\substack{j,k,l\\j+l,k+l\le n<j+k+l}}\binom{j+k+l}{j,k,l}\frac{(-1)^lz^{j+l}w^{k+l}}{j+k+l}.\]
By Lemma~\ref{lem:abs-bd},
\begin{align*}
\Abs[big]{\lambda(z,n)+\lambda(w,n)-\lambda(z+w-zw,n)}
&\le\sum_{\substack{j,k,l\\j+l,k+l\le n<j+k+l}}\binom{j+k+l}{j,k,l}\frac{r^{j+l}s^{k+l}}{j+k+l}\\
&\le\frac1{n+1}\sum_{\substack{j,k,l\\n<j+k+l\le2n}}\binom{j+k+l}{j,k,l}r^{j+l}s^{k+l}\\
&=\frac1{n+1}\sum_{a=n+1}^{2n}(r+s+rs)^a\\
&\le\frac{(r+s+rs)^{n+1}}{(n+1)(1-r-s-rs)}.
\end{align*}
By assumption, we can fix $h\in\ML$ such that $r+s+rs\le1-h^{-1}$. Then for all $n,t\in\ML$,
\[n+1\ge ht\implies
\Abs[big]{\lambda(z,n)+\lambda(w,n)-\lambda(z+w-zw,n)}
\le\frac h{n+1}(1-h^{-1})^{n+1}\le2^{-t}\]
using Lemma~\ref{lem:exp-est}. The result follows by taking the limit $n\to\infty$.
\end{Pf}
\begin{Lem}\label{lem:log-unif}
Let $h\in\ML_{>0}$ and $r\in\MQ_{>0}$. Then for all $z,w\in\ob D_{1-h^{-1}}(0)\cap\MQ(i)$,
\[\abs{z-w}\le r\implies\abs{\Lambda(z)-\Lambda(w)}\le hr.\]
\end{Lem}
\begin{Pf}
For any $n\in\ML$, we have
\begin{align*}
\abs{\lambda(z,n)-\lambda(w,n)}
=\Abs[bigg]{\sum_{j=1}^n\frac{z^j-w^j}j}
&=\Abs[bigg]{(z-w)\sum_{j=1}^n\frac1j\sum_{k<j}z^kw^{j-1-k}}\\
&\le r\sum_{j=1}^n(1-h^{-1})^{j-1}\le hr.
\end{align*}
The result follows by taking the limit $n\to\infty$.
\end{Pf}

The next lemma (and definition) is the main result of Section~\ref{sec:logarithm-near-1}.
\begin{Lem}\label{lem:log-disc}
There is a unique continuous function
\[\log_D\colon D^*_1(1)\to\MC\]
such that $\log_Dz=-\Lambda(1-z)$ for all $z\in D^*_1(1)\cap\MQ(i)$. It satisfies
\begin{equation}\label{eq:3}
\Abs[big]{\log_D(1+z)-z}\le\abs z^2
\end{equation}
for $z\in\ob D_{1/2}(0)$. If $r,s\in\MQ_{>0}$ are such that $(1+r)(1+s)<^*2$, then
\begin{equation}\label{eq:4}
\log_Dzw=\log_Dz+\log_Dw
\end{equation}
for all $z\in\ob D_r(1)$ and $w\in\ob D_s(1)$. In particular, \eqref{eq:4} holds for all $z,w\in\ob D_{2/5}(1)$.
\end{Lem}
\begin{Pf}
As in the proof of Lemma~\ref{lem:exp-CL}, the existence and uniqueness of $\log_D$ follows from the uniform
continuity of $\Lambda\res\ob D_{1-h^{-1}}(0)\cap\MQ(i)$ for every $h\in\ML_{>0}$, which was proved in
Lemma~\ref{lem:log-unif}. Properties \eqref{eq:3} and~\eqref{eq:4} for $z,w\in\MQ(i)$ follow from
Lemmas \ref{lem:log-small} and~\ref{lem:log-ser-hom}; this implies the general case using the density of $\MQ(i)$ in~$\MC$,
similarly to the proof of Lemma~\ref{thm:exp-homo}.
\end{Pf}

As in the case of $\exp$, we will define several versions of $\log$ on various domains, distinguished by subscripts.
(The $D$ in $\log_D$ refers to the disk $D^*_1(1)$.) Again, we will sometimes drop the subscript if it can be inferred
from the context, but we have to be more careful than before, because the variants of $\log$ do not a
priori agree on their common domains (though we will eventually prove they do, in Lemma~\ref{lem:log-all}).

\subsection{Real logarithm}\label{sec:real-logarithm}

We now start extending $\log_D$ to further domains, first to $\MR_{>0}$. The idea is simple: using the function
$2^n\colon\ML\to\MN$ (or rather, $\MZL\to\MQ$), we can write any $x\in\MR_{>0}$ as $x=2^nx'$ with $n\in\MZL$ and
$x'\in\bigl[\frac12,1\bigr]$, and put $\log_\MR x=\log_Dx'+n\log2$ for a suitably defined $\log2$. We will actually
ensure this defining identity to hold for $x'$ from a larger interval; the resulting interval overlap will assist us in
proving that $\log_\MR$ is well behaved, such as that it satisfies the identity $\log_\MR xy=\log_\MR x+\log_\MR
y$. We begin with a definition of $\ell_2=\log2$.
\begin{Lem}\label{lem:log2}
The constant $\ell_2=-\log_D\frac12=\Lambda\bigl(\frac12\bigr)$ satisfies
\[\log_D2x=\log_Dx+\ell_2\]
for all $x\in\bigl(\frac13,\frac34\bigr)^*$.
\end{Lem}
\begin{Pf}
First, if $x\in\bigl(\frac13,\frac23\bigr)^*$, we have
\[\log_D2x+\log_D\tfrac12=\log_Dx\]
by eq.~\eqref{eq:4} in Lemma~\ref{lem:log-disc} with $r=\abs{1-2x}<^*\frac13$ and $s=\frac12$.
Next, if $x\in\bigl[\frac12,\frac34\bigr)^*$, let $y=\frac23-h^{-1}$ for sufficiently large $h\in\ML$. Then
\[\log_D2x+\log_Dy=\log_D2xy=\log_Dx+\log_D2y=\log_Dx+\log_Dy+\ell_2\colon\]
the first equality follows from~\eqref{eq:4} with $r=2x-1<^*\frac12$ and $s=1-y=\frac13+h^{-1}$; we can choose $h$ such
that $(1+r)(1+s)<^*2$. The second equality follows from \eqref{eq:4} with $r=1-x\le\frac12$ and $s=2y-1<^*\frac13$.
\end{Pf}
\begin{Lem}\label{lem:log-r}
There is a unique continuous function $\log_{\MR}\colon\MR_{>0}\to\MRL$ such that
\[\log_{\MR}2^nx=\log_Dx+n\ell_2\]
for all $n\in\MZL$ and $x\in\bigl(\frac13,\frac32\bigr)^*$.
\end{Lem}
\begin{Pf}
Every $x\in\MR_{>0}$ can be written as $2^nx'$ for some $n\in\MZL$ and
$x'\in\bigl[\frac12,1\bigr]$. Since $\lambda(z,n)\in\MQ$ for $z\in\MQ$, $\Lambda$ maps
$D^*_1(0)\cap\MQ$ to $\MR$, hence $\log_D$ maps $D^*_1(1)\cap\MR$ to $\MR$. Moreover, Lemma~\ref{lem:log-disc}
(eq.~\eqref{eq:3}) implies that
$\log_D$ maps $\bigl[\frac12,\frac32\bigr]$ to $\bigl[-\frac34,\frac34\bigr]$. Thus, $\log_Dx'+n\ell_2\in\MRL$.

Assume $2^nx=2^my$ for some $n,m\in\MZL$ and $x,y\in\bigl(\frac13,\frac32\bigr)^*$, $x<y$. Then
$2^{n-m}=y/x\in\bigl(1,\frac92\bigr)$, hence $n=m+1$, $y=2x$, and $x\in\bigl(\frac13,\frac34\bigr)^*$, or $n=m+2$,
$y=4x$, and $x\in\bigl(\frac13,\frac38\bigr)^*$. Thus,
\[\log_Dx+n\ell_2=\log_Dy+m\ell_2\]
using Lemma~\ref{lem:log2}.

The continuity of $\log_D$ ensures that $\log_{\MR}$ is continuous on $\bigl(\frac132^n,\frac322^n)^*$ for each
$n\in\MZL$. These open sets cover $\MR_{>0}$, hence $\log_{\MR}$ is continuous.
\end{Pf}

While we eventually want to show that $\log_\MR$ is an inverse of $\exp_{\MRL}$, we will be content for now with
proving that it is an ordered group embedding $\p{\MR_{>0},\cdot,<}\to\p{\MRL,+,<}$.
\begin{Lem}\label{lem:log-r-incr}
The function $\log_{\MR}$ is strictly increasing (hence injective).
\end{Lem}
\begin{Pf}
We first show that $\log_D$ is strictly increasing on $\bigl[\frac12,1\bigr]$. Let $\frac12\le x<y\le1$, and consider
$u=x/y\in\bigl[\frac12,1\bigr)$. We have $\log_Du\le(u-1)+(u-1)^2=u(u-1)<0$ by Lemma~\ref{lem:log-disc} (eq.~\eqref{eq:3}), thus
\[\log_Dx=\log_Du+\log_Dy<\log_Dy\]
by \eqref{eq:4}, as long as $u>^*\frac23$: then $r=\abs{1-y}\le\frac12$ and $s=1-u<^*\frac13$, thus
$(1+r)(1+s)<^*2$. It follows that $\log_D$ is strictly increasing on $\bigl[\frac12,\frac34\bigr)^*$ and on
$\bigl(\frac23,1\bigr]^*$, hence on $\bigl[\frac12,1\bigr]$.

Now, let $x,y\in\MR_{>0}$, and assume $x<y$. Write $x=2^nx'$ and $y=2^my'$ with $n,m\in\MZL$ and
$x',y'\in\bigl(\frac12,1\bigr]$. The previous part ensures that $\log_{\MR}$ is strictly increasing on $[2^{n-1},2^n]$,
hence $\log_{\MR} x<\log_{\MR} y$ if $m=n$. If $m>n$, we have
\[\log_{\MR} y>(m-1)\ell_2\ge n\ell_2\ge\log_{\MR} x\]
as $\log_D$ maps $\bigl(\frac12,1\bigr]$ to $(-\ell_2,0]$.
\end{Pf}
\begin{Lem}\label{lem:log-r-homo}
For all $x,y\in\MR_{>0}$,
\[\log_{\MR} xy=\log_{\MR} x+\log_{\MR} y.\]
\end{Lem}
\begin{Pf}
Write $x=2^nx'$ and $y=2^my'$, where $n,m\in\MZL$ and
$x',y'\in\bigl[\frac35,\frac65\bigr]\sset\bigl(\frac13,\frac32\bigr)^*$. Then $xy=2^{n+m}x'y'$, where
$x'y'\in\bigl[\frac9{25},\frac{36}{25}\bigr]\sset\bigl(\frac13,\frac32\bigr)^*$. It follows that
\[\log_{\MR} xy=(n+m)\ell_2+\log_Dx'y'=n\ell_2+\log_Dx'+m\ell_2+\log_Dy'=\log_{\MR} x+\log_{\MR} y\]
using~\eqref{eq:4}, as $\abs{1-x'},\abs{1-y'}\le\frac25$.
\end{Pf}

\subsection{Logarithm in a sector}\label{sec:logarithm-sector}

The next extension of $\log$ is to an angular sector by means of $\log z=\log_\MR\abs z+\log_D\bigl(z/\abs z\bigr)$, as long as $z/\abs
z$ is close enough to~$1$.
\begin{Def}\label{def:log-s}
We define the \emph{complex sign} function $\sgn\colon\MC_{\ne0}\to\bigl\{z:\abs z=1\bigr\}$
by $\sgn z=z/\abs z$. We consider the sector
\[S=\bigl\{z\in\MC_{\ne0}:\abs{\sgn z-1}<^*1\bigr\}.\]
We define a continuous function $\log_S\colon S\to\MCL$ by
\[\log_Sz=\log_{\MR}\abs z+\log_D\sgn z.\]
\end{Def}

Let us clarify the geometry of~$S$:
\begin{Lem}\label{lem:sect}
Let $z=x+iy\in\MC_{\ne0}$. Then
\[z\in S\iff\frac x{\abs z}>^*\frac12\iff x>0\land\frac{\abs y}{\abs z}<^*\frac{\sqrt3}2\iff x>0\land\frac{\abs y}x<^*\sqrt3.\]
\end{Lem}
\begin{Pf}
We have $z\in S$ iff $\abs{\sgn z-1}^2<^*1$, where
\[\abs{\sgn z-1}^2=\frac{\bigl(z-\abs z\bigr)\bigl(\ob z-\abs z\bigr)}{\abs z^2}
=\frac{2\abs z^2-(z+\ob z)\abs z}{\abs z^2}=2\left(1-\frac x{\abs z}\right),\]
which proves the first equivalence. The rest follows easily, using $x^2+y^2=\abs z^2$.
\end{Pf}

Basic properties of $\log_S$ are easy to establish by combining the properties of $\log_D$ and $\log_\MR$:
\begin{Lem}\label{lem:log-s}
\ \begin{enumerate}
\item\label{item:13}
If $z,w\in S$ and $\bigl(1+\abs{\sgn z-1}\bigr)\bigl(1+\abs{\sgn w-1}\bigr)<^*2$, then
\[\log_Szw=\log_Sz+\log_Sw.\]
In particular, this holds if $z=x+iy$ and $w=u+iv$ satisfy $\abs{y/z},\abs{v/w}\le\frac25$.
\item\label{item:24}
If $z=x+iy\in S$ satisfies $\abs{y/z}\le\frac25$, then $\log_Sz^{-1}=-\log_Sz$.
\item\label{item:14}
For any $x\in\MR_{>0}$, $\log_Sx=\log_{\MR} x$.
\item\label{item:15}
If $z\in\ob D_{2/5}(1)$, then $\log_Sz=\log_Dz$.
\end{enumerate}
\end{Lem}
\begin{Pf}

\ref{item:13}: We have $\sgn zw=\sgn z\sgn w$, hence the identity follows from eq.~\eqref{eq:4} in
Lemma~\ref{lem:log-disc} and Lemma~\ref{lem:log-r-homo}.

If $\abs{y/z}\le0{.}4$, then $x^2/\abs z^2\ge1-0{.}4^2=0{.}84$, hence $x/\abs z\ge0{.}916$, and
$\abs{\sgn z-1}^2=2\bigl(1-x/\abs z\bigr)\le0{.}168$, thus $\abs{\sgn z-1}\le0{.}41$. Likewise,
$\abs{v/w}\le0{.}4$ implies $\abs{\sgn w-1}\le0{.}41$, hence
$\bigl(1+\abs{\sgn z-1}\bigr)\bigl(1+\abs{\sgn w-1}\bigr)\le1{.}41^2<^*2$.

\ref{item:24}: Write $z^{-1}=u+iv$. Since $z^{-1}=\ob z/\abs z^2$, we have $v/\abs{z^{-1}}=-y/\abs z$, hence
$\log_Sz+\log_Sz^{-1}=0$ by~\ref{item:13}.

\ref{item:14} is immediate, as $\sgn x=1$ for $x\in\MR_{>0}$.

\ref{item:15}: Assume $\abs{z-1}\le\frac25$. The triangle inequality (Lemma~\ref{lem:abs-bd}) implies
$\abs z\in\bigl[\frac35,\frac75\bigr]\sset\bigl(\frac13,\frac32\bigr)^*$, thus
\[\log_Sz=\log_{\MR}\abs z+\log_D\sgn z=\log_D\abs z+\log_D\sgn z,\]
and $\Abs[big]{\abs z-1}\le\frac25$. Moreover,
\[\frac25\ge\abs{z-1}=\Abs{\frac z{\abs z}\left(\abs z-\frac{\ob z}{\abs z}\right)}=\Abs{\abs z-\frac{\ob z}{\abs z}}
\ge\frac{\abs y}{\abs z},\]
hence $\abs{\sgn z-1}\le0{.}41$ by the proof of~\ref{item:13}. Thus,
\[\log_D\abs z+\log_D\sgn z=\log_Dz\]
by~\eqref{eq:4}.
\end{Pf}

\subsection{Complex square root}\label{sec:complex-square-root}

The idea for our final extension of $\log$ is to widen the domain of $\log_S$ by several iterations of $\log
z=2\log\sqrt z$ (each of which doubles the angle of the sector) until it covers all of $\MC_{\ne0}$. To do that, we
need first to define carefully the complex square root function with the right branch cut, and establish its properties.
\begin{Def}\label{def:sqrt}
We define the \emph{lopsided sign} function $\sgn^+\colon\MR\to\{-1,1\}$ by
\[\sgn^+y=\begin{cases}\phantom+1&\text{if $y\ge0$,}\\-1&\text{if $y<0$,}\end{cases}\]
and the \emph{complex square root} function $\sqrt{\phantom-}\colon\MC\to\MC$ for $z=x+iy$ by
\[\sqrt z=\sqrt{\frac{\abs z+x}2}+i\sqrt{\frac{\abs z-x}2}\sgn^+y.\]
\end{Def}
\begin{Lem}\label{lem:sqrt}
\ \begin{enumerate}
\item\label{item:16}
For all $z\in\MC$, $\bigl(\sqrt z\bigr)^2=z$ and $\sgn^+\Im\sqrt z=\sgn^+\Im z$.
\item\label{item:25}
For any $z\in\MC\bez\MR_{<0}$, $\sqrt{\ob z}=\ob{\sqrt z}$.
\item\label{item:17}
The restrictions $\sqrt{\phantom-}\res\MC\bez\MR_{<0}$ and $\sqrt{\phantom-}\res\{z:\Im z\ge0\}$ are continuous.
\end{enumerate}
\end{Lem}
\begin{Pf}

\ref{item:16}: Write $z=x+iy$ and $\sqrt z=u+iv$. We have $u^2-v^2=x$ and
\[2uv=2\sqrt{\frac{\abs z^2-x^2}4}\sgn^+y=\abs y\sgn^+y=y,\]
thus $(u+iv)^2=z$. Clearly, $\sgn^+v=\sgn^+y$ if $v\ne0$; otherwise, $y=v=0$.

\ref{item:25}: For $z\in\MR_{\ge0}$, $\sqrt z\in\MR$; for $z\notin\MR$, the only difference between $\sqrt z$ and
$\sqrt{\ob z}$ is that $\sgn^+y$ is negated.

\ref{item:17}: Being a composition of continuous functions, $\sqrt{\phantom-}\res\{z:\Im z\ge0\}$ is continuous. Thus,
in view of~\ref{item:25}, $\sqrt{\phantom-}\res\{z\notin\MR_{<0}:\Im z\le0\}$ is continuous, which we can glue with
$\sqrt{\phantom-}\res\{z\notin\MR_{<0}:\Im z\ge0\}$ to get the continuity of $\sqrt{\phantom-}\res\MC\bez\MR_{<0}$.
\end{Pf}

The next lemma elucidates how we can use $\sqrt{\phantom-}$ to enlarge the defining sector of $\log$.
\begin{Lem}\label{lem:sqrt-iter}
Let $z\in\MC$ and $w=\sqrt z$.
\begin{enumerate}
\item\label{item:18}
$\Re w\ge0$, with strict inequality if $z\notin\MR_{\le0}$.
\item\label{item:19}
If $\Re z\ge0$, then $\abs{\Im w}\le\Re w$, with strict inequality if $\Re z>0$.
\item\label{item:20}
If $\abs{\Im z}\le\Re z$, then $\abs{\Im w}\le\frac25\abs w$.
\end{enumerate}
\end{Lem}
\begin{Pf}
\ref{item:18} and~\ref{item:19} are immediate from the definition.

\ref{item:20}: Write $z=x+iy$ and $w=u+iv$. We have $x\ge0$ and $2x^2\ge\abs z^2$, hence $x\ge\frac7{10}\abs z$. It
follows that $v^2=\frac12\bigl(\abs z-x\bigr)\le\frac3{20}\abs z=\frac3{20}\abs w^2$, thus $\abs v\le\frac25\abs w$.
\end{Pf}

As we already mentioned, a crucial property of $\log$ we need to show is $\log zw=\log z+\log w$ under suitable
restrictions on $z,w$. Extending this identity from $\log_S$ to full $\log$ will require multiplicativity of the square
root function, thus let us establish some convenient conditions under which the latter holds.
\begin{Lem}\label{lem:sqrt-mul}
If $z,w\in\MC$ are such that $\Re z\ge0$ and $\Re w>0$, then $\sqrt{zw}=\sqrt z\sqrt w$.
\end{Lem}
\begin{Pf}
Since $(\sqrt z\sqrt w)^2=zw$, we have $\sqrt z\sqrt w=\pm\sqrt{zw}$; we only need to check that the sign is correct.
Write $\sqrt z=x+iy$, $\sqrt w=u+iv$. We have $x\ge\abs y$ and $u>\abs v$ by Lemma~\ref{lem:sqrt-iter}, hence
$\Re\bigl(\sqrt z\sqrt w\bigr)=xu-yv>0$ (unless $x=y=0$, thus $z=0$), which means, in view of
Lemma \ref{lem:sqrt-iter}~\ref{item:18}, that $\sqrt{zw}\ne-\sqrt z\sqrt w$. Thus, $\sqrt{zw}=\sqrt z\sqrt w$ as required.
\end{Pf}

We can, in fact, formulate a comprehensive criterion for multiplicativity of $\sqrt{\phantom-}$. (One can check that
the sufficient condition below is also necessary up to exchanging $z$ and~$w$, though we will not need this.)
\begin{Lem}\label{lem:sqrt-mul-full}
Let $z,w\in\MC$ be such that $\sgn^+\Im zw\in\{\sgn^+\Im z,\sgn^+\Im w\}$ and $z\notin\MR_{<0}$.
Then $\sqrt{zw}=\sqrt z\sqrt w$.
\end{Lem}
\begin{Pf}
Write $\sqrt z=x+iy$, $\sqrt w=u+iv$. Assume first that $\sgn^+\Im z\ne\sgn^+\Im w$; say, $\Im z\ge0>\Im w$. Then
$y\ge0>v$ by Lemma~\ref{lem:sqrt}, and $x\ge0$, $u>0$ by Lemma~\ref{lem:sqrt-iter}, hence $\Re\bigl(\sqrt z\sqrt
w\bigr)=xu-yv>0$ unless $x=y=0$ (in which case $z=0=zw$). Thus, $\sqrt{zw}=\sqrt z\sqrt w$ using Lemma~\ref{lem:sqrt-iter}.

Next, assume $\Im z,\Im w,\Im(zw)<0$. Then $x,u>0>y,v$ by Lemmas \ref{lem:sqrt} and~\ref{lem:sqrt-iter}, thus $\Im\bigl(\sqrt z\sqrt
w\bigr)=xv+yu<0$, which implies $\sqrt{zw}=\sqrt z\sqrt w$ using Lemma~\ref{lem:sqrt}.

Finally, if $\Im z,\Im w,\Im zw\ge0$, we have $x,u,y,v\ge0$, hence $\Im\bigl(\sqrt z\sqrt w\bigr)=xv+yu\ge0$. If the
inequality is strict, we get $\sqrt{zw}=\sqrt z\sqrt w$ using Lemma~\ref{lem:sqrt} again. Otherwise $xv=yu=0$: thus, $x=0$
(in which case either $z=0=zw$ or $z\in\MR_{<0}$, but the latter is ruled out), or $u=v=0$ (in which case $w=0=zw$), or
$y=v=0$ (in which case $z,w,zw\in\MR_{\ge0}$).
\end{Pf}
\begin{Cor}\label{cor:sqrt-inv}
If $z\in\MC\bez\MR_{<0}$, then $\sqrt{z^{-1}}=\bigl(\sqrt z\bigr)^{-1}$.
\end{Cor}
\begin{Pf}
$\Im z$ and $\Im z^{-1}$ have opposite signs.
\end{Pf}
\begin{Rem}\label{rem:sqrt-3}
We can restate Lemma~\ref{lem:sqrt-mul-full} in the following symmetric form: if $z_0,z_1,z_2\in\MC$ are such that
$z_0z_1z_2=1$, then $\sqrt{z_0}\sqrt{z_1}\sqrt{z_2}=1$, unless $\sgn^+z_0=\sgn^+z_1=\sgn^+z_2$ and at most one
$z_j$ is in $\MR_{>0}$. Lemma \ref{lem:log-c-homo}~\ref{item:22} below can also be stated like this.
\end{Rem}

\subsection{Full complex logarithm}\label{sec:full-compl-logar}

In this section, we are going to finish the definition of $\log$ (and, eventually, $\exp$), and prove the remaining
desired properties of $\log$ and $\exp$, such as the right inverse property $\exp\log z=z$.
\begin{Def}\label{def:log-c}
We put $\sqrt[4]z=\sqrt{\sqrt z}$, $\sqrt[8]z=\sqrt{\sqrt[4]z}$, and define $\log_{\MC}\colon\MC_{\ne0}\to\MCL$ by
\[\log_{\MC} z=8\log_S\sqrt[8]z.\]
Note that $\sqrt[8]z\in S$ by Lemmas \ref{lem:sqrt-iter} and~\ref{lem:sect}.
\end{Def}

In fact, we have even $\sqrt[4]z\in S$ for all $z\ne0$, hence already $4\log_S\sqrt[4]z$ would define $\log$
on all of $\MC_{\ne0}$. The purpose of the extra iteration of $\sqrt{\phantom-}$ is to facilitate our proof of $\log
zw=\log z+\log w$ below.

We start with a few basic properties that follow either directly from the definition, or from properties of $\log_S$
and $\sqrt{\phantom-}$.
\begin{Lem}\label{lem:log-cont}
The restrictions ${\log_{\MC}}\res\MC\bez\MR_{<0}$ and ${\log_{\MC}}\res\{z:\Im z\ge0\}$ are continuous.
\end{Lem}
\begin{Pf}
This follows from Lemma \ref{lem:sqrt}~\ref{item:17} and the continuity of $\log_S$.
\end{Pf}
\begin{Lem}\label{lem:log-c-s}
If $z\in\MC_{\ne0}$ and $\abs{\Im z}\le\Re z$, then $\log_{\MC} z=\log_Sz$.

Consequently, $\log_{\MC} z=\log_{\MR} z$ for $z\in\MR_{>0}$, and $\log_{\MC} z=\log_Dz$ for $z\in\ob D_{2/5}(1)$.
\end{Lem}
\begin{Pf}
By Lemma~\ref{lem:sqrt-iter}, $w=\sqrt z$ satisfies $\abs{\Im w}\le\frac25\abs w$, hence $\log_Sz=2\log_Sw$ by
Lemma~\ref{lem:log-s}. Iterating this argument, we obtain $\log_Sz=8\log_S\sqrt[8]z=\log_{\MC} z$. The rest
follows by Lemma \ref{lem:log-s} \ref{item:14} and~\ref{item:15}.
\end{Pf}

We will improve Lemma~\ref{lem:log-c-s} in Lemma~\ref{lem:log-all}.
\begin{Lem}\label{lem:log-c-homo}
\ \begin{enumerate}
\item\label{item:21}
If $z,w\in\MC$ are such that $\Re z\ge0$ and $\Re w>0$, then
\begin{equation}\label{eq:5}
\log_{\MC} zw=\log_{\MC} z+\log_{\MC} w.
\end{equation}
\item\label{item:22}
If $z,w\in\MC_{\ne0}$ satisfy $\sgn^+\Im zw\in\{\sgn^+\Im z,\sgn^+\Im w\}$ and
$z\notin\MR_{<0}$, then \eqref{eq:5} holds.
\item\label{item:27}
For all $z\in\MC_{\ne0}$, $\log_{\MC} z=2\log_{\MC}\sqrt z$.
\item\label{item:23}
If $z\in\MC\bez\MR_{\le0}$, then $\log_{\MC} z^{-1}=-\log_{\MC} z$ and $\log_{\MC}\ob z=\ob{\log_{\MC} z}$.
\item\label{item:26}
If $\abs z=1$, then $\log_{\MC} z\in i\MRL$.
\end{enumerate}
\end{Lem}
\begin{Pf}

\ref{item:22}: The assumption implies $\sqrt{zw}=\sqrt z\sqrt w$ by Lemma~\ref{lem:sqrt-mul}. Moreover,
$\sgn^+\Im\sqrt z=\sgn^+\Im z$ and $\sqrt z\notin\MR_{<0}$, and similarly for $w$ and~$zw$, hence the
original assumptions continue to hold with $\sqrt z$, $\sqrt w$ in place of $z,w$; thus, we can iterate the argument,
eventually obtaining $\sqrt[8]{zw}=\sqrt[8]z\sqrt[8]w$. By Lemma~\ref{lem:sqrt-iter},
$\Abs[big]{\Im\sqrt[8]z}\le\frac25\Abs[big]{\sqrt[8]z}$, and similarly for $\sqrt[8]w$, hence
\[\log_S\sqrt[8]z+\log_S\sqrt[8]w=\log_S\bigl(\sqrt[8]z\sqrt[8]w\bigr)
=\log_S\sqrt[8]{zw}\]
by Lemma~\ref{lem:log-s}. Multiplying by~$8$ yields \eqref{eq:5}.

\ref{item:27} follows from~\ref{item:22} using Lemma \ref{lem:sqrt}~\ref{item:16}.

The proofs of \ref{item:21} and the first identity in~\ref{item:23} are similar to~\ref{item:22}, using
Lemma~\ref{lem:sqrt-mul} and Corollary~\ref{cor:sqrt-inv} in place of Lemma~\ref{lem:sqrt-mul-full}.

In order to prove $\log_{\MC}\ob z=\ob{\log_{\MC} z}$ for $z\notin\MR_{\le0}$, we first observe that $\lambda(\ob
z,n)=\ob{\lambda(z,n)}$ for all $z\in\MQ(i)$ and $n\in\ML$, hence $\Lambda(\ob z)=\ob{\Lambda(z)}$ for $z\in
D_1^*(0)\cap\MQ(i)$, and $\log_D\ob z=\ob{\log_Dz}$ for all $z\in D^*_1(1)$ by density. Since $\log_{\MR}$ is real and
$\sgn\ob z=\ob{\sgn z}$, we obtain $\log_S\ob z=\ob{\log_Sz}$ for all $z\in S$. This implies $\log_{\MC}\ob
z=\ob{\log_{\MC} z}$ for all $z\in\MC\bez\MR_{\le0}$ using Lemma \ref{lem:sqrt}~\ref{item:25}.

\ref{item:26}: If $z\ne-1$, we have $\ob{\log_{\MC} z}=\log_{\MC}\ob z=\log_{\MC} z^{-1}=-\log_{\MC} z$, hence
$\log_{\MC} z$ is purely imaginary. The case $z=-1$ follows using~\ref{item:27}.
\end{Pf}

We now make a short detour: we define the argument function and establish its monotonicity properties. Besides being
useful in its own right, our immediate goal here is to prove that $\log_\MC$ is injective (which will be instrumental
in deriving $\exp\log z=z$ from $\log\exp z=z$): the injectivity of $\log_\MR$ ensures that $\Re\log_\MC z=\log_\MR\abs
z$ distinguishes numbers with different absolute values, and $\arg$ will give us a handle on numbers with the same
absolute value. It will also help us establish the image of $\log_\MC$.
\begin{Def}\label{def:arg}
If $z\in\MC_{\ne0}$, we define $\arg z=\Im\log_{\MC} z$. Let $\pi=\arg(-1)$.
\end{Def}
\begin{Lem}\label{lem:arg}
For any $z\in\MC_{\ne0}$, $\arg z=\arg\sgn z$, and $\log_{\MC} z=\log_{\MR}\abs z+i\arg z$.
\end{Lem}
\begin{Pf}
We have $\log_{\MC} z=\log_{\MR}\abs z+\log_{\MC}\sgn z$ by Lemma \ref{lem:log-c-homo}~\ref{item:22}, which implies
$\Im\log_{\MC} z=\Im\log_{\MC}\sgn z$, and $\Re\log_{\MC} z=\log_{\MR}\abs z$ using
Lemma \ref{lem:log-c-homo}~\ref{item:26}.
\end{Pf}
\begin{Lem}\label{lem:arg-mon}
Let $z,w\in\MC_{\ne0}$.
\begin{enumerate}
\item\label{item:36}
If $z\notin\MR_{<0}$, then $\arg z^{-1}=\arg\ob z=-\arg z$.
\item\label{item:43}
If $\Re z\ge0$ or $\Im z<0$, then $\arg iz=\arg z+\frac\pi2$.
\item\label{item:37}
If $\Im z\ge0$ and $z\notin\MR_{>0}$, then $\arg z>0$.
\item\label{item:38}
If $\Re z,\Re w\ge0$, then $\arg z<\arg w$ iff $\Im\sgn z<\Im\sgn w$.
\item\label{item:39}
If $\Im z,\Im w\ge0$, then $\arg z<\arg w$ iff $\Re\sgn z>\Re\sgn w$.
\item\label{item:44}
$\arg z=\arg w$ iff $\sgn z=\sgn w$.
\end{enumerate}
\end{Lem}
\begin{Pf}

\ref{item:36} is immediate from \ref{lem:log-c-homo}~\ref{item:23}.

\ref{item:43}: We observe that $\arg i=\frac\pi2$ by Lemma~\ref{lem:log-c-homo}~\ref{item:27}, and apply
Lemma~\ref{lem:log-c-homo}~\ref{item:22}: we have $\Im i>0$ and $i\notin\MR_{<0}$; if $\Im z<0$, then
$\sgn^+\Im iz\in\{1,-1\}$ trivially, and if $\Re z\ge0$, then $\sgn^+\Im iz=1$.

\ref{item:37}: We may assume $\abs z=1$. Then $i\arg z=\log_{\MC} z=8\log_Sw=8\log_Dw$, where $w=\sqrt[8]z=u+iv$
satisfies $0<v\le\frac25$ by Lemma~\ref{lem:sqrt}~\ref{item:16} and Lemma~\ref{lem:sqrt-iter}, thus $\abs{w-1}\le0{.}41$
by the proof of Lemma~\ref{lem:log-s}~\ref{item:13}, whence $\Im\log_Dw\ge v-\abs{w-1}^2=v+2u-2$ using
Lemma~\ref{lem:log-disc}. Since $v\ge\frac52v^2$, we have $(2-v)^2=4-4v+v^2\le4-9v^2<4u^2$, thus $2-v<2u$, i.e.,
$v+2u-2>0$.

\ref{item:38}: We may assume $\abs z=\abs w=1$. Write $z=x+iy$ and $w=u+iv$, thus $x,u\ge0$.

If $v>y\ge0$, then $x^2=1-y^2>1-v^2=u^2$, thus $x>u\ge0$, and $\Im w\ob z=xv-yu>0$. It follows that
\[\arg w-\arg z=\arg w+\arg\ob z=\arg w\ob z>0\]
using \ref{item:36}, \ref{item:37}, and Lemma~\ref{lem:log-c-homo}~\ref{item:21}.

If $0\ge v>y$, then $\arg w=-\arg\ob w>-\arg\ob z=\arg z$ using the previous part (with $z$ and~$w$ swapped)
and~\ref{item:36}.

If $v>0>y$, then $\arg w>\arg1>\arg z$ by the previous two cases. This completes the proof of the right-to-left
implication in~\ref{item:38}.

If $y>v$, then $\arg z>\arg w$ by what we have already proved. If $y=v$, then $x^2=u^2$, thus $x=u$, i.e., $z=w$ and
$\arg z=\arg w$.

\ref{item:39}: We have $z=iz'$, $w=iw'$ for some $z',w'$, which satisfy $\Re z',\Re w'\ge0$,
$\Im\sgn z'=-\Re\sgn z$, $\Im\sgn w'=-\Re\sgn w$, and $\arg z<\arg w$ iff $\arg z'<\arg w'$
by~\ref{item:43}, hence the result follows from~\ref{item:38}.

\ref{item:44}: We have $\sgn^+\arg z=\sgn^+\Im z$ by \ref{item:37} and~\ref{item:36}. Thus, if $\arg z=\arg w$,
then either $\Im z,\Im w\ge0$, in which case $\sgn z=\sgn w$ by~\ref{item:39}, or $\Im z,\Im w<0$, which reduces to
the previous case by~\ref{item:36}.
\end{Pf}
\begin{Cor}\label{cor:log-inj}
The function $\log_{\MC}$ is injective.
\end{Cor}
\begin{Pf}
If $\log_{\MC} z=\log_{\MC} w$, then $\log_{\MR}\abs z=\log_{\MR}\abs w$ and $\arg z=\arg w$ by
Lemma~\ref{lem:arg}. The former implies $\abs z=\abs w$ by Lemma~\ref{thm:exp-homo}, while the latter implies
$\sgn z=\sgn w$ by Lemma~\ref{lem:arg-mon}. Thus, $z=w$.
\end{Pf}
\begin{Cor}\label{cor:log-range}
For any $z\in\MC_{\ne0}$, $\arg z\in(-\pi,\pi]$ and $\log_{\MC} z\in\MRL+i(-\pi,\pi]$.
\end{Cor}
\begin{Pf}
If $\Im z\ge0$, we have $0\le\arg z\le\pi$ by Lemma~\ref{lem:arg-mon}~\ref{item:39}. If $\Im z<0$, then
$0<\arg\ob z=-\arg z<\pi$ using Lemma~\ref{lem:arg-mon}~\ref{item:36}.
\end{Pf}

Let us prove a yet another version of eq.~\eqref{eq:5} in Lemma~\ref{lem:log-c-homo}, this time indicating exactly how much it may be off in cases where
it does not hold.
\begin{Lem}\label{lem:log-homo-all}
For all $z,w\in\MC_{\ne0}$,
\[\log_{\MC} z+\log_{\MC} w-\log_{\MC} zw\in\{-2\pi i,0,2\pi i\}.\]
\end{Lem}
\begin{Pf}
We first observe that $\log_{\MC}(-z)=\log_{\MC} z+\pi i$ when $\Im z<0$ or $z\in\MR_{>0}$ by
Lemma~\ref{lem:log-c-homo} \ref{item:22}, hence
\[\log_{\MC}(-z)=\log_{\MC} z\pm\pi i\]
for all $z\in\MC_{\ne0}$. Now, given $z,w\in\MC_{\ne0}$, let $z'=\pm z$ and $w'=\pm w$ be such that $\Re z'>0$, or $\Re
z'=0$ and $\Im z'>0$, and similarly for $w'$. Then
\[\log_{\MC} z'+\log_{\MC} w'=\log_{\MC} z'w'\]
by Lemma~\ref{lem:log-c-homo}~\ref{item:21} (unless $\Re z'=\Re w'=0$, i.e., $z',w'\in i\MR_{>0}$, in which case the
identity holds as well). Since $0$ or~$2$ of $z'$, $w'$, $z'w'$ are negated as compared to $z$, $w$, $zw$ (resp.),
$\log_{\MC} z+\log_{\MC} w-\log_{\MC} zw$ is a sum of $0$ or~$2$ terms of the form $\pm\pi i$, which gives the result.
\end{Pf}

One application is to estimate the value of~$\pi$, which we indicate below (without actually carrying out
the computation with standard rationals in the proof). The same argument can establish in $\vtc$ all true
inequalities of the form $q<\pi<r$ where $q,r\in\Q$ are standard.
\begin{Prop}\label{prop:pi-apx}
$3{.}1<\pi<3{.}2$.
\end{Prop}
\begin{Pf}
Let $z=1+\frac1{100}i$. We have $\Abs{\log_{\MC} z-\frac1{100}i}\le\frac1{10000}$ by Lemmas \ref{lem:log-c-s} and~\ref{lem:log-disc},
hence $\arg z\in[0{.}0099,0{.}0101]$. One can check $\Im z^n>0$ for $n=1,\dots,314$, thus
$\log_{\MC} z^{314}=314\log_{\MC} z$ by repeated use of Lemma~\ref{lem:log-c-homo}~\ref{item:22}, which implies
$\pi\ge314\arg z\ge3{.}1086$ by Corollary~\ref{cor:log-range}. On the other hand, $\Im z^{315}<0$, thus $\arg z^{315}<0$ by
Lemma~\ref{lem:arg-mon}; in view of Lemma~\ref{lem:log-homo-all} and Corollary~\ref{cor:log-range}, this means
$-\pi<\log_{\MC} z^{315}=315\arg z-2\pi$, thus $\pi<315\arg z\le3{.}1815$. 
\end{Pf}

We now come to the crucial Cauchy functional equation for $\log\exp z$.
\begin{Lem}\label{lem:logexp-hom}
If $z\in\MRL+i(-1,1)$, then $\Re\exp_{\MCL} z>0$. Consequently,
\begin{equation}\label{eq:6}
\log_{\MC}\exp_{\MCL}(z+w)=\log_{\MC}\exp_{\MCL} z+\log_{\MC}\exp_{\MCL} w
\end{equation}
for all $z,w\in\MRL+i(-1,1)$.
\end{Lem}
\begin{Pf}
Let $z=x+iy$ with $x\in\MRL$ and $y\in(-1,1)$. We have $\Abs[big]{\exp iy-(1+iy)}\le y^2$ by Lemma~\ref{thm:exp-homo},
hence $\abs{\Re\exp iy-1}\le y^2<1$. Thus, $\Re\exp z=\exp x\Re\exp iy>0$.

Assuming the same holds for $w$, we have
\[\log\exp(z+w)=\log(\exp z\exp w)=\log\exp z+\log\exp w\]
by Lemmas \ref{thm:exp-homo} and~\ref{lem:log-c-homo}.
\end{Pf}

We are heading towards a proof of $\log\exp z=z$ when $z$ is sufficiently small. To this end, we intend to use the
identity $\log\exp2^{-n}z=2^{-n}\log\exp z$ for $n\in\ML$. This appears to follow ``by induction on $n$''
from~\eqref{eq:6}, but there is an obstacle to formalizing this idea: $z$, $\exp z$, and $\log\exp z$ are
elements of the completion $\MC^\sM$, which is too big to be definable in~$\sM$, hence we cannot use induction
directly. Instead, we need to argue about Gaussian rational approximations of $\exp z$ and $\log\exp z$ for
$z\in\MQL(i)$; moreover, we need to make sure these approximations are computable by $\tc$ functions so that the
induction formula has the right complexity for $\vtc$. On a more fundamental level, we need such approximations so that
the facts we prove about $\exp$, $\log$, and other functions $\MC\to\MC$ can be transferred back to
the language of $\vtc$. This is the goal of the next lemma.
\begin{Lem}\label{lem:log-exp-apx}
We can construct $\tc$~functions $E_{\MCL}(z,r,n)$, $\M{SR}_{\MR}(x,n)$, $A_\times(z,n)$, $A_+(z,n)$, $\M{SR}_{\MC}(z,n)$,
$L_D(z,r,n)$, $L_{\MR}(x,n)$, $L_{\MC}(z,n)$, and $\M{LE}(z,r,n)$ with the following properties.
\begin{enumerate}
\item\label{item:11}
$\E_{\MCL}(z,r,n)$ is a multiplicative approximation of $\exp_{\MCL} z$ for $z\in\MQL(i)$, parametrized by $r\in\ML$ such
that $\abs z\le r$.
\item\label{item:31}
$\M{SR}_{\MR}(x,n)$ is a multiplicative approximation of $\sqrt x$ for $x\in\MQ_{>0}$.
\item\label{item:71}
$A_\times(z,n)$ and $A_+(z,n)$ are multiplicative and additive (respectively) approximations of $\abs z\in\MR$ for
$z\in\MQ(i)$.
\item\label{item:30}
$\M{SR}_{\MC}(z,n)$ is a multiplicative approximation of $\sqrt z$ for $z\in\MQ(i)\bez\{0\}$.
\item\label{item:28}
$L_D(z,r,n)$ is an additive approximation of $\log_Dz$ for $z\in D^*_1(1)\cap\MQ(i)$, parametrized by $r\in\ML$ such
that $\abs{z-1}\le1-r^{-1}$.
\item\label{item:29}
$L_{\MR}(x,n)$ is an additive approximation of $\log_{\MR} x$ for $x\in\MQ_{>0}$.
\item\label{item:10}
$L_{\MC}(z,n)$ is an additive approximation of $\log_{\MC} z$ for $z\in\MQ(i)\bez\{0\}$.
\item\label{item:12}
$\M{LE}(z,r,n)$ is an additive approximation of $\log_{\MC}\exp_{\MCL} z$ for $z\in\MQL(i)$ with $\abs{\Im z}<1$,
parametrized by $r\in\ML$ such that $\abs z\le r$.
\end{enumerate}
\end{Lem}
\begin{Pf}[sketch]
We employ the $e(z,n)$ function to construct $E_{\MCL}(z,r,n)$. The results of \cite{ej:vtc0iopen} give
$\M{SR}_{\MR}(x,n)$, which we use to construct $A_\times$, $A_+$, and $\M{SR}_{\MC}(z,n)$. We define $L_D(z,r,n)$ using the $\lambda(z,n)$
function, combine it with the integer length function to get $L_{\MR}(x,n)$, and we construct $L_{\MC}(z,n)$ from $L_D$,
$L_{\MR}$, and $\M{SR}_{\MC}$. Finally, we compose $E_{\MCL}$ and $L_{\MC}$ to get $\M{LE}(z,r,n)$.

The tedious but mostly unenlightening details have been moved to the appendix: see Lemma~\ref{lem:A-log-exp-apx}.
\end{Pf}
\begin{Lem}\label{lem:log-2n}
For all $z\in\MRL+i(-1,1)$ and $n\in\ML$,
\[\log_{\MC}\exp_{\MCL}2^{-n}z=2^{-n}\log_{\MC}\exp_{\MCL} z.\]
\end{Lem}
\begin{Pf}
In view of Lemmas \ref{lem:logexp-hom} and~\ref{lem:log-cont}, both sides are continuous in~$z$ (for fixed~$n$), hence it suffices to
prove the result for $z\in\MQL(i)$ by density. Fix $z\in\MQ(i)$ and $r\in\ML$ such that $\abs z\le r$ and $\abs{\Im
z}<1$, and $t\in\ML$; we will prove the $\tc$~formula
\begin{equation}\label{eq:7}
\Abs[big]{\M{LE}(2^{-n}z,r,t)-2^{-n}\M{LE}(z,r,t)}\le3\cdot2^{-t}
\end{equation}
by induction on $n\in\ML$. The statement for $n=0$ is trivial. Assuming \eqref{eq:7} holds for~$n$, we have
\[\Abs[big]{\log\exp2^{-n}z-2^{-n}\M{LE}(z,r,t)}\le4\cdot2^{-t}\]
by Lemma~\ref{lem:log-exp-apx}. Since $\log\exp2^{-n}z=2\log\exp2^{-(n+1)}z$ by Lemma~\ref{lem:logexp-hom},
\[\Abs[big]{\log\exp2^{-(n+1)}z-2^{-(n+1)}\M{LE}(z,r,t)}\le2\cdot2^{-t},\]
hence
\[\Abs[big]{\M{LE}(2^{-(n+1)}z,r,t)-2^{-(n+1)}\M{LE}(z,r,t)}\le3\cdot2^{-t}\]
by Lemma~\ref{lem:log-exp-apx}.

In view of Lemma~\ref{lem:log-exp-apx}, \eqref{eq:7} implies
\[\Abs{\log\exp2^{-n}z-2^{-n}\log\exp z}\le4\cdot2^{-t}+2^{-t-n}\le5\cdot2^{-t}.\]
Since $t\in\ML$ is arbitrary, we obtain $\log\exp2^{-n}z=2^{-n}\log\exp z$.
\end{Pf}

In the real world, all continuous solutions of Cauchy's functional equation are linear. Armed with Lemma~\ref{lem:log-2n},
we use a similar argument to derive $\log\exp z=z$ from the asymptotic estimate $\log\exp z=z+O(z^2)$ for small~$z$.
\begin{Lem}\label{lem:log-exp-id}
For all $z\in\MRL+i(-1,1)$, $\log_{\MC}\exp_{\MCL} z=z$.
\end{Lem}
\begin{Pf}
For any $n\in\ML$, we have
\[\log\exp z-z=2^n\bigl(\log\exp2^{-n}z-2^{-n}z\bigr)\]
by Lemma~\ref{lem:log-2n}. Assume $\abs z\le r$. If $2^n\ge4r$, then
\[\Abs{\exp2^{-n}z-(1+2^{-n}z)}\le2^{-2n}r^2\le\tfrac142^{-n}r\]
by Lemma \ref{thm:exp-homo}~\ref{item:6}, hence
\[\Abs{\exp2^{-n}z-1}\le\tfrac542^{-n}r\le\tfrac5{16}<\tfrac25,\]
whence
\[\Abs{\log\exp2^{-n}z-(\exp2^{-n}z-1)}\le\abs{\exp2^{-n}z-1}^2\le\tfrac{25}{16}2^{-2n}r^2\]
by Lemmas \ref{lem:log-c-s} and~\ref{lem:log-disc}. Consequently,
\[\Abs{\log\exp2^{-n}z-2^{-n}z}\le\tfrac{25}{16}2^{-2n}r^2+2^{-2n}r^2=\tfrac{41}{16}2^{-2n}r^2,\]
and
\[\Abs{\log\exp z-z}\le\tfrac{41}{16}2^{-n}r^2.\]
Since $n\in\ML$ can be arbitrarily large, this implies $\log\exp z=z$.
\end{Pf}
\begin{Cor}\label{cor:log-exp-id}
For all $z\in\MRL+i(-\pi,\pi]$, $\log_{\MC}\exp_{\MCL} z=z$.
\end{Cor}
\begin{Pf}
Using Proposition~\ref{prop:pi-apx}, we have $w=\frac14z\in\MRL+i(-1,1)$, hence
\[\log\exp z=\log\bigl((\exp w)^4\bigr)=4\log\exp w+2\pi i\,n=z+2\pi i\,n\]
for some $n\in\{-3,\dots,3\}$ by Lemmas \ref{thm:exp-homo}, \ref{lem:log-homo-all}, and~\ref{lem:log-exp-id}. Since both $\log\exp z$ and
$z$ are in $\MRL+i(-\pi,\pi]$ due to Corollary~\ref{cor:log-range}, the only possibility is $n=0$.
\end{Pf}

We have everything ready to derive the crucial right inverse property:
\begin{Cor}\label{thm:exp-log-id}
For all $z\in\MC_{\ne0}$, $\exp_{\MCL}\log_{\MC} z=z$.
\end{Cor}
\begin{Pf}
We have $\log_\MC z\in\MRL+i(-\pi,\pi]$ by Corollary~\ref{cor:log-range}, hence
\[\log_{\MC}\exp_{\MCL}\log_{\MC} z=\log_{\MC} z\]
by Corollary~\ref{cor:log-exp-id}, and $\exp_{\MCL}\log_{\MC} z=z$ by Corollary~\ref{cor:log-inj}.
\end{Pf}

A useful property to know is that we can compute powers by means of $\exp$ and $\log$, namely $z^n=\exp(n\log z)$. It
follows ``by induction on~$n$'' from Corollary~\ref{thm:exp-log-id}, but again, this takes a bit of work to formalize, as we
have to carry out the induction argument using $\tc$ approximations. In Section~\ref{sec:compl-powers-iter}, we will
use this property to justify the definition of complex exponentiation via $z^w=\exp(w\log z)$.
\begin{Lem}\label{lem:exp-qi-dense}
$\{z:\exp_{\MCL} z\in\MQ(i)\}$ is dense in $\MCL$.
\end{Lem}
\begin{Pf}
For any $z\in\MCL$ and $0<\delta\le\frac45$, the image of $D_\delta(z)$ under $\exp$ includes
\[\{\exp z\exp w:w\in D_\delta(0)\}\Sset\{w\exp z:\log w\in D_\delta(0)\}\Sset\{w\exp z:w\in D_{\delta/2}(1)\}\]
using Lemma~\ref{thm:exp-homo}, Corollary~\ref{thm:exp-log-id}, and Lemmas \ref{lem:log-disc} and~\ref{lem:log-c-s}.
\end{Pf}
\begin{Lem}\label{lem:exp-pow}
If $z\in\MCL$ and $n\in\MZL$, then $\exp_{\MCL} nz=(\exp_{\MCL} z)^n$. In particular, we have
$\exp_{\MCL}(n\log_{\MC} w)=w^n$ for all $w\in\MC_{\ne0}$.
\end{Lem}
\begin{Pf}
Without loss of generality, fix $n>0$. In view of Lemma~\ref{lem:exp-qi-dense}, we may assume $w:=\exp z\in\MQ(i)$ as
both sides are continuous in~$z$. For any $t\in\ML$ such that $2^t\ge8n+42$, fix $z'\in\MQ(i)$ and $r\in\ML$ such that
$\abs{z'-z}\le2^{-t}$ and $n\abs{z'}\le r$; we will prove
\begin{equation}\label{eq:8}
\Abs{\frac{E_{\MCL}(mz',r,t)}{w^m}-1}\le(6m+1)2^{-t}
\end{equation}
by induction on~$m\le n$. The statement holds for $m=0$. For the induction step, we have
\begin{align*}
\Abs{\frac{\exp mz'}{E_{\MCL}(mz',r,t)}-1}&\le\frac1{2^t-1}\le2^{-t}+2\cdot2^{-2t},\\
\Abs{\frac{\exp((m+1)z')}{w^{m+1}}\frac{w^m}{\exp mz'}-1}
&=\Abs{\frac{\exp z'}w-1}=\Abs[big]{\exp(z'-z)-1}\le2^{-t}+2^{-2t},\\
\Abs{\frac{E_{\MCL}((m+1)z',r,t)}{\exp((m+1)z')}-1}&\le2^{-t}
\end{align*}
by Lemma~\ref{lem:log-exp-apx}~\ref{item:11}, Lemma~\ref{lem:mult-apx}~\ref{item:34}, and Lemma~\ref{thm:exp-homo}, hence assuming \eqref{eq:8} for $m<n$,
\begin{align*}
\Abs{\frac{\exp mz'}{w^m}-1}&\le(6m+2)2^{-t}+(6m+3)2^{-2t}+(12m+2)2^{-3t}\le(6m+3)2^{-t},\\
\Abs{\frac{\exp((m+1)z')}{w^{m+1}}-1}&\le(6m+4)2^{-t}+(6m+4)2^{-2t}+(6m+3)2^{-3t}\le(6m+5)2^{-t},\\
\Abs{\frac{E_{\MCL}((m+1)z',r,t)}{w^{m+1}}-1}&\le(6m+6)2^{-t}+(6m+5)2^{-2t}\le(6m+7)2^{-t}
\end{align*}
using Lemma~\ref{lem:mult-apx}~\ref{item:32}.

By the first part of the induction step, \eqref{eq:8} for $m=n$ gives
\[\Abs{\frac{\exp nz'}{w^n}-1}\le(6n+3)2^{-t}.\]
Since
\[\Abs{\frac{\exp nz}{\exp nz'}-1}=\Abs{\exp\bigl(n(z-z')\bigr)-1}\le n2^{-t}+n^22^{-2t},\]
Lemma~\ref{lem:mult-apx}~\ref{item:32} implies
\[\Abs{\frac{\exp nz}{w^n}-1}\le(7n+3)2^{-t}+(7n^2+3n)2^{-2t}+(6n^3+3n^2)2^{-3t}\le8n2^{-t}.\]
Since $t\in\ML$ can be arbitrarily large, we obtain $\exp nz=w^n$.
\end{Pf}

Apart from its intrinsic value, we have a few applications for Lemma~\ref{lem:exp-pow}: it enables us to extend the
definition of $\exp$ to $\MRL+i\MR$ by exploiting its $2\pi i$-periodicity, and it implies numerical bounds on $e$ based
on the approximation $\bigl(1+\frac1n\bigr)^n$. We start with the latter.
\begin{Def}\label{def:e}
Let $e=\exp1$.
\end{Def}
\begin{Lem}\label{cor:e-apx}
Let $n\in\ML$, $n>0$.
\begin{enumerate}
\item\label{item:45} $\bigl(1+\frac1n\bigr)^n\le e\le\bigl(1+\frac1n\bigr)^{n+1}$.
\item\label{item:47} $2^n\le\exp n\le4^n$.
\item\label{item:46} $2{.}7<e<2{.}8$.
\end{enumerate}
\end{Lem}
\begin{Pf}

\ref{item:45}: We have $1+\frac1n\le\exp\frac1n$ by Lemma~\ref{prop:exp-conv}, thus
$\bigl(1+\frac1n\bigr)^n\le\exp1=e$ by Lemma~\ref{lem:exp-pow} and the monotonicity of $x^n$. Likewise,
Lemma~\ref{prop:exp-conv} gives $1-\frac1{n+1}\le\exp\bigl(-\frac1{n+1}\bigr)$, thus
$\bigl(1+\frac1n\bigr)=\bigl(1-\frac1{n+1}\bigr)^{-1}\ge\exp\frac1{n+1}$, which implies
$\bigl(1+\frac1n\bigr)^{n+1}\ge e$.

\ref{item:47} follows from $2=(1+1)^1\le e\le(1+1)^2=4$ and Lemma~\ref{lem:exp-pow}.

\ref{item:46}: One can check that $2{.}7048<1{.}01^{100}\le e\le1{.}01^{101}<2{.}7319$.
\end{Pf}

We can generalize Lemma \ref{cor:e-apx}~\ref{item:45} to a form of the alternative definition
\[\exp z=\lim_{n\to\infty}\left(1+\frac zn\right)^n.\]
Notice that we could not have actually used this expression to \emph{define} $\exp$, as it has relative error
proportional to $n^{-1}$, hence it only determines logarithmically many most significant bits of $\exp z$
rather than its precise value.
\begin{Prop}\label{prop:exp-limit}
If $z\in\MC$ and $n\in\ML_{>0}$ are such that $n\ge\max\bigl\{2\abs z,\abs z^2\bigr\}$, then
\[\Abs{\frac{\left(1+\frac zn\right)^n}{\exp_{\MCL} z}-1}\le\frac{2\abs z^2}n.\]
\end{Prop}
\begin{Pf}
Let $w=n\log\bigl(1+\frac zn\bigr)-z$. We have $\abs w\le\frac1n\abs z^2\le1$ by Lemma~\ref{lem:log-disc}, hence
$\abs{\exp w-1}\le\abs w+\abs w^2\le2\abs w$ by Lemma~\ref{thm:exp-homo}, and
\[\left(1+\frac zn\right)^n=\exp\left(n\log\left(1+\frac zn\right)\right)=\exp(z+w)=\exp z\exp w\]
by Lemma~\ref{lem:exp-pow}.
\end{Pf}

We now define our final extension of $\exp$; since it will not be modified any further, it will not carry any subscript.
Recall that $\MR_{\DML}=\{x\in\MR:\exists n\in\ML\,x\le n\}=\MRL\cup\MR_{<0}$.
\begin{Lem}\label{lem:exp-ir}
There is a unique function $\exp\colon\MR_{\DML}+i\MR\to\MC$ such that
\[\exp(z+2\pi in)=\exp_{\MCL} z\]
for all $z\in\MCL$ and $n\in\MZ$, and $\exp z=0$ when $\Re z\notin\MRL$. It satisfies
\begin{equation}\label{eq:9}
\exp(z+w)=\exp z\exp w
\end{equation}
for all $z,w\in\MR_{\DML}+i\MR$. For each $r\in\ML$, $\exp$ is uniformly continuous on $(-\infty,r]+i\MR$.
\end{Lem}
\begin{Pf}
Any $z\in\MRL+i\MR$ can be written as $z'+2\pi in$ for $n\in\MZ$ and $z'\in\MRL+i(-\pi,\pi]\sset\MCL$; on the other
hand, if $z+2\pi in=z'+2\pi in'$, where $z,z'\in\MCL$ and $n,n'\in\MZ$, we have $n-n'\in\MZL$, thus
\[\exp_{\MCL} z'=\exp_{\MCL}\bigl(2\pi i(n-n')\bigr)\exp_{\MCL} z=(-1)^{2(n-n')}\exp_{\MCL} z=\exp_{\MCL} z\]
using Lemmas \ref{thm:exp-homo} and~\ref{lem:exp-pow} (note that $\log_{\MC}(-1)=\pi i$ by Lemma~\ref{lem:arg}). This shows the existence
and uniqueness of $\exp$.

If $\Re z\notin\MRL$ or $\Re w\notin\MRL$, then the same holds for $z+w$, hence both sides of~\eqref{eq:9} equal~$0$.
Otherwise, \eqref{eq:9} follows from Lemma~\ref{thm:exp-homo}.

If $\Re z,\Re w\le r$ and $\abs{w-z}\le1$, we have
\[\abs{\exp w-\exp z}=\abs{\exp z}\abs{\exp_{\MCL}(w-z)-1}\le2\exp_{\MRL}(r)\abs{w-z}\]
by \eqref{eq:9} and Lemma \ref{thm:exp-homo}~\ref{item:6}, using that $\abs{\exp_{\MCL} z}=\exp_{\MRL}\Re
z\le\exp_{\MRL} r$ by Lemma \ref{thm:exp-homo} \ref{item:42} and~\ref{item:5} (if $\Re z\notin\MRL$, this bound holds
trivially).
\end{Pf}

We also extend our $\tc$ approximation to $\exp$. Notice that we cannot define a multiplicative approximation of $\exp$
by a $\tc$ formula on a domain with real part crossing the $\MRL$ boundary, as this would give us a $\tc$ definition of
$\MQL$ inside $\MQ$, and therefore of $\ML$ inside $\MN$, contradicting induction.
\begin{Lem}\label{lem:exp-apx}
There are $\tc$ functions $E_\times(z,r,n)$ and $E_+(z,r,n)$ with the following properties.
\begin{enumerate}
\item\label{item:40}
$E_\times(z,r,n)$ is a multiplicative approximation of $\exp z$ for $z\in\MQL+i\MQ$, parametrized by $r\in\ML$ such
that $\abs{\Re z}\le r$.
\item\label{item:41}
$E_+(z,r,n)$ is an additive approximation of $\exp z$ for $z\in\MQ_{\DML}+i\MQ$, parametrized by $r\in\ML$ such
that $\Re z\le r$.
\end{enumerate}
\end{Lem}
\begin{Pf}[sketch]
We use $L_{\MC}(-1,\dots)$ to compute $P\approx\pi$, and $N\in\MZ$ close to $\frac1{2\pi}\Im z$. Then we define
$E_\times(z,r,n)$ as $E_{\MCL}(z-2PNi,\dots)$. For $E_+(z,r,n)$, we use $E_\times(z,\dots)$ if $\Re z$ is not too
negative, and $0$ otherwise. Details are again presented in Lemma~\ref{lem:A-log-exp-apx} in the appendix.
\end{Pf}

By Lemma~\ref{lem:log-c-s}, $\log_{\MR}\sset\log_{\MC}$, but we do not know yet whether $\log_{\MC}$ extends $\log_D$ and $\log_S$
on the entirety of their domains. Let us remedy this now; since this establishes that all our $\log_X$ functions agree
whenever they are defined, it justifies that we officially rename $\log_{\MC}$ to just $\log$.
\begin{Lem}\label{lem:log-all}
$\log_D\sset\log_{\MC}$ and $\log_S\sset\log_{\MC}$.
\end{Lem}
\begin{Pf}
It is enough to prove the claim for $\log_D$: then for all $z\in S$,
\[\log_Sz=\log_{\MR}\abs z+\log_D\sgn z=\log_{\MC}\abs z+\log_{\MC}\sgn z=\log_{\MC} z\]
using Lemma~\ref{lem:log-c-homo}. Moreover, by the continuity of $\log_{\MC}$ and $\log_D$, it suffices to show
that $\log_{\MC}$ agrees with $\log_D$ on $\MQ(i)$.

Thus, let $z\in\MQ(i)\cap\ob D_{1-h^{-1}}(0)$, where $h\in\ML$, $h\ge2$; we will show $\log_D(1+z)=\log_{\MC}(1+z)$. Put
$n=2h^2$. For each $j<n$, $\Abs[big]{\frac jnz}\le1-h^{-1}$, thus $\Abs[big]{1+\frac jnz}\ge h^{-1}$, and
\[\Abs{\frac{1+\frac{j+1}nz}{1+\frac jnz}-1}=\Abs{\frac z{n+jz}}\le\frac hn\abs z\le\frac hn=\frac1{2h}.\]
Since $(2-h^{-1})\bigl(1+(2h)^{-1}\bigr)=2-(2h^2)^{-1}<^*2$ and $(2h)^{-1}\le\frac25$, we obtain
\begin{align*}
\log_D\bigl(1+\tfrac{j+1}nz\bigr)-\log_D\bigl(1+\tfrac jnz\bigr)
&=\log_D\left(1+\frac z{n+jz}\right)\\
&=\log_{\MC}\left(1+\frac z{n+jz}\right)\\
&=\log_{\MC}\bigl(1+\tfrac{j+1}nz\bigr)-\log_{\MC}\bigl(1+\tfrac jnz\bigr)
\end{align*}
using Lemmas \ref{lem:log-disc}, \ref{lem:log-c-s}, and~\ref{lem:log-c-homo}, thus
\begin{equation}\label{eq:11}
\log_D\bigl(1+\tfrac{j+1}nz\bigr)-\log_{\MC}\bigl(1+\tfrac{j+1}nz\bigr)
=\log_D\bigl(1+\tfrac jnz\bigr)-\log_{\MC}\bigl(1+\tfrac jnz\bigr).
\end{equation}

Fix $t\in\ML$; we will prove
\begin{equation}\label{eq:10}
\Abs[big]{L_D\bigl(1+\tfrac jnz,h,t\bigr)-L_{\MC}\bigl(1+\tfrac jnz,t\bigr)}\le(4j+2)2^{-t}
\end{equation}
by induction on~$j\le n$. Since $\log_D1=0=\log_{\MC}1$, the base case $j=0$ follows immediately from
Lemma~\ref{lem:log-exp-apx}. Assuming \eqref{eq:10} holds for some $j<n$, Lemma~\ref{lem:log-exp-apx} and~\eqref{eq:11} give
\begin{align*}
\Abs[big]{\log_D\bigl(1+\tfrac{j+1}nz\bigr)-\log_{\MC}\bigl(1+\tfrac{j+1}nz\bigr)}
&=\Abs[big]{\log_D\bigl(1+\tfrac jnz\bigr)-\log_{\MC}\bigl(1+\tfrac jnz\bigr)}\\
&\le(4j+4)2^{-t},\\
\Abs[big]{L_D\bigl(1+\tfrac{j+1}nz,h,t\bigr)-L_{\MC}\bigl(1+\tfrac{j+1}nz,t\bigr)}
&\le(4j+6)2^{-t}.
\end{align*}

Taking \eqref{eq:10} for $j=n$, Lemma~\ref{lem:log-exp-apx} implies
\[\Abs[big]{\log_D(1+z)-\log_{\MC}(1+z)}\le(4n+4)2^{-t}.\]
Since $t\in\ML$ can be chosen arbitrarily large, we obtain $\log_D(1+z)=\log_{\MC}(1+z)$.
\end{Pf}

To tie up one more loose end, Lemma~\ref{lem:exp-ir} gives a useful result on uniform continuity of $\exp$, but for $\log$,
we only have the rather unsatisfactory Lemma~\ref{lem:log-unif}. Let us state a better result for the sake of
completeness.
\begin{Lem}\label{lem:log-unif-opt}
For every $\ep\in\MR_{>0}$, $\log$ is uniformly continuous on
\[\bigl\{z\in\MC:\abs z\ge\ep\land(\Re z\ge0\lor\Im z\ge0\lor\Im z\le-\ep)\bigr\}.\]
\end{Lem}
\begin{Pf}
We claim that if $z,w$ belong to the indicated set, then for all $0<\delta\le1$,
\[\abs{z-w}\le\frac\ep2\delta\implies\abs{\log z-\log w}\le\delta.\]
First, we observe
\[\Abs{\frac wz-1}=\frac{\abs{z-w}}{\abs z}\le\frac{\ep\delta/2}\ep\le\frac\delta2,\qquad
\Abs{\log\frac wz}\le\delta,\]
using Lemma~\ref{lem:log-disc}. Thus, it suffices to show
\begin{equation}\label{eq:12}
\log z+\log\frac wz=\log w.
\end{equation}

If $\Re z\ge0$, then \eqref{eq:12} is true by Lemma~\ref{lem:log-c-homo}~\ref{item:21}, as $\Re(w/z)>0$; if $\Re w\ge0$,
we may swap $z$ and~$w$, noting that $\log(z/w)=-\log(w/z)$ by Lemma~\ref{lem:log-c-homo}~\ref{item:23}. Thus, it remains
to consider the case $\Re z,\Re w<0$. Then either $\Im z,\Im w\ge0$, or $\Im z,\Im w\le-\ep$: we cannot have, say,
$\Im z\ge0$ and $\Im w\le-\ep$, as $\abs{z-w}<\ep$.

Assume $\Im z,\Im w\ge0$. Then $\sgn^+\bigl(z\frac wz\bigr)=\sgn^+z$, hence \eqref{eq:12} follows from
Lemma~\ref{lem:log-c-homo}~\ref{item:22}, unless $z\in\MR_{<0}$. Again, we may swap $z$ and~$w$ if necessary, obtaining
\eqref{eq:12} unless $w\in\MR_{<0}$ as well; but if $z,w\in\MR_{<0}$ and $w/z\in\MR_{>0}$, then \eqref{eq:12} is also
true.

If $\Im z,\Im w\le-\ep$, we obtain \eqref{eq:12} by applying the previous part to $\ob z,\ob w$ in place of $z,w$,
using Lemma~\ref{lem:log-c-homo}~\ref{item:23}.
\end{Pf}

\begin{Pf}[of Theorem~\ref{thm:main}]
We refer to item numbers from the statement of Theorem~\ref{thm:main}.

Lemmas \ref{thm:exp-homo} and~\ref{lem:exp-ir} ensure that \ref{item:50} and~\ref{item:52} hold (by definition), and that $\exp$ is a group
homomorphism $\p{\MRL+i\MR,+}\to\p{\MC_{\ne0},\cdot}$ whose kernel includes $2\pi i\MZ$. By Corollary~\ref{cor:log-range},
$\log$ maps $\MC_{\ne0}$ to $\MRL+i(-\pi,\pi]$, and $\exp\log z=z$ by Corollary~\ref{thm:exp-log-id}, which also implies
that $\exp\colon\MRL+i\MR\to\MC_{\ne0}$ is surjective, and $\log$ is injective. Conversely, $\log\exp z=z$ for
$z\in\MRL+i(-\pi,\pi]$ by Corollary~\ref{cor:log-exp-id}, hence $\log\colon\MC_{\ne0}\to\MRL+i(-\pi,\pi]$ is
surjective, and $\exp$ is injective on $\MRL+i(-\pi,\pi]$, which implies the kernel of $\exp$ is \emph{exactly}
$2\pi i\MZ$: if $\exp(z)=1$, we may write $z=w+2\pi i\,n$ for some $n\in\MZ$ and $w\in\MRL+i(-\pi,\pi]$; then
$\exp w=1$, hence $w=0$. This gives \ref{item:51} and~\ref{item:53}.

\ref{item:54} follows from Lemma~\ref{thm:exp-homo} and~\ref{item:53}, using the fact that ${\log}\res\MR_{>0}$ is
real-valued due to Lemma~\ref{lem:log-c-s}. \ref{item:55} is stated in Lemmas \ref{lem:exp-ir}, \ref{lem:log-cont}, and~\ref{lem:log-unif-opt}.
\ref{item:56} follows from Lemmas \ref{thm:exp-homo} and~\ref{lem:arg}.

\ref{item:57} follows from Lemma~\ref{lem:arg-mon}: e.g., if $z_0,z_1\in\MC_{\ne0}$ are such that $\Re z_j\ge0\ge\Im z_j$,
then $\arg z_0<\arg z_1$ iff $\Re\sgn z_0<\Re\sgn z_1$ iff $\Im\sgn z_0<\Im\sgn z_1$ by
Lemma~\ref{lem:arg-mon} \ref{item:36}, \ref{item:38}, and~\ref{item:39}; in particular,
$-\frac\pi2=\arg(-i)\le\arg z_j\le\arg1=0$.

\ref{item:58} follows from Lemmas \ref{thm:exp-homo}, \ref{lem:log-disc}, and~\ref{lem:log-all}; \ref{item:60} is Lemma~\ref{lem:exp-pow},
\ref{item:61} is Proposition~\ref{prop:exp-limit}, and \ref{item:62} is Lemma~\ref{prop:exp-conv} (with trivial extension to
$\MR_{\DML}$).

\ref{item:63}: Lemmas \ref{lem:log-exp-apx} and~\ref{lem:exp-apx} (proved in Lemma~\ref{lem:A-log-exp-apx}) give $E_+$,
$E_\times$, and $L_+$ (called $L_\MC$ there). The existence of $L_\times$ follows from Lemma~\ref{lem:mult-apx}: we only
need to exhibit a $\tc$ function $h\colon\MQ(i)\bez\{0\}\to\ML$ such that $\abs{\log z}\ge2^{-h(z)}$ for all $z\ne0,1$.
If $0<\abs{z-1}\le\frac12$, we have $\abs{\log z}\ge\abs{z-1}-\abs{z-1}^2\ge\abs{z-1}^2$ by~\ref{item:58}, hence it
suffices to put $h(z)=\Dlh{\CL{\abs{z-1}^{-2}}}$ using integer division and the length function. We claim that if
$\abs{z-1}\ge\frac12$, then $\abs{\log z}>\frac13$, hence we can just take $h(z)=2$: indeed, if
$\abs{\log z}\le\frac13$, then
\[\abs{z-1}=\abs{\exp\log z-1}\le\abs{\log z}+\abs{\log z}^2\le\tfrac49<\tfrac12\]
by \ref{item:53} and~\ref{item:58}.
\end{Pf}

\section{Complex powers and iterated multiplication}\label{sec:compl-powers-iter}

Having completed our treatment of $\exp$ and $\log$, we come to applications. The first one is a definition of powering.
We have so far defined the powering function $z^n$ for $z\in\MC_{\ne0}$ and $n\in\MZL$ (and for $z=0$ and $n\in\ML$);
we can now extend it to all exponents in $\MCL$.
\begin{Def}\label{def:pow}
If $z\in\MC_{\ne0}$ and $w\in\MCL$, let $z^w=\exp(w\log z)$.
\end{Def}

Notice that this definition provides an alternative notation for exponentiation: $e^w=\exp w$. We will use this
notation also when $\Re w\in\MR_{\DML}\bez\MRL$ (even though we did not bother to define $z^w$ in such circumstances).
\begin{Prop}\label{prop:pow}
Let $z,z'\in\MC_{\ne0}$ and $w,w'\in\MCL$.
\begin{enumerate}
\item\label{item:64}
If $w\in\MZL$, then $z^w$ agrees with the previous definition. In particular, $z^0=1$ and $z^1=z$.
\item\label{item:65}
$z^{w+w'}=z^wz^{w'}$ and $z^{-w}=1/z^w$.
\item\label{item:66}
If $\arg z+\arg z'\in(-\pi,\pi]$, then $(zz')^w=z^wz'^w$.
\item\label{item:67}
If $\Im(w\log z)\in(-\pi,\pi]$, then $(z^w)^{w'}=z^{ww'}$. In particular, this holds when $w\in(-1,1]$, or when
$z\in\MR_{>0}$ and $w\in\MRL$.
\end{enumerate}
\end{Prop}
\begin{Pf}

\ref{item:64} and \ref{item:65} follow from Theorem \ref{thm:main} \ref{item:50}, \ref{item:53}, and~\ref{item:60}.

\ref{item:66} follows from Corollary~\ref{cor:log-homo-arg}.

\ref{item:67}: If $\Im(w\log z)\in(-\pi,\pi]$, then $\log z^w=w\log z$ by Theorem \ref{thm:main}~\ref{item:53},
thus $(z^w)^{w'}=\exp(w'w\log z)=z^{ww'}$. If $w\in\MRL$, then $\Im(w\log z)=w\arg z$, which is
in $(-\pi,\pi]$ if $w\in(-1,1]$ or $\arg z=0$.
\end{Pf}

The usefulness of the general definition of $z^w$ is limited, as $\log z$ is conceptually a multivalued function (only
defined up to integer multiples of $2\pi i$), thus $z^w$ should be only defined up to multiplying by integer powers of
$\exp(2\pi iw)$; this explains why Proposition \ref{prop:pow} \ref{item:66} and~\ref{item:67} hold only under unsightly side
conditions. The definition is better-behaved for:
\begin{itemize}
\item $w\in\MZL$, when it is independent of the branch of $\log$, as $\exp(2\pi iw)=1$; in this case, it coincides with
the previous definition using iterated multiplication.
\item $z\in\MR_{>0}$, in which case the choice of the branch $\log z\in\MR$ is canonical.
\end{itemize}
Another interesting case is $w=1/n$, $n\in\ML_{>0}$:
\begin{Def}\label{def:root}
If $z\in\MC_{\ne0}$ and $n\in\ML_{>0}$, let $\sqrt[n]z=z^{1/n}$. We also put $\sqrt[n]0=0$.
\end{Def}
\begin{Prop}\label{prop:root}
Let $z,z'\in\MC$, $n,m\in\ML_{>0}$, and $w\in\MCL$.
\begin{enumerate}
\item\label{item:68}
$\sqrt[1]z=z$ and $\sqrt[2]z=\sqrt z$.
\item\label{item:69}
$(\sqrt[n]z)^w=z^{w/n}$. In particular, $(\sqrt[n]z)^n=z$ and $\sqrt[m]{\!\!\sqrt[n]z}=\sqrt[nm]z$.
\item\label{item:70}
If $zz'=0$ or $\arg z+\arg z'\in(-\pi,\pi]$, then $\sqrt[n]{zz'}=\sqrt[n]z\sqrt[n]{z'}$.
\end{enumerate}
\end{Prop}
\begin{Pf}
\ref{item:69} and \ref{item:70} follow from Proposition~\ref{prop:pow}.
\ref{item:68}: $\sqrt[1]z=z$ is clear, and
\[\sqrt z=\exp\log\sqrt z=\exp\bigl(\tfrac12\log z\bigr)=\sqrt[2]z\]
by Lemma~\ref{lem:log-c-homo}~\ref{item:27}.
\end{Pf}

The second application we promised in Section~\ref{sec:preliminaries} is an extension of the definition of iterated
multiplication $\prod_{j<n}z_j$ to coded sequences $\p{z_j:j<n}$ of Gaussian rationals $z_j\in\MQ(i)$. Basically, we
want to put
\[\prod_{j<n}z_j=\exp\Bigl(\sum_{j<n}\log z_j\Bigr),\]
but we cannot do this directly as $\log z_j\notin\MQ(i)$ in general, hence the sum is meaningless. We can use an
approximation $L_+(z_j,t)$ instead of $\log z_j$, but then we have another problem---how to determine the exact result.
For this reason, we first define $\prod_{j<n}z_j$ for Gaussian \emph{integers} $z_j\in\MZ[i]$, in which case we can
round a sufficiently good approximation to an exact result.
\begin{Def}\label{def:prod-c}
We put $\fcl x=\Fl{x+\frac12}$ for $x\in\MR$, and $\fcl z=\fcl{\Re z}+i\fcl{\Im z}$ for $z\in\MC$.

If $\p{z_j:j<n}$ is a sequence of Gaussian integers $z_j=x_j+iy_j$, $x_j,y_j\in\MZ$, we define
\[\prod_{j<n}z_j=\begin{cases}
0,&\text{if $z_j=0$ for some $j<n$,}\\
\FCL[big]{E_\times\bigl(\sum_{j<n}L_+(z_j,t),r,t\bigr)},&\text{otherwise,}
\end{cases}\]
where $r=1+\sum_j\Dlh[big]{\abs{x_j}+\abs{y_j}}$ and $t=r+3+\dlh n$.

If $\p{z_j:j<n}$ is a sequence of Gaussian rationals $z_j=w_j/x_j$, where $w_j\in\MZ[i]$ and
$x_j\in\MZ_{>0}$, we define
\[\prod_{j<n}z_j=\frac{\prod_{j<n}w_j}{\prod_{j<n}x_j}.\]
\end{Def}

Observe that $\prod_{j<n}z_j$ is defined by a $\tc$ function, and $\prod_{j<0}z_j=1$, as
$\abs{E_\times(0,r,t)-1}<\frac12$.
\begin{Thm}\label{thm:prod-qi}
For any sequence $\p{z_j:j\le n}$ of Gaussian rationals,
\[\prod_{j<n+1}z_j=z_n\prod_{j<n}z_j.\]
\end{Thm}
\begin{Pf}
It suffices to prove the result for $z_j\in\MZ[i]\bez\{0\}$. Let $z_j=x_j+iy_j$, $r_j=\Dlh[big]{\abs{x_j}+\abs{y_j}}$,
$r=1+\sum_{j\le n}r_j$, and $t=r+3+\dlh{n+1}$ as in the definition of $\prod_{j<n+1}z_j$; notice that
$1\le\abs{z_j}\le\abs{x_j}+\abs{y_j}\le2^{r_j}$, hence $0\le\Re\log(z_j)\le r_j$ using Theorem \ref{thm:main}~\ref{item:56}
and Lemma~\ref{cor:e-apx}. Thus,
\[\Abs[Big]{\Re\sum_{j<m}L_+(z_j,t)}\le\sum_{j<m}r_j+m2^{-t}\le r\]
for all $m\le n+1$, which ensures that the usage of $E_\times$ in the definition of $\prod_{j<n+1}z_j$ is sound, and
more generally, that it makes sense to define
\[w_m=E_\times\Bigl(\sum_{j<m}L_+(z_j,t),r,t\Bigr),\qquad m\le n+1.\]
We will prove
\begin{align}
\label{eq:14}m>0\implies\fcl{w_m}&=z_{m-1}\fcl{w_{m-1}},\\
\label{eq:15}\abs{\fcl{w_m}}&\le2^{\sum_{j<m}r_j},\\
\label{eq:16}\Abs{\frac{w_m}{\fcl{w_m}}-1}&\le(4m+1)2^{-t}
\end{align}
for all $m\le n+1$ by induction on~$m$. For $m=0$, $\abs{w_0-1}\le2^{-t}<\frac12$, thus $\fcl{w_0}=1$, which
gives the claim. Assume
\eqref{eq:14}--\eqref{eq:16} hold for $m\le n$, we will prove them for $m+1$. We have
\[\Abs{\frac{\exp\sum_{j<m}L_+(z_j,t)}{\fcl{w_m}}-1}\le(4m+2)2^{-t}+O(m2^{-2t})\]
by \eqref{eq:16}, the properties of $E_\times$, and Lemma~\ref{lem:mult-apx}, while $\abs{L_+(z_m,t)-\log(z_m)}\le2^{-t}$
gives
\[\Abs{\frac{\exp L_+(z_m,t)}{z_m}-1}=\Abs{\exp\bigl(L_+(z_m,t)-\log z_m\bigr)-1}\le2^{-t}+O(2^{-2t}),\]
using Theorem \ref{thm:main}~\ref{item:58}. Thus,
\begin{align*}
\Abs{\frac{\exp\sum_{j\le m}L_+(z_j,t)}{z_m\fcl{w_m}}-1}&\le(4m+3)2^{-t}+O(m2^{-2t}),\\
\Abs{\frac{w_{m+1}}{z_m\fcl{w_m}}-1}&\le(4m+4)2^{-t}+O(m2^{-2t})\le(4m+5)2^{-t},
\end{align*}
using again Lemma~\ref{lem:mult-apx} and the approximation property of $E_\times$. Since \eqref{eq:15} and
$\abs{z_m}\le2^{r_m}$ imply
\[\Abs[big]{z_m\fcl{w_m}}\le2^{\sum_{j\le m}r_j}\le2^{r-1},\]
we obtain
\[\Abs[big]{w_{m+1}-z_m\fcl{w_m}}\le(4n+5)2^{r-1-t}\le2^{r+1+\dlh{n+1}-t}\le2^{-2},\]
thus, in view of $z_m\fcl{w_m}\in\MZ[i]$,
\[\fcl{w_{m+1}}=z_m\fcl{w_m}.\]
This gives \eqref{eq:14}, \eqref{eq:15}, and~\eqref{eq:16} for $m+1$.

Now, \eqref{eq:14} for $m=n+1$ shows
\[\prod_{j<n+1}z_j=\fcl{w_{n+1}}=z_n\fcl{w_n}.\]
Moreover, putting $r'=1+\sum_{j<n}r_j$, $t'=r'+3+\dlh n$, and
\[w'_m=E_\times\Bigl(\sum_{j<m}L_+(z_j,t'),r',t'\Bigr),\qquad m\le n,\]
the same argument as above shows
\[\fcl{w'_0}=1,\qquad\forall 0<m\le n\:\fcl{w'_m}=z_{m-1}\fcl{w'_{m-1}},\]
which implies $\fcl{w_m}=\fcl{w'_m}$ by induction on~$m\le n$. Thus,
\[\fcl{w_n}=\fcl{w'_n}=\prod_{j<n}z_j,\]
completing the proof.
\end{Pf}

\section{Trigonometric and hyperbolic functions}\label{sec:trig-hyperb-funct}

Armed with complex exponential and logarithm, we can easily define trigonometric and hyperbolic functions and their
inverses in the usual way. We present the definitions and a few basic properties below, mostly to indicate the
effects of our setup with (possibly) $\ML\ne\MN$ on domains of the functions, but we will skip many routine details
(which are generally easy to verify using Theorem~\ref{thm:main}); since we deal with 24 functions here, we cannot afford to
give each the same level of attention we spent on $\exp$ and $\log$.
\begin{Def}\label{def:hyp-trig}
We introduce the following functions, where we write $x=\Re z$:
\begin{align*}
\sinh\colon&\MRL+i\MR\to\MC,&\sinh z&=\tfrac12\bigl(e^z-e^{-z}\bigr),\\
\sin\colon&\MR+i\MRL\to\MC,&\sin z&=\tfrac1i\sinh iz,\\
\cosh\colon&\MRL+i\MR\to\MC,&\cosh z&=\tfrac12\bigl(e^z+e^{-z}\bigr),\\
\cos\colon&\MR+i\MRL\to\MC,&\cos z&=\cosh iz,\\
\tanh\colon&\MC\bez i\pi\bigl(\tfrac12+\MZ\bigr)\to\MC,
&\tanh z&=\left\{\begin{aligned}
&\frac{\sinh z}{\cosh z},&x\in\MRL\\[\medskipamount]
&\sgn x,&x\notin\MRL
\end{aligned}\right\}
=\left\{\begin{aligned}
&\frac{1-e^{-2z}}{1+e^{-2z}},&x\ge0,\\[\medskipamount]
&\frac{e^{2z}-1}{e^{2z}+1},&x\le0,
\end{aligned}\right.\\
\tan\colon&\MC\bez\pi\bigl(\tfrac12+\MZ\bigr)\to\MC,
&\tan z&=\tfrac1i\tanh iz,\\
\coth\colon&\MC\bez i\pi\MZ\to\MC,
&\coth z&=\left\{\begin{aligned}
&\frac{\cosh z}{\sinh z},&x\in\MRL\\[\medskipamount]
&\sgn x,&x\notin\MRL
\end{aligned}\right\}
=\left\{\begin{aligned}
&\frac{1+e^{-2z}}{1-e^{-2z}},&x\ge0,\\[\medskipamount]
&\frac{e^{2z}+1}{e^{2z}-1},&x\le0,
\end{aligned}\right.\\
\cot\colon&\MC\bez\pi\MZ\to\MC,
&\cot z&=i\coth iz,\\
\sech\colon&\MC\bez i\pi\bigl(\tfrac12+\MZ\bigr)\to\MC,
&\sech z&=\left\{\begin{aligned}
&\frac1{\cosh z},&x\in\MRL\\[\medskipamount]
&0,&x\notin\MRL
\end{aligned}\right\}
=\left\{\begin{aligned}
&\frac{2e^{-z}}{1+e^{-2z}},&x\ge0,\\[\medskipamount]
&\frac{2e^z}{e^{2z}+1},&x\le0,
\end{aligned}\right.\\
\sec\colon&\MC\bez\pi\bigl(\tfrac12+\MZ\bigr)\to\MC,
&\sec z&=\sech iz,\\
\csch\colon&\MC\bez i\pi\MZ\to\MC,
&\csch z&=\left\{\begin{aligned}
&\frac1{\sinh z},&x\in\MRL\\[\medskipamount]
&0,&x\notin\MRL
\end{aligned}\right\}
=\left\{\begin{aligned}
&\frac{2e^{-z}}{1-e^{-2z}},&x\ge0,\\[\medskipamount]
&\frac{2e^z}{e^{2z}-1},&x\le0,
\end{aligned}\right.\\
\csc\colon&\MC\bez\pi\MZ\to\MC,
&\csc z&=i\csch iz.
\end{align*}
\end{Def}
\begin{Def}\label{def:per}
A function $f$ is \emph{$p$-periodic} if $f(z+pn)=f(z)$ for all $z\in\dom(f)$ and $n\in\MZ$, and
\emph{$p$-antiperiodic} if $f(z+pn)=(-1)^nf(z)$ for all $z\in\dom(f)$ and $n\in\MZ$ (which implies it is $2p$-periodic).
\end{Def}
\begin{Prop}\label{prop:hyp-trig}
Each of the 12 functions $f$ from Definition~\ref{def:hyp-trig} satisfies $f(\ob z)=\ob{f(z)}$, whence $f$ maps
$\MR\cap\dom(f)$ to $\MR$. The functions $\sin$, $\cos$, $\sec$, and $\csc$ are $\pi$-antiperiodic; $\sinh$, $\cosh$,
$\sech$, and $\csch$ are $\pi i$-antiperiodic; $\tan$ and $\cot$ are $\pi$-periodic; $\tanh$ and $\coth$ are $\pi
i$-periodic. The functions $\cos$, $\cosh$, $\sec$, and $\sech$ are even, while the remaining 8 functions are odd.
\end{Prop}
\begin{Pf}
Straightforward consequence of $\exp\ob z=\ob{\exp z}$ and the $\pi i$-antiperiodicity of $\exp$, which follow
from Theorem~\ref{thm:main}.
\end{Pf}

We now turn to inverse trigonometric and hyperbolic functions. Similar to $\log$, these functions are multivalued,
and it is a somewhat arbitrary decision how to choose their branch cuts; we will define them in such a way that they
extend the most natural choices of inverse \emph{real} trigonometric and hyperbolic functions.
\begin{Def}\label{def:inv-hyp-trig}
We introduce the functions below:
\begin{align*}
\arsinh\colon&\MC\to\MRL+i\bigl[-\tfrac\pi2,\tfrac\pi2\bigr],&\arsinh z&=\log\bigl(z+\sqrt{z^2+1}\bigr),\\
\arcsin\colon&\MC\to\bigl[-\tfrac\pi2,\tfrac\pi2\bigr]+i\MRL,&\arcsin z&=\tfrac1i\arsinh iz,\\
\arcosh\colon&\MC\to\MR_{\ML,\ge0}+i(-\pi,\pi],&\arcosh z&=\log\bigl(z+\sqrt{z+1}\sqrt{z-1}\bigr),\\
\arccos\colon&\MC\to[0,\pi]+i\MRL,&\arccos z&=\tfrac\pi2-\arcsin z,\\
\artanh\colon&\MC\bez\{\pm1\}\to\MRL+i\bigl(-\tfrac\pi2,\tfrac\pi2\bigr],&\artanh z&=\tfrac12\log\bigl(\tfrac{1+z}{1-z}\bigr),\\
\arctan\colon&\MC\bez\{\pm i\}\to\bigl(-\tfrac\pi2,\tfrac\pi2\bigr]+i\MRL,&\arctan z&=\tfrac1i\artanh iz,\\
\arcoth\colon&\MC\bez\{\pm1\}\to\MRL+i\bigl(-\tfrac\pi2,\tfrac\pi2\bigr],&\arcoth z&=\tfrac12\log\bigl(\tfrac{z+1}{z-1}\bigr),\\
\arccot\colon&\MC\bez\{\pm i\}\to[0,\pi)+i\MRL,&\arccot z&=\tfrac\pi2-\arctan z,\\
\arsech\colon&\MC_{\ne0}\to\MR_{\ML,\ge0}+i(-\pi,\pi],&\arsech z&=\arcosh z^{-1},\\
\arcsec\colon&\MC_{\ne0}\to[0,\pi]+i\MRL,&\arcsec z&=\arccos z^{-1},\\
\arcsch\colon&\MC_{\ne0}\to\MRL+i\bigl[-\tfrac\pi2,\tfrac\pi2\bigr],&\arcsch z&=\arsinh z^{-1},\\
\arccsc\colon&\MC_{\ne0}\to\bigl[-\tfrac\pi2,\tfrac\pi2\bigr]+i\MRL,&\arccsc z&=\arcsin z^{-1}.
\end{align*}
\end{Def}
To see that the indicated codomains of these functions are valid, we need the following:
\begin{Lem}\label{lem:inv-hyp-trig-cod}
For all $z\in\MC$, $\Re\bigl(z+\sqrt{z^2+1}\bigr)\ge0$ and $\Abs{z+\sqrt{z+1}\sqrt{z-1}}\ge1$.
\end{Lem}
\begin{Pf}
Putting $w_\pm=\sqrt{z^2+1}\pm z$ for $\pm\in\{+,-\}$, we have $w_+w_-=1$, i.e., $w_-=\ob{w_+}/\abs{w_+}^2$. We obtain
$\sgn^+\Re w_+=\sgn^+\Re w_-=\sgn^+\Re(w_++w_-)=\sgn^+\Re\sqrt{z^2+1}=1$, using Lemma~\ref{lem:sqrt-iter}.

Observe that if $z_0$ and~$z_1$ belong to the same quadrant, then $\abs{z_0+z_1}\ge\abs{z_0-z_1}$: writing
$z_j=x_j+iy_j$, we have $\abs{z_0+z_1}^2=(x_0+x_1)^2+(y_0+y_1)^2\ge(x_0-x_1)^2+(y_0-y_1)^2=\abs{z_0-z_1}^2$ as
$x_0x_1+y_0y_1\ge0$.

Lemmas \ref{lem:sqrt} and~\ref{lem:sqrt-iter} give $\Re\sqrt{z\pm1}\ge0$ and $\sgn^+\Im\sqrt{z\pm1}=\sgn^+\Im z$, hence
\[\Abs{\sqrt{z+1}+\sqrt{z-1}}\ge\Abs{\sqrt{z+1}-\sqrt{z-1}}\]
by the observation above. Since $\bigl(\sqrt{z+1}\pm\sqrt{z-1}\bigr)^2=2z\pm2\sqrt{z+1}\sqrt{z-1}$, we obtain
\[\Abs{z+\sqrt{z+1}\sqrt{z-1}}\ge\Abs{z-\sqrt{z+1}\sqrt{z-1}}.\]
In view of $\bigl(z+\sqrt{z+1}\sqrt{z-1}\bigr)\bigl(z-\sqrt{z+1}\sqrt{z-1}\bigr)=1$, this shows
$\Abs{z+\sqrt{z+1}\sqrt{z-1}}\ge1$.
\end{Pf}

The functions as presented in Definition~\ref{def:inv-hyp-trig} are not quite surjective. Their precise ranges as well as the
complete structure of preimages of trigonometric and hyperbolic functions in the complex and real cases are described
below.
\begin{Prop}\label{prop:inv-hyp-trig}
Let $f\colon\dom(f)\to\cod(f)$ be a hyperbolic or trigonometric function from Definition~\ref{def:hyp-trig},
$g\colon\dom(g)\to\cod(g)$ the corresponding function from Definition~\ref{def:inv-hyp-trig}, and $B(g)$, $X(g)$, $\dom_{\MR}(g)$,
$\im_{\MR}(g)$ the corresponding sets in Table~\ref{tab:hyp-trig}. Let $\dom_{\MR}(f)=\dom(f)\cap\MR$ if $f$ is trigonometric, and
$\dom_{\MR}(f)=\dom(f)\cap\MRL$ if it is hyperbolic.
\begin{enumerate}
\item The function $g$ is continuous in $\MC\bez B(g)$.
\item The image of $g$ is $\cod(g)\bez X(g)$. We have $f(g(z))=z$ for all $z\in\dom(g)$, and $g(f(z))=z$ for all
$z\in\cod(g)\bez X(g)$.
\item $f$ maps $\dom_{\MR}(f)$ onto $\dom_{\MR}(g)$, and $g$ maps $\dom_{\MR}(g)$ onto $\im_{\MR}(g)$.
\item If $g(z)=w$, then $f^{-1}(z)$ is the set described in the last column of Table~\ref{tab:hyp-trig}.
\end{enumerate}
\end{Prop}
\begin{Pf}
Left to the reader.
\end{Pf}
\begin{table}
\def\arraystretch{1.4}\def\AS{\def\arraystretch{1}}
\[\begin{array}{cccccc}
g&B(g)&X(g)&\dom_{\MR}(g)&\im_{\MR}(g)&f^{-1}(z)\\
\hline
\arsinh&\pm i(1,\infty)&\pm\bigl(\MR_{\ML,<0}+i\frac\pi2\bigr)&\MR&\MRL
&\AS\begin{array}[t]{r}w+2\pi i\MZ\\\pi i-w+2\pi i\MZ\end{array}\\
\arcsin&\pm(1,\infty)&\pm\bigl(\frac\pi2+i\MR_{\ML,>0}\bigr)&[-1,1]&\bigl[-\frac\pi2,\frac\pi2\bigr]
&\AS\begin{array}[t]{r}w+2\pi\MZ\\\pi-w+2\pi\MZ\end{array}\\
\arcosh&(-\infty,1)&i(-\pi,0)&[1,\infty)&\MR_{\ML,\ge0}&\pm w+2\pi i\MZ\\
\arccos&\pm(1,\infty)&\AS\begin{array}[t]{r}(\pi+i\MR_{\ML,>0})\\{}\cup i\MR_{\ML,<0}\end{array}
&[-1,1]&[0,\pi]&\pm w+2\pi\MZ\\
\artanh&\pm[1,\infty)&i\frac\pi2&(-1,1)&\MRL&w+\pi i\MZ\\
\arctan&\pm i[1,\infty)&\frac\pi2&\MR&\bigl(-\frac\pi2,\frac\pi2\bigr)&w+\pi\MZ\\
\arcoth&[-1,1]&0&\pm(1,\infty)&\MR_{\ML,\ne0}&w+\pi i\MZ\\
\arccot&\pm i[1,\infty)&0&\MR&(0,\pi)&w+\pi\MZ\\
\arsech&\AS\begin{array}[t]{l}(-\infty,0]\\{}\cup(1,\infty)\end{array}&i(-\pi,0)\cup\bigl\{i\frac\pi2\bigr\}
&(0,1]&\MR_{\ML,\ge0}&\pm w+2\pi i\MZ\\
\arcsec&(-1,1)&\AS\begin{array}[t]{l}(\pi+i\MR_{\ML,>0})\\{}\cup i\MR_{\ML,<0}\cup\bigl\{\frac\pi2\bigr\}\end{array}
&\pm[1,\infty)&[0,\pi]\bez\bigl\{\frac\pi2\bigr\}&\pm w+2\pi\MZ\\
\arcsch&i(-1,1)&\AS\begin{array}[t]{r}\pm\bigl(\MR_{\ML,<0}+i\frac\pi2\bigr)\\{}\cup\{0\}\end{array}
&\MR_{\ne0}&\MR_{\ML,\ne0}&\AS\begin{array}[t]{r}w+2\pi i\MZ\\\pi i-w+2\pi i\MZ\end{array}\\
\arccsc&(-1,1)&\AS\begin{array}[t]{r}\pm\bigl(\frac\pi2+i\MR_{\ML,>0}\bigr)\\{}\cup\{0\}\end{array}
&\pm[1,\infty)&\bigl[-\frac\pi2,\frac\pi2\bigr]\bez\{0\}&\AS\begin{array}[t]{r}w+2\pi\MZ\\\pi-w+2\pi\MZ\end{array}
\end{array}\]
\caption{Properties of inverse hyperbolic and trigonometric functions (see Proposition~\ref{prop:inv-hyp-trig})}
\label{tab:hyp-trig}
\end{table}

Using Theorem~\ref{thm:main}, it is routine to prove all standard trigonometric identities. The following is just an
example.
\begin{Prop}\label{prop:trigid}
For all $z,w\in\MR+i\MRL$,
\begin{align}
\sin^2z+\cos^2z&=1,\label{eq:13}\\
\sin(z+w)&=\sin z\cos w+\cos z\sin w.\label{eq:17}
\end{align}
\end{Prop}
\begin{Pf}
\begin{align*}
\sin z\cos w+\cos z\sin w
&=\frac{(e^{iz}-e^{-iz})(e^{iw}+e^{-iw})+(e^{iz}+e^{-iz})(e^{iw}-e^{-iw})}{4i}\\
&=\frac{2e^{iz}e^{iw}-2e^{-iz}e^{-iw}}{4i}\\
&=\frac{e^{i(z+w)}-e^{-i(z+w)}}{2i}=\sin(z+w).
\end{align*}
We leave \eqref{eq:13} to the reader.
\end{Pf}

In order to work with the functions introduced in this section in inductive arguments and other reasoning within
$\vtc$, we need their $\tc$~approximations. We start with the inverse functions which are relatively straightforward
to approximate.
\begin{Prop}\label{prop:inv-trig-hyp-apx}
Each of the 12 functions $g(z)$ from Definition~\ref{def:inv-hyp-trig} has additive and multiplicative $\tc$ approximations for
$z\in\MQ(i)\cap\dom(g)$.
\end{Prop}
\begin{Pf}
We only sketch the arguments, leaving details to the reader. Let us start with additive approximations.

For $\artanh z$, we can just take $\frac12L_+\bigl(\frac{1+z}{1-z},n\bigr)$; similarly for $\arcoth z$.

Using an additive approximation for $\sqrt z$ (which exists by Lemmas \ref{lem:log-exp-apx} and~\ref{lem:add-mult-apx}), we can
compute an additive approximation to $z+\sqrt{z^2+1}$. In view of
$\bigl(z+\sqrt{z^2+1}\bigr)\bigl(z-\sqrt{z^2+1}\bigr)=-1$, we can bound the value of $z+\sqrt{z^2+1}$ away from~$0$ by
a $\tc$~function of~$z$, hence we can compute a multiplicative approximation of $z+\sqrt{z^2+1}$ using
Lemma~\ref{lem:add-mult-apx}. Plugging it into $L_+$, we obtain an additive approximation of $\arsinh z$. A similar
argument applies to $\arcosh z$: we apply $L_+$ to a multiplicative approximation of $z+\sqrt{z+1}\sqrt{z-1}$, making
sure that its $\sgn^+\Im$ is correct%
\footnote{We have $\sgn^+\Im\bigl(z+\sqrt{z+1}\sqrt{z-1}\bigr)=\sgn^+\Im z$ by a similar argument as in
Lemma~\ref{lem:inv-hyp-trig-cod}; if $\sgn^+\Im$ of the computed multiplicative approximation is wrong, we adjust its
imaginary part slightly by moving across the real axis, which can only improve the accuracy of the approximation. This
issue does not arise for $\arsinh$ as $\Re\bigl(z+\sqrt{z^2+1}\bigr)\ge0$ by Lemma~\ref{lem:inv-hyp-trig-cod}.}
so that we do not inadvertently cross the branch cut of $\log$.

The remaining functions reduce to these four, either directly, or using an additive approximation for $\frac\pi2$ such
as $\Im L_+(i,n)$.

In order to construct multiplicative approximations using Lemma~\ref{lem:add-mult-apx}, it suffices to bound $g(z)$ away
from~$0$ (except for the at most one $z$ where $g(z)$ is exactly~$0$). Let $f$ be the hyperbolic or trigonometric
function whose inverse $g$ is. Since $f(g(z))=z$, $g(z)$ is close to~$0$ only if $z$ is in the $f$-image of a small
neighbourhood of~$0$. If $f$ is $\sinh$, $\sin$, $\tanh$, or $\tan$, then $g(z)=0$ only if $z=0$, and
Theorem \ref{thm:main}~\ref{item:58} implies that there is a constant $c>0$ such that $\abs{f(w)-w}\le c\abs w^2$ for $\abs
w\le c$. Thus, $\abs{g(z)}\ge\min\{c_1\abs z,c_2\}$ for some constants $c_1,c_2>0$. If $f$ is $\coth$, $\cot$, $\csch$,
or $\csc$, then likewise $\abs{g(z)}\ge\min\{c_1\abs z^{-1},c_2\}$ (here, $g(z)$ is never~$0$).

For $\cosh$, we can use the definitions of $\exp_{\MQL(i)}$ and $\exp_{\MCL}$ to prove there is a constant $c>0$ such
that $\Abs[big]{e^w-(1+w+\frac12w^2)}\le c\abs w^3$ for $\abs w\le c$, thus
$\Abs[big]{\cosh w-(1+\frac12w^2)}\le c\abs w^3$, and $\abs{\arcosh z}\ge\min\{c_1\abs{z-1}^{1/2},c_2\}$ for some
constants $c_1,c_2>0$. Similarly for $\arccos$, $\arsech$, and $\arcsec$.
\end{Pf}

We now turn to approximation of hyperbolic and trigonometric functions. Many of these functions run into rough spots
(singularities or zeros) near integer multiples of $\pi$, $i\pi$, $\frac\pi2$, or $i\frac\pi2$, hence we will need to
parameterize their approximations to bound them away from the problematic points. Let us introduce some notation to this
end.
\begin{Def}\label{def:dist-piz}
For any $z\in\MC$ and $\alpha\in\MC_{\ne0}$, we put
\begin{align*}
\dist(z,\alpha\MZ)&=\min\bigl\{\abs{z-\alpha n}:n\in\MZ\bigr\}=\Abs[big]{z-\alpha\fcl{\Re(z/\alpha)}},\\
\dist(z,\alpha\MZ_{\ne0})&=\min\bigl\{\abs{z-\alpha n}:n\in\MZ_{\ne0}\bigr\}=
\begin{cases}\min\bigl\{\abs{z-\alpha},\abs{z+\alpha}\bigr\},&\Abs{\Re(z/\alpha)}\le\frac12,\\
\dist(z,\alpha\MZ),&\text{otherwise.}
\end{cases}
\end{align*}
(The expressions on the right-hand side exhibit that the minima exist.) More generally, if $\beta\in\MC$, let
$\dist(z,\beta+\alpha\MZ)=\dist(z-\beta,\alpha\MZ)$.
\end{Def}
\begin{Lem}\label{lem:dist-apx}
For any $\alpha\in\MQ(i)\bez\{0\}$ and $\beta\in\MQ(i)$, there are additive $\tc$ approximations
$D_{\pi(\beta+\alpha\MZ)}(z)$ of $\dist\bigl(z,\pi(\beta+\alpha\MZ)\bigr)$, and $D_{\pi\alpha\MZ_{\ne0}}(z)$ of
$\dist(z,\pi\alpha\MZ_{\ne0})$, for $z\in\MQ(i)$.
\end{Lem}
\begin{Pf}
Similarly to Lemma~\ref{lem:A-log-exp-apx}~\ref{item:A-40}, let $m\in\ML$ be such that
$2^{m-c}\ge\abs{\Re z}+\abs{\Im z}+1$ for a suitable constant~$c$ (depending on $\alpha$, $\beta$), and put
\begin{align*}
P&=\Im L_+(-1,n+m),\\
N&=\FCL{\Re\left(\alpha^{-1}\left(\frac zP-\beta\right)\right)},\\
D_{\pi(\beta+\alpha\MZ)}(z,n)&=\min\bigl\{A_+\bigl(z-P(\beta+\alpha(N+k)),n+1\bigr):k\in\{-1,0,1\}\bigr\}.
\end{align*}
Write $N'=\FCL{\Re\bigl(\alpha^{-1}\bigl(\frac z\pi-\beta\bigr)\bigr)}$ so that
$\dist\bigl(z,\pi(\beta+\alpha\MZ)\bigr)=\abs{z-\pi(\beta+\alpha N')}$. Since $\abs{P-\pi}\le2^{-(n+m)}$, we have
\[\Abs{\Re\left(\alpha^{-1}\left(\frac zP-\beta\right)\right)-\Re\left(\alpha^{-1}\left(\frac z\pi-\beta\right)\right)}
=\Abs{\Re\frac{(P-\pi)z}{\alpha\pi P}}\le1\]
as long as $2^c\abs\alpha\pi^2\ge1$ or so. Thus, $N-N'\in\{-1,0,1\}$. Moreover, for $k\in\{-1,0,1\}$,
\[\Abs[big]{\beta+\alpha(N+k)}\le\abs\beta+\abs\alpha\left(\Abs{\alpha^{-1}\left(\frac z\pi-\beta\right)}+\frac52\right)
\le2\abs\beta+\frac52\abs\alpha+\Abs{\frac z\pi}\le2^{m-1}\]
assuming $2^{c-2}\ge2\abs\beta+\frac52\abs\alpha$, hence
\begin{align*}
\bigl|A_+\bigl(z-P(\beta+\alpha(N+k)),n+1\bigr)-{}&\abs{z-\pi(\beta+\alpha(N+k))}\bigr|\\
&\le2^{-(n+1)}+\abs{P-\pi}\Abs[big]{\beta+\alpha(N+k)}\\
&\le2^{-(n+1)}+2^{-(n+m)}2^{m-1}=2^{-n}.
\end{align*}
Consequently, $\Abs{\dist(z,\pi(\beta+\alpha\MZ))-D_{\pi(\beta+\alpha\MZ)}(z,n)}\le2^{-n}$.

The argument for $\dist(z,\pi\alpha\MZ_{\ne0})$ is similar.
\end{Pf}

The reader may wonder why here and below, we work with the distance of $z$ to \emph{nonzero} integer multiples of $\pi$
($\pi i$, $\pi/2$, \dots) rather than \emph{all} integer multiplies. The reason is simply that the distance to the zero
multiple is trivial to estimate, being $\abs z$, thus we do not need a parameter to bound it. In the same vein, if we
are given a $z\in\MQ(i)\bez\MQ$, a suitable lower bound on $\dist(z,\pi\MZ)$ is given by $\abs{\Im z}$; thus, we only
need a parameter to bound $\dist(z,\pi\MZ)$ when $z\in\MQ$ (similarly for other configurations such as $\pi i\MZ$).
However, the actual arguments below are a bit more complicated, as we really need to bound values of $f(z)$ rather than
the distance of~$z$ to zeros or poles of~$f$. Let us also consider that a priori it is possible that $\pi\in\MQ$ (we
will discuss this in more detail below).
\begin{Prop}\label{prop:trig-hyp-apx}
If $f$ is any of the 12 functions from Definition~\ref{def:hyp-trig}, then $f$ has an additive $\tc$ approximation $F_+(z,r,n)$
and a multiplicative $\tc$ approximation $F_\times(z,r,n)$, parametrized by $r\in\ML$ that satisfies the conditions
specified in Table~\ref{tab:hyp-trig-apx} (the parameter of~$F_\times$ is subject to conditions from both the additive
and multiplicative columns). The additive approximation is defined for all $z\in\MQ(i)\cap\dom(f)$, while the
multiplicative approximation excludes $z\ne0$ such that $f(z)=0$, if any.
\end{Prop}
\begin{table}
\def\arraystretch{1.4}
\[\begin{array}{ccc}
f&\text{additive parameter}&\text{multiplicative parameter}\\
\hline
\sinh&\abs{\Re z}\le r&z\notin i\MQ\text{ or }D_{\pi i\MZ_{\ne0}}(z,r+1)\ge2^{-r}\\
\sin&\abs{\Im z}\le r&z\notin\MQ\text{ or }D_{\pi\MZ_{\ne0}}(z,r+1)\ge2^{-r}\\
\cosh&\abs{\Re z}\le r&z\notin i\MQ\text{ or }D_{\pi i(\frac12+\MZ)}(z,r+1)\ge2^{-r}\\
\cos&\abs{\Im z}\le r&z\notin\MQ\text{ or }D_{\pi(\frac12+\MZ)}(z,r+1)\ge2^{-r}\\
\tanh&z\notin i\MQ\text{ or }D_{\pi i(\frac12+\MZ)}(z,r+1)\ge2^{-r}
&z\notin i\MQ\text{ or }D_{\frac\pi2i\MZ_{\ne0}}(z,r+1)\ge2^{-r}\\
\tan&z\notin\MQ\text{ or }D_{\pi(\frac12+\MZ)}(z,r+1)\ge2^{-r}
&z\notin\MQ\text{ or }D_{\frac\pi2\MZ_{\ne0}}(z,r+1)\ge2^{-r}\\
\coth&z\notin i\MQ\text{ or }D_{\pi i\MZ_{\ne0}}(z,r+1)\ge2^{-r}
&z\notin i\MQ\text{ or }D_{\frac\pi2i\MZ_{\ne0}}(z,r+1)\ge2^{-r}\\
\cot&z\notin\MQ\text{ or }D_{\pi\MZ_{\ne0}}(z,r+1)\ge2^{-r}
&z\notin\MQ\text{ or }D_{\frac\pi2\MZ_{\ne0}}(z,r+1)\ge2^{-r}\\
\sech&z\notin i\MQ\text{ or }D_{\pi i(\frac12+\MZ)}(z,r+1)\ge2^{-r}&\abs{\Re z}\le r\\
\sec&z\notin\MQ\text{ or }D_{\pi(\frac12+\MZ)}(z,r+1)\ge2^{-r}&\abs{\Im z}\le r\\
\csch&z\notin i\MQ\text{ or }D_{\pi i\MZ_{\ne0}}(z,r+1)\ge2^{-r}&\abs{\Re z}\le r\\
\csc&z\notin\MQ\text{ or }D_{\pi\MZ_{\ne0}}(z,r+1)\ge2^{-r}&\abs{\Im z}\le r\\
\end{array}\]
\caption{$\tc$ approximation of hyperbolic and trigonometric functions}
\label{tab:hyp-trig-apx}
\end{table}
\begin{Pf}
An additive approximation of $\sinh z$ is given by $\frac12\bigl(E_+(z,r,n)-E_+(-z,r,n)\bigr)$. We can construct
a multiplicative approximation using Lemma~\ref{lem:add-mult-apx}---it suffices to bound $\sinh z$ away from~$0$ by a
$\tc$~function of $z=x+iy$ and~$r$. If, say, $\abs x\ge\frac14$, then
\[\abs{\sinh z}\ge\frac{\Abs[big]{\abs{e^z}-\abs{e^{-z}}}}2=\frac{e^{\abs x}-e^{-\abs x}}2\ge\frac{e^{1/4}-e^{-1/4}}2>0.\]
Write $y=\pi N+y_0$ and $z=\pi iN+z_0$, where $N\in\MZ$ and $\abs{y_0}\le\frac\pi2$. (Note that we cannot directly
compute $N$, $y_0$, or $z_0$.) We have
\[\abs{\Im\sinh z}=\frac{e^x+e^{-x}}2\sin\abs{y_0}\ge\sin\abs{y_0}.\]
Since $\sin t=\Im e^{it}$ is increasing on $\bigl[0,\frac\pi2\bigr]$ by Theorem \ref{thm:main}~\ref{item:57}, 
$\abs{\Im\sinh z}\ge\sin\frac14>0$ if, say, $\abs{y_0}\ge\frac14$.

In the remaining case, we have $\abs{z_0}=\dist(z,\pi i\MZ)\le\frac12$. Theorem \ref{thm:main}~\ref{item:58} easily implies
$\abs{\sinh z_0-z_0}\le\abs{z_0}^2$, thus $\abs{\sinh z}=\abs{\sinh z_0}\ge\frac12\abs{z_0}$. Now, if $x\ne0$, then
$\abs{z_0}\ge\abs x>0$, while if $x=0$ and $N=0$, then $\abs{z_0}=\abs y>0$ (unless $z=0$, in which case $\sinh z=0$).
Finally, if $x=0$ and $N\ne0$, then our assumption $D_{\pi i\MZ_{\ne0}}(z,r+1)\ge2^{-r}$ on the parameter~$r$ implies
$\abs{z_0}\ge2^{-(r+1)}$. Thus, all in all, if $z\ne0$, we can lower bound $\abs{\sinh z}$ by a constant or one of
$\frac12\abs x$, $\frac12\abs y$, or $2^{-(r+2)}$.

The arguments for $\sin$, $\cosh$, and $\cos$ are similar.

In view of Lemma~\ref{lem:mult-apx}~\ref{item:34}, the reciprocal of a suitable multiplicative approximation of $\cosh z$
gives a multiplicative approximation of $\sech z$; by Lemma~\ref{lem:add-mult-apx}, this also gives an additive
approximation with the same parametrization. For additive approximation, we can get rid of the $\abs{\Re z}\le r$
condition on~$r$ by observing that if $\abs{\Re z}\ge n$ (including the case $\Re z\notin\MRL$), then
$\abs{\sech z}\le2^{-n}$, hence we can take $0$ for the approximation; otherwise, we can use the original approximation
with $\max\{r,n\}$ in place of~$r$. A similar argument applies to $\sec$, $\csch$, and $\csc$.

A multiplicative approximation of $\tanh z$ can be computed by dividing multiplicative approximations of $\sinh z$
and $\cosh z$. We get rid of the $\abs{\Re z}\le r$ condition in the same way as for $\sech$: if $\abs{\Re z}\le n$, we
proceed with $\max\{r,n\}$ in place of~$r$, otherwise $\sgn\Re z$ can serve as approximation (since this is $\pm1$
rather than~$0$, it works as a multiplicative approximation as well as additive, unlike the case of $\sech$). For
additive approximation, we can relax the condition on~$r$ near $\pi i\MZ$: if $D_{\pi i\MZ}(z,n+2)\le2^{-(n+1)}$, then
$\abs{\tanh z}\le2^{-n}$, hence we can approximate it with~$0$; otherwise, we can proceed with $\max\{r,n+1\}$ in place
of~$r$. Thus, for $z\in i\MQ$, we only need to assume $D_{\pi i(\frac12+\MZ)}(z,r+1)\ge2^{-r}$ rather than
$D_{\frac\pi2i\MZ}(z,r+1)\ge2^{-r}$. Again, the arguments for $\tan$, $\coth$, and $\cot$ are analogous.
\end{Pf}

In the standard model, the picture becomes much simpler: multiplicative approximations are defined on full domains as
$f(z)=0$ with $z\ne0$ is impossible (i.e., $\pi$ is irrational and $\ML=\MN$), and more importantly, we do not need the
$D_{\dots}(z,r+1)\ge2^{-r}$ conditions on the parameters, as we can \emph{compute} $r\in\ML$ with these properties from
$z$ alone. The reason is that $\pi$ has a finite \emph{irrationality measure}: i.e., there is a constant~$\mu$ such that
\[\Abs{\frac pq-\pi}\ge\frac1{q^\mu}\]
for all but finitely many pairs $\p{p,q}\in\Z_{>0}^2$, which ensures that
\[\dist\left(\frac pq,\pi\Z_{\ne0}\right)=N\Abs{\frac p{qN}-\pi}\ge\frac N{(qN)^\mu}
\approx\frac1q\left(\frac\pi p\right)^{\mu-1},\]
where $N\in\Z_{>0}$ is the integer closest to $p/(q\pi)$. This was originally proved by Mahler~\cite{mahl:pi}; the
current best bound $\mu\le7{.}1032\dots$ is due to Zeilberger and Zudilin~\cite{zeil-zud:pi}. We do not know whether
$\vtc$ can prove that $\pi$ has a finite irrationality measure, or even the simpler and more fundamental property that
$\pi$ is irrational.

In fact, it turns out that the latter seemingly weaker property would be sufficient. First, observe that
in the argument above, we do not quite need the finiteness of the irrationality measure: it would be enough if $\pi$
has a finite ``quasipolynomial irrationality measure'', i.e., a constant $\nu$ such that
\[\Abs{\frac pq-\pi}=\Omega\bigl(2^{-(\log q)^\nu}\bigr).\]
Below, ``$\vtc$ proves \dots'' means, more precisely, ``\dots{} holds for all models of $\vtc$''.
\begin{Prop}\label{prop:irr-meas}
If $\vtc$ proves that $\pi\notin\MQ$, then there is a constant $\nu$ such that $\vtc$ proves
\[\forall p,q\in\MZ_{\ge2}\:\Abs{\frac pq-\pi}\ge2^{-\dlh q^\nu},\]
and consequently, that there is a $\tc$-computable lower bound on $\dist(z,\pi\MZ_{\ne0})$.
\end{Prop}
\begin{Pf}
The irrationality of $\pi$ is equivalent to
\[\forall p,q\in\MZ_{>0}\:\exists n\in\ML\:\Abs{\frac pq-P(n)}\ge2^{1-n},\]
where $P(n)=\Im L_+(-1,n)$ is an additive $\tc$ approximation of~$\pi$. This is a $\forall\exists\Sig0$ statement in
the language of~$\vtc$, thus if it is provable in $\vtc$, then $\vtc$ proves
\[\forall p,q\in\MZ_{>0}\:\exists n\le\bigl(\dlh p+\dlh q\bigr)^c\:\Abs{\frac pq-P(n)}\ge2^{1-n}\]
for some constant $c$ by Parikh's theorem, which implies
\[\forall p,q\in\MZ_{>0}\:\Abs{\frac pq-\pi}\ge2^{-(\dlh p+\dlh q)^c}.\]
Now, if $p\ge4q$, then $\Abs[big]{\frac pq-\pi}\ge4-\pi\ge\frac12$; otherwise, $\dlh p\le\dlh q+2$. This allows the
bound to be restated in terms of $q$ alone, using a possibly larger constant $\nu>c$.
\end{Pf}
\begin{Que}\label{que:pi}
Does $\vtc$ prove that $\pi$ is irrational?
\end{Que}

There are some fairly elementary proofs of the irrationality of~$\pi$, in particular the proof of Niven~\cite{niv:pi}.
This proof can be formalized in $\thry{I\Delta_0+EXP}$ with no difficulty, but it essentially relies on the totality of
exponentiation, and it is unclear how to prove the result in any weaker theory.

We observe that $\vtc$ easily proves that \emph{some} reals are irrational: e.g., the irrationality of $\sqrt2\in\MR$
follows in the usual way from the fact that any $u\in\MN$ can be written uniquely as $2^nv$ with $n\in\ML$ and
odd $v\in\MN$.

\section{Conclusion}\label{sec:conclusion}

Even though it has taken us some effort, we have successfully formalized in $\vtc$ the construction of complex $\exp$
and $\log$ as well as other elementary analytic functions, and we have shown that they share basic properties enjoyed
by the prototypes of these functions in the real world, adjusted in expected ways to an environment where integer
exponentiation is not necessarily total. We also managed to extend the definition of iterated multiplication to
Gaussian rationals. We may consider these results as further evidence that $\vtc$ is a robust and somewhat unexpectedly
powerful theory.

This is not to say that no problems remain. We already identified one missing fundamental piece of the puzzle, namely
Question~\ref{que:pi}: can $\vtc$ prove that $\pi$ is irrational? In view of Proposition~\ref{prop:irr-meas}, this is really asking
for a feasible proof that $\pi$ has a certain Diophantine inapproximability property a little weaker than finiteness of
irrationality measure.

This paper is a modest start of investigation of analytic functions in models of $\vtc$, and it opens various
possibilities of how it could be extended. We treated the elementary analytic functions which are an important but
small group of functions; there are many other functions of interest (``special functions'') that might deserve similar
attention, such as $\Gamma(z)$, $\zeta(z)$, Bessel functions, the error function, elliptic functions, etc.\
\cite{abr-steg,nist:fun}. An intriguing problem is whether we can formulate in $\vtc$ some form of a general
theory of analytic functions, i.e., basic results of complex analysis. Can $\vtc$ understand differentiation and
integration (or even simple differential equations such as Pfaffian chains)? We leave these open-ended questions for
possible future work.

\section*{Acknowledgements}

I am grateful to the anonymous referees for many helpful suggestions to improve the presentation of the paper.

The research was supported by the Czech Academy of Sciences (RVO 67985840) and GA~\v CR project 23-04825S.

\appendix

\section{Detailed construction of $\tc$ approximations}\label{sec:deta-constr-tc}

Here are the full proofs of Lemmas \ref{lem:log-exp-apx} and~\ref{lem:exp-apx}.
\begin{Lem}\label{lem:A-log-exp-apx}
We can construct $\tc$~functions $E_{\MCL}(z,r,n)$, $\M{SR}_{\MR}(x,n)$, $A_\times(z,n)$, $A_+(z,n)$, $\M{SR}_{\MC}(z,n)$, $L_D(z,r,n)$, $L_{\MR}(x,n)$,
$L_{\MC}(z,n)$, $\M{LE}(z,r,n)$, $E_\times(z,r,n)$, and $E_+(z,r,n)$ with the following properties.
\begin{enumerate}
\item\label{item:A-11}
$\E_{\MCL}(z,r,n)$ is a multiplicative approximation of $\exp_{\MCL} z$ for $z\in\MQL(i)$, parametrized by $r\in\ML$ such
that $\abs z\le r$.
\item\label{item:A-31}
$\M{SR}_{\MR}(x,n)$ is a multiplicative approximation of $\sqrt x$ for $x\in\MQ_{>0}$.
\item\label{item:A-71}
$A_\times(z,n)$ and $A_+(z,n)$ are multiplicative and additive (respectively) approximations of $\abs z\in\MR$ for
$z\in\MQ(i)$.
\item\label{item:A-30}
$\M{SR}_{\MC}(z,n)$ is a multiplicative approximation of $\sqrt z$ for $z\in\MQ(i)\bez\{0\}$.
\item\label{item:A-28}
$L_D(z,r,n)$ is an additive approximation of $\log_Dz$ for $z\in D^*_1(1)\cap\MQ(i)$, parametrized by $r\in\ML$ such
that $\abs{z-1}\le1-r^{-1}$.
\item\label{item:A-29}
$L_{\MR}(x,n)$ is an additive approximation of $\log_{\MR} x$ for $x\in\MQ_{>0}$.
\item\label{item:A-10}
$L_{\MC}(z,n)$ is an additive approximation of $\log_{\MC} z$ for $z\in\MQ(i)\bez\{0\}$.
\item\label{item:A-12}
$\M{LE}(z,r,n)$ is an additive approximation of $\log_{\MC}\exp_{\MCL} z$ for $z\in\MQL(i)$ with $\abs{\Im z}<1$,
parametrized by $r\in\ML$ such that $\abs z\le r$.
\item\label{item:A-40}
$E_\times(z,r,n)$ is a multiplicative approximation of $\exp z$ for $z\in\MQL+i\MQ$, parametrized by $r\in\ML$ such
that $\abs{\Re z}\le r$.
\item\label{item:A-41}
$E_+(z,r,n)$ is an additive approximation of $\exp z$ for $z\in\MQ_{\DML}+i\MQ$, parametrized by $r\in\ML$ such
that $\Re z\le r$.
\end{enumerate}
\end{Lem}
\begin{Pf}

\ref{item:A-11}: We know that $\Abs[big]{e(z,\max\{8r,n\})-\exp z}\le2^{-n}$ from the proof of Lemma~\ref{lem:exp-cauch},
hence
\[\Abs{\frac{e(z,\max\{8r,n\})}{\exp z}-1}\le2^{-n}\abs{\exp(-z)}\le2^{-n}\exp r\]
using Lemmas \ref{lem:exp-bdabs} and~\ref{lem:exp-homo}. Thus, the crude bound
\[\exp r\le2^{-8r}+e(r,8r)\le2^{-8r}+\sum_{j<8r}r^j\le r^{8r}\le2^{8r^2}\]
shows that it suffices to take $E_{\MCL}(z,r,n)=e(z,n+8r^2)$.

\ref{item:A-31}: In view of \cite[Thm.~6.8]{ej:vtc0iopen}, the existence of $\M{SR}_{\MR}$ is a consequence of
\cite[Prop.~3.7]{ej:vtc0iopen}. An explicit description can be given as follows. Let $x\in\MQ_{>0}$. Similarly to the
proof of Lemma~\ref{lem:add-mult-apx}, we can compute $m\in\MZL$ (in unary) such that $\frac25\le2^{-2m}x\le\frac85$ by a
$\tc$ function. Putting $u=1-2^{-2m}x$, we apply \cite[Thm.~5.5]{ej:vtc0iopen} to the polynomial $h(y)=-y^2+y-\frac u4$
(writing $y$ instead of~$x$ for the indeterminate to avoid clash with our~$x$), whose root is
$y=\frac12\bigl(1-\sqrt{1-u}\bigr)$, i.e., $\sqrt x=2^m(1-2y)$. Let $\alpha$, $b_j$, and $y_N$ be as in
\cite[Thm.~5.5]{ej:vtc0iopen}. Since $\alpha=\abs u\le\frac35$, we obtain
\[\abs{y-y_N}\le\frac52\left(\frac35\right)^N,\]
where
\[y_N=\sum_{j=1}^Nb_j\left(\frac u4\right)^N=\sum_{j=1}^N\binom{2(j-1)}{j-1}\frac1j\left(\frac u4\right)^j.\]
Since $(1-2y)^2\ge\frac25$, we have $\abs{1-2y}\ge\frac58$, hence
\[\Abs{\frac{2^m(1-2y_N)}{\sqrt x}-1}=\Abs{\frac{2(y-y_N)}{1-2y}}\le8\left(\frac35\right)^N.\]
Thus, we may take $\M{SR}_{\MR}(x,n)=2^m(1-2y_{2(n+3)})$.

\ref{item:A-71}: We can put $A_\times(z,n)=\M{SR}_{\MR}(x^2+y^2,n)$ and $A_+(z,n)=A_\times(z,n+m)$, where $z=x+iy$, and
$m\in\ML$ is such that $2^m\ge\abs x+\abs y$.

\ref{item:A-30}: Write $z=x+iy$, and assume first $x\ge0$. Note that
\[\sqrt z=\sqrt{\frac{\abs z+x}2}+i\frac y{2\sqrt{\frac{\abs z+x}2}}\]
as $y=\Im\bigl((\sqrt z)^2\bigr)=2\Re\sqrt z\Im\sqrt z$. Put
\begin{align*}
r&=A_\times(z,n+1),\\
u&=\M{SR}_{\MR}\bigl(\tfrac12(r+x),n+1\bigr),\\
\M{SR}_{\MC}(z,n)&=u+i\frac y{2u}.
\end{align*}
We have
\[\Abs{\frac{\frac12(r+x)}{\frac12(\abs z+x)}-1}=\Abs{\frac{r-\abs z}{\abs z+x}}
=\frac{\abs z}{\abs z+x}\Abs{\frac r{\abs z}-1}\le2^{-(n+1)}.\]
Put $n'=n+2$ and $\ep=2^{-n'}+2^{-2n'}$. Since
\[2\ep-\ep^2=2\cdot2^{-n'}+2^{-2n'}-2\cdot2^{-3n'}-2^{-4n'}\ge2\cdot2^{-n'}=2^{-(n+1)},\]
Lemma~\ref{lem:mult-apx} implies
\begin{align*}
\Abs{\frac{\sqrt{\frac12(r+x)}}{\sqrt{\frac12(\abs z+x)}}-1}&\le\ep,\\
\Abs{\frac u{\sqrt{\frac12(\abs z+x)}}-1}&\le2^{-(n+1)}+\ep+2^{-(n+1)}\ep\\
&=3\cdot2^{-n'}+3\cdot2^{-2n'}+2\cdot2^{-3n'}\le4\cdot2^{-n'}=2^{-n}.
\end{align*}
If $y=0$, we are done. Otherwise,
\[(1+2^{-n})(3\cdot2^{-n'}+3\cdot2^{-2n'}+2\cdot2^{-3n'})
=3\cdot2^{-n'}+15\cdot2^{-2n'}+14\cdot2^{-3n'}+8\cdot2^{-4n'}\le2^{-n}\]
as long as $n'\ge4$, i.e., $n\ge2$; thus,
\[\Abs{\frac{y/2u}{y\Big/2\sqrt{\frac12(\abs z+x)}}-1}=\Abs{\frac{\sqrt{\frac12(\abs z+x)}}u-1}\le2^{-n}\]
by Lemma~\ref{lem:mult-apx}~\ref{item:34}, and
\[\Abs{\frac{\M{SR}_{\MC}(z,n)}{\sqrt z}-1}\le2^{-n}\]
by Lemma~\ref{lem:mult-apx}~\ref{item:35}.

If $\Re z<0$, we may take $\M{SR}_{\MC}(z,n)=i\M{SR}_{\MC}(-z,n)\sgn^+y$.

\ref{item:A-28}: By the proof of Lemma~\ref{lem:log-cauch}, $L_D(z,r,n)=-\lambda(1-z,nr)$ works.

\ref{item:A-29}:
Given $x\in\MQ_{>0}$, we can compute $m\in\MZL$ such that $2^m\ge x>2^{m-1}$ as in~\ref{item:A-31}. Put $x'=2^{-m}x\in\bigl(\frac12,1\bigr]$, so that
\[\log_{\MR} x=\log_Dx'+m\ell_2=\log_Dx'-m\log_D\tfrac12.\]
Thus, it suffices to take $L_{\MR}(x,n)=L_D(x',2,n+1)-mL_D\bigl(\frac12,2,n+\abs m\bigr)$.

\ref{item:A-10}: Let $z\in\MQ(i)\bez\{0\}$. We have
\[\log_{\MC} z=8\log_S\sqrt[8]z=8\log_{\MR}\Abs{\sqrt[8]z}+8\log_D\sgn\sqrt[8]z.\]
Put $z_1=\M{SR}_{\MC}(z,n+7)$, $z_2=\M{SR}_{\MC}(z_1,n+7)$, $z_3=\M{SR}_{\MC}(z_2,n+7)$,
$x=\M{SR}_{\MR}(z_3\ob{z_3},n+7)$, $w=z_3/x$, and
\[L_{\MC}(z,n)=8L_{\MR}(x,n+5)+8L_D(w,2,n+5).\]
Write $n'=n+7$, $\ep=2^{-n'}+4\cdot2^{-2n'}$, and $\delta=2\cdot2^{-n'}+6\cdot2^{-2n'}$. Notice that
\[2^{-n'}+\ep+2^{-n'}\ep=2\cdot2^{-n'}+5\cdot2^{-2n'}+4\cdot2^{-3n'}\le\delta\]
and
\[2\ep-\ep^2=2\cdot2^{-n'}+7\cdot2^{-2n'}-8\cdot2^{-3n'}-16\cdot2^{-4n'}\ge\delta.\]
Thus, using Lemma~\ref{lem:mult-apx},
\[\Abs{\frac{z_1}{\sqrt z}-1}\le2^{-n'}\le\delta\le2\ep-\ep^2\]
implies
\[\Abs{\frac{\sqrt{z_1}}{\sqrt[4]z}-1}\le\ep,\]
whence
\[\Abs{\frac{z_2}{\sqrt[4]z}-1}\le2^{-n'}+\ep+2^{-n'}\ep\le\delta.\]
Repeating the same argument, we obtain
\[\Abs{\frac{z_3}{\sqrt[8]z}-1}\le\delta.\]
Using $\Abs[big]{\abs z-1}\le\abs{z-1}$ and Lemma~\ref{lem:mult-apx}, we get
\[\Abs{\frac x{\abs{\sqrt[8]z}}-1}\le2^{-n'}+\delta+2^{-n'}\delta\le3\cdot2^{-n'}+9\cdot2^{-2n'},\]
hence
\begin{align*}
\Abs{\log_{\MR} x-\log_{\MR}\abs{\sqrt[8]z}}
&=\Abs{\log_{\MR}\bigl(x/\abs{\sqrt[8]z}\bigr)}\\
&\le3\cdot2^{-n'}+9\cdot2^{-2n'}+(3\cdot2^{-n'}+9\cdot2^{-2n'})^2\\
&\le4\cdot2^{-n'}=2^{-n-5}
\end{align*}
using Lemmas \ref{lem:log-r-homo} and~\ref{lem:log-disc} (eq.~\eqref{eq:3}), thus
\[\Abs{L_{\MR}(x,n+5)-\log_{\MR}\abs{\sqrt[8]z}}\le2^{-n-4}.\]
Using similar arguments, we obtain $\Abs[big]{w/\sgn z_3-1}\le2^{-n'}+O(2^{-2n'})$,
$\Abs[big]{w/\sgn\sqrt[8]z-1}\le3\cdot2^{-n'}+O(2^{-2n'})$ (which implies $\abs{w-1}\le\frac12$ as
$\abs{\sgn\sqrt[8]z-1}\le0{.}41$ using Lemma~\ref{lem:sqrt-iter} and the proof of Lemma~\ref{lem:log-s}~\ref{item:13}), and
$\Abs[big]{\log_Dw-\log_D\sgn\sqrt[8]z}\le3\cdot2^{-n'}+O(2^{-2n'})\le2^{-n-5}$, whence
\[\Abs{L_D(w,2,n+5)-\log_D\sgn\sqrt[8]z}\le2^{-n-4},\]
which implies $\abs{L_{\MC}(z,n)-\log_{\MC} z}\le8(2^{-n-4}+2^{-n-4})=2^{-n}$.

\ref{item:A-12}: We put $\M{LE}(z,r,n)=L_{\MC}\bigl(E_{\MCL}(z,r,n+2),n+1\bigr)$. By~\ref{item:A-11}, we have
\[E_{\MCL}(z,r,n+2)=(1+w)\exp z\]
for some $w$ such that $\abs w\le2^{-(n+2)}$. We have $\Re\exp z>0$ by Lemma~\ref{lem:logexp-hom}, and trivially
$\Re(1+w)>0$, thus
\[\Abs[big]{\log_{\MC} E_{\MCL}(z,r,n+2)-\log_{\MC}\exp z}=\abs{\log_D(1+w)}\le2^{-(n+2)}+2^{-2(n+2)}\le2^{-(n+1)}\]
by Lemmas \ref{lem:log-c-homo}, \ref{lem:log-c-s}, and~\ref{lem:log-disc} (eq.~\eqref{eq:3}), while
\[\Abs[big]{\M{LE}(z,r,n)-\log_{\MC} E_{\MCL}(z,r,n+2)}\le2^{-(n+1)}\]
by~\ref{item:A-10}.

\ref{item:A-40}: We compute $m\in\ML$ such that $m\ge3$ and $2^m\ge\abs{\Im z}$, and put
\begin{align*}
P&=\Im L_{\MC}(-1,n+m+1),\\
N&=\FCL{\frac{\Im z}{2P}},\\
E_\times(z,r,n)&=E_{\MCL}(z-2PNi,r+4,n+2).
\end{align*}
Put $w=z-2PNi$. We have $\abs{\pi-P}\le2^{-n-m-1}$ by \ref{item:A-10}, thus $\abs{\Im w}\le P\le4$ using
Proposition~\ref{prop:pi-apx}, and $\abs w\le r+4$. Consequently,
\[\Abs{\frac{E_\times(z,r,n)}{\exp w}-1}\le2^{-n-2}\]
by~\ref{item:A-11}. Moreover,
\[\Abs[big]{2(\pi-P)N}\le2^{-n-m}\left(\frac{\abs{\Im z}}{2P}+\frac12\right)
\le2^{-n-m}\frac{2^m}4=2^{-n-2}\]
using Proposition~\ref{prop:pi-apx}, hence
\begin{align*}
\Abs{\frac{\exp w}{\exp z}-1}&=\Abs[big]{\exp(-2PNi)-1}=\Abs{\exp_{\MCL}\bigl(2(\pi-P)Ni\bigr)-1}\\
&\le\Abs[big]{2(\pi-P)N}+\Abs[big]{2(\pi-P)N}^2\le2^{-n-1}
\end{align*}
using Lemma \ref{thm:exp-homo}~\ref{item:6},
and
\[\Abs{\frac{E_\times(z,r,n)}{\exp z}-1}\le2^{-n-1}+2^{-n-2}+2^{-2n-3}\le2^{-n}\]
by Lemma~\ref{lem:mult-apx}~\ref{item:32}.

\ref{item:A-41}: We put
\[E_+(z,r,n)=\begin{cases}
E_\times\bigl(z,\max\{n,r\},n+2r\bigr)&\text{if $-n\le\Re z\le r$,}\\
0&\text{otherwise.}
\end{cases}\]
If $\Re z\le-n$, we have
\[\abs{\exp z}=\exp\Re z\le\exp(-n)\le2^{-n}\]
by Lemmas \ref{thm:exp-homo} and~\ref{cor:e-apx} (this bound holds trivially if $\Re z\notin\MRL$). If $-n\le\Re z\le r$,
\ref{item:A-40} gives
\[\Abs{E_\times\bigl(z,\max\{n,r\},n+2r\bigr)-\exp z}\le2^{-n-2r}\abs{\exp z}
\le2^{-n-2r}\exp r\le2^{-n}\]
using Lemmas \ref{thm:exp-homo} and~\ref{cor:e-apx} again.
\end{Pf}

\bibliographystyle{mybib}
\bibliography{mybib}
\end{document}